\documentclass[aps,12pt,eqsecnum,tightenlines,showpacs,nofootinbib,endfloats,amsmath,amssymb]{revtex4}
\usepackage{graphicx}
\usepackage{bm}

\begin{document}

\title{Three dimensional stationary cyclic symmetric
Einstein--Maxwell solutions; black holes\footnote{This work has been
published as: Alberto A. Garc\'{\i}a, Annals of Physics {\bf 324}
(2009) 2004--2050.}} \topmargin -2cm

\author{Alberto A. Garcia--Diaz}
\altaffiliation{aagarcia@fis.cinvestav.mx}
 \affiliation{
Departamento~de~F\'{\i}sica,
~Centro~de~Investigaci\'on~y~de~Estudios~Avanzados~del~IPN, Apdo.
Postal 14-740, 07000 M\'exico DF, M\'exico, and \\
Department of Physics, University of California, Davis, CA 95616,
USA.\\}

\begin{abstract}
From a general metric for stationary cyclic symmetric gravitational
fields coupled to Maxwell electromagnetic fields within the
$(2+1)$--dimensional gravity the uniqueness of wide families of
exact solutions is established, among them, all uniform
electromagnetic solutions possessing electromagnetic fields with
vanishing covariant derivatives, all fields having constant
electromagnetic invariants $F_{\mu\nu}\,F^{\mu\nu}$ and
$T_{\mu\nu}\,T^{\mu\nu}$, the whole classes of hybrid
electromagnetic solutions, and also wide classes of stationary
solutions are derived for a third order nonlinear key equations.
Certain of these families can be thought of as black hole solutions.
For the most general set of Einstein--Maxwell equations, reducible
to three non--linear equations for the three unknown functions, two
new classes of solutions--having anti-de Sitter spinning metric
limit--are derived. The relationship of various families with those
reported by different authors' solutions has been established. Among
the classes of solutions with cosmological constant a relevant place
occupy: the electrostatic and magnetostatic Peldan solutions, the
stationary uniform and spinning Clement classes, the constant
electromagnetic invariant branches with the particular
Kamata--Koikawa solution, the hybrid cyclic symmetric stationary
black hole fields, and the non--less important solutions generated
via $SL(2,R)$ transformations where the Clement spinning charged
solution, the Martinez--Teitelboim--Zanelli black hole solution, and
Dias--Lemos metric merit mention.

 \vspace{0.5cm}
\end{abstract}

\date{January 23, 2009}

 \maketitle

 \tableofcontents


\section{Introduction}
During the last two decades three--dimensional gravity has received
some attention, in particular, in topics such as: black hole
physics, search of exact solutions, quantization of fields coupled
to gravity, cosmology, topological aspects, and others. This
interest in part has been motivated by the discovery, in 1992, of
the $2+1$ stationary circularly symmetric black hole solution by
Ba\~nados, Teitelboim and Zanelli ~\cite{BTZ}--the BTZ black hole--
see also \cite{BTZ93,CangemiLM93,Carlip-cqg-95}, which possesses
certain features inherent to $3+1$ black holes. On the other hand,
it is believed that $2+1$ gravity may provide new insights towards a
better understanding of the physics of $3+1$ gravity. In the
framework of exact solutions in $2+1$ gravity the list of references
on the topic is extremely vast; one finds works on point masses,
cosmological and perfect fluid solutions, dilaton and string fields,
and on electromagnetic fields coupled to gravity, among others.

The purpose of this contribution is to provide a new approach on the
search of electromagnetic--gravitational solutions to the
Einstein--Maxwell fields of the $2+1$ gravity in the presence of a
cosmological constant, allowing for stationary and cyclic
symmetries. The search and interpretation of this kind of solutions
has been the goal and realm of several authors' investigations
starting from quite different perspectives and using a variety of
approaches, which sometime have brought about duplication of results
and efforts. The main objective of this work is to derive general
families of stationary (static) cyclic symmetric solutions to the
Einstein--Maxwell field equations, establishing their relationship
with known to-date solutions, and to point out the families allowing
for black hole interpretation.

The outline of this work is as follows:{\newline}
Sec.~\ref{electromagnetic} contains the theorem on the existence of
possible classes of electromagnetic fields for stationary cyclic
symmetric 2+1 spacetimes; $\ast
\bm{F}=a\bm{dt}+b\bm{d\phi}+c{g_{rr}}/{\sqrt{-g}}\bm{dr}$ fully
characterizes the families of Maxwell electromagnetic fields. In
Sec.~\ref{general} the canonical metrics to be used and the
corresponding Einstein--Maxwell equations are explicitly given.
Sec.~\ref{static} is devoted to the determination of the general
static solutions. Sec.~\ref{uniform} deals with the determination of
the uniform electromagnetic fields, i.e., those which possess
vanishing covariant derivatives $F_{\alpha\beta;\gamma}=0$. The set
of stationary solutions for constant invariant
$F_{\mu\nu}\,F^{\mu\nu}$, and consequently, due to the structure of
the electromagnetic fields, with constant energy momentum invariants
$T_{\mu}^{\mu}$ and $T_{\mu\nu}\,T^{\mu\nu}$, are derived in
Sec.~\ref{constant}. The so called self (anti)--dual fields are
derived in Sec.~\ref{AsdMf}. The determination of the general
stationary solution for the electromagnetic field ${\ast {\bm{F}}}=
c{g_{rr}}/{\sqrt{-g}}\bm{dr}$ is accomplished in Sec.~\ref{hybrid}.
A master equation--a single nonlinear fourth order ordinary equation
subsequently reducible to a third order one--is established for the
determination of the stationary fields having pure electric or
magnetic features and particular solutions to it are reported in
Sec.\ref{StaMagab}. In Sec.~\ref{GENSTATIONARY} we search for
general stationary solutions for the electromagnetic field ${\ast
{\bm{F}}}= a\bm{dt}+b\bm{d\phi}$; solutions within a wide class of
structural functions allowing for logarithms are derived and their
uniqueness demonstrated. Sec.~\ref{GENERATING} is devoted to
stationary solutions generated via $SL(2,R)$--transformations.
Finally, we end with some concluding remarks in Sec.~\ref{Remarks}.

As it has been stated above, the main objective of this report is
to demonstrate, via straightforward integration of the field
equations, the completeness of electromagnetic classes of stationary
cyclic symmetric solutions. A full characterization of the physical
contents of these solutions would require a considerable more
extension of this paper, for this reason, some short comments are
made in this respect close to those contained in the related
references if there are any, and also about the new found families
with special emphasis on their black hole feature.

\section{Electromagnetic field for stationary cyclic
 symmetric $2+1$ spacetimes}\label{electromagnetic}

To begin with, we consider a stationary cyclic symmetric spacetime
with signature (-,+,+), i.e., a space endowed with stationary
symmetry $\bm {k}=\bm{\partial_{t}}$, $\bm {k}\cdot \bm {k} <0$,
such that $\pounds_{\bm {k}}\bm {g}=0$, and cyclic symmetry
$\bm{m=\partial_{\phi}}$, $\bm {m}\cdot \bm {m} >0$, such that
$\pounds_{\bm{m}}\bm {g}=0$, with closed integral curves from $0$ to
$2\pi$, which in turn commute $[\bm {k},\bm {m}]=0$. Hence the
Killing vector fields $\bm {k}$ and $\bm {m}$ generate the group
$SO(2)\times {R}$. The electromagnetic field, described by the
antisymmetric tensor field $\bm
{F}=\frac{1}{2}F_{\mu\nu}\bm{dx}^{\mu}\wedge\bm{dx}^{\nu}$, is
assumed to be stationary cyclic symmetric, i.e., $\pounds_{\bm
{k}}\bm {F}=0=\pounds_{\bm {m}}\bm {F}$. It should be pointed out
that, in contrast to the general $3+1$ stationary cyclic symmetric
spacetime, any $2+1$ stationary cyclic symmetric spacetime is
necessarily circular, i.e., the circularity conditions
\begin{equation}\label{fr41}
\bm {k}\wedge \bm {m}\wedge \bm {dk}=0=\bm {k}\wedge \bm {m}\wedge
\bm {dm}
\end{equation}
are identically fulfilled because of their 4--form character and
hence there exists the discrete symmetry when simultaneously
$t\rightarrow -t$ and $\phi \rightarrow -\phi$. One may find a
coordinate system such that the metric tensor components $\bm
{g}(\bm{k\, dr})=0$ and $\bm {g}(\bm{m \,dr})=0$, where the
coordinate direction $\bm{dr}$ is orthogonal to the surface spanned
by $\bm {k}\wedge \bm {m}$. Commonly one introduces the
coordinate system $\{t,\phi,r\}$ in
(2+1)-dimensional gravity .\\
The main goal of this section is to demonstrate of the following
theorem.

{\bf Theorem:}\label{theorem} The general form of stationary cyclic
symmetric electromagnetic fields in 2+1 dimensions is given by
\begin{equation}\label{field1}
\ast \bm{F}=a\bm{dt}+b\bm{d\phi}+c\frac{g_{rr}}{\sqrt{-g}}\bm{dr},
\end{equation}
where the  constants $a,b$ and $c$ are subjected, by virtue of the
Ricci circularity conditions, to the equations
\begin{equation}\label{cond1}
a\,c=0=b\,c,
\end{equation}
which gives rise to two disjoint branches
\begin{equation}\label{field1a}
c\neq0,\ast\bm{F}=c\frac{g_{rr}}{\sqrt{-g}}\bm{dr},
\end{equation}
and
\begin{equation}\label{field1b}
c=0,\ast\bm{F}=a\bm{dt}+b\bm{d\phi},
\end{equation}
with its own sub-classes $a=0$ or $b=0$.\\
To establish that the field $\ast\bm{F}$ possesses the form given
by Eq.~(\ref{field1}) one uses the source--free Maxwell equations
\begin{equation}\label{field1ba}
\bm{d}\bm{F}=0=\bm{d}\ast\bm{F},
\end{equation}
where $\ast$ denotes the Hodge star operation.\\
Let us evaluate the
exterior derivative of the $t$--component $\ast \bm{F(\bm{k})}$ of
$\ast \bm{F}$,
\begin{eqnarray}
&&\bm{d{\ast \bm{F(\bm{k})}}}=\bm{d}{\,i_{\bm{k}}\ast{\bm{F}}}
=\pounds_{\bm{k}}\ast \bm{F}-i_{\bm{k}}\bm{d}\ast \bm{F}=
0-0\rightarrow \ast \bm{F(\bm{k})}=:a={\rm constant},
\end{eqnarray}
the first zero arises from the stationary character of the field
$\bm{F}$, while the second one corresponds to the Maxwell equation.
Similarly, for the $\phi$--component $\ast \bm{F(\bm{m})}$ one has
\begin{eqnarray}
&&\bm{d{\ast\bm{F(\bm{m})}}}=\bm{d}{\,i_{\bm{m}}\ast{\bm{F}}}
=\pounds_{\bm{m}}\ast \bm{F}-i_{\bm{m}}\bm{d}\ast \bm{F}=
0-0\rightarrow \ast \bm{F(\bm{m})}=:b={\rm constant}.
\end{eqnarray}
In this manner we have established that the $t$ and $\phi$
components of the dual field $\ast\bm{F}$ are constants given
correspondingly by $a$ and $b$. The
component of $\ast\bm{F}$ along the vector direction
$\bm{\partial_{r}}$ remains to be determined. For this purpose, consider the
$t\phi$--component $\bm{F(k,m)}$ of the field $\bm{F}$, which can be
expressed as $\bm{F(k,m)}=i_{\bm{m}}i_{\bm{k}}\bm{F}
=(-i_{\bm{m}}i_{\bm{k}}\ast\ast\bm{F}
=i_{\bm{m}}\ast(\bm{k}\wedge\ast\bm{F})
=\ast(\bm{m}\wedge\bm{k}\wedge\ast\bm{F}))
=-\ast\bm{F}(\ast(\bm{k}\wedge\bm{m}))$,
thus its derivative yields
\begin{eqnarray}
\bm{d}\bm{F}(\bm{k},\bm{m})&&=\bm{d}(i_{\bm{m}}i_{\bm{k}}\bm{F})
=\bm{d}\,i_{\bm{m}}\,(i_{\bm{k}}\bm{F}) \nonumber\\
&&=(\pounds_{\bm{m}}-i_{\bm{m}}\bm{d})(i_{\bm{k}}\bm{F})
=i_{\bm{k}}\pounds_{\bm{m}}\bm{F}+i_{[\bm{k},\bm{m}]}\bm{F}-
i_{\bm{k}}(\pounds_{\bm{m}}-i_{\bm{k}}\bm{d})\bm{F}\nonumber\\
&&=0\rightarrow \bm{F(\bm{k},\bm{m})}=:c={\rm constant.}
\end{eqnarray}
Since the constant $c$ can be written as
$c=-\ast\bm{F}(\ast(\bm{k}\wedge\bm{m}))$, to determine it, one
evaluates $\ast(\bm{k}\wedge\bm{m})$. Identify the Killing
vectors accordingly with $\bm{k}=\bm{\partial_{t}}$ and
$\bm{m}=\bm{\partial_{\phi}}$, then
\begin{eqnarray}
\ast(\bm{k}\wedge\bm{m})=-\sqrt{-g}\bm{dr}
=-\sqrt{-g}\,g^{rr}\bm{\partial_{r}}\nonumber\\
\end{eqnarray}
thus
\begin{eqnarray}
c&&=-\ast\bm{F}(-\sqrt{-g}\,g^{rr}\bm{\partial_{r}})=
\sqrt{-g}\,g^{rr}\ast\bm{F(\bm{\partial_{r}})}.
\end{eqnarray}
Conversely, from the above--mentioned relation one determines the
$r$--component of the field $\ast\bm{F}$, namely
$\ast\bm{F(\bm{\partial_{r}})}=\frac{c}{\sqrt{-g}}g_{rr}$. In this
manner, the structure  of $\bm{F}$, explicitly given by
~(\ref{field1}), has been established.

The vanishing conditions (\ref{cond1}) straightforwardly arise from
the Ricci circularity conditions $\bm{m}\wedge \bm{k}\wedge
\bm{R}(\bm{k})=0$ and $\bm{k}\wedge \bm{m}\wedge \bm{R}(\bm{m})=0$.
Correspondingly, the vanishing conditions $a\,c=0=b\,c$ can be
established immediately, as we shall see in the next section, from
the Einstein equations $R_{\mu\nu}-\frac{1}{2}R\,g_{\mu\nu}=\kappa
T_{\mu\nu}-\Lambda\,g_{\mu\nu}$, where the electromagnetic energy--
momentum tensor components are defined through the electromagnetic
field $F_{\mu\nu}=-F_{\nu\mu}$ as $4\pi\,T_{\mu
\nu}=F_{\mu\sigma}{F_{\nu}}^{\sigma}
-\frac{1}{4}g_{\mu\nu}F_{\alpha\sigma}F^{\alpha\sigma}$. A first
formulation of this theorem with an outline of its demonstration has
been reported in \cite{AyonCG04}.

\section{General metric and Einstein
equations}\label{general}

In general, in $(2+1)$--dimensional gravity any stationary cyclic
symmetric metric can be given as
\begin{eqnarray}\label{genericmetr}
\bm{g}&=&g_{tt}{\bm{dt}}^2+2g_{t\phi}{\bm{dt}}\,{\bm{d\phi}}
+g_{\phi\phi}\,{\bm{d\phi}}^2 +g_{rr}{\bm{dr}}^2.\nonumber\\
\end{eqnarray}
When a Maxwell electromagnetic field is present, the field tensor,
as we established previously, possesses the structure
\begin{equation}\label{elecmag1}
F^{{\alpha}{\beta}}=\frac{1}{{\sqrt{-g}}}\left[
\begin {array}{ccc} 0&{b}&
-\frac {{c} g_{rr}}{\sqrt{-g}}\\
\noalign{\medskip}-{b}&0&{a}\\
\noalign{\medskip}\frac {{c} g_{rr}}{\sqrt{-g}}&-{a}&0\end {array}
\right],
\end{equation}
where $g:=\det(g_{\mu\nu})$, which makes apparent the fulfillment of
the divergence equation
$$ ({\sqrt{-\det(g_{\mu\nu})}}F^{{\alpha}{\beta}})_{;\beta}=0 $$ for constants
${a, b,}$ and $c$. The Maxwell electromagnetic
energy--momentum tensor is given as usual as
\begin{equation}\label{elenergy}
{T_{\mu}}^{\nu}=\frac{1}{4\pi}(F_{\mu\sigma}\,F^{\nu\sigma}
-\frac{1}{4}\delta_{\mu}^{\nu}F_{\tau\sigma}F^{\tau\sigma}).
\end{equation}

\subsection{Canonical metrics and
Einstein--Maxwell equations}\label{canonical}

Without loss of generality one can choose the coordinates for a
stationary cyclic symmetric $2+1$ metric, developed with respect
to the cyclic symmetry $\bm{m=\partial_{\phi}}$, in such a way
that it becomes
\begin{eqnarray}\label{metricsc-i}
\bm{g}&&=-\frac{F(r)}{H(r)}\bm{dt^2}+\frac{\bm{dr^2}}{F(r)}
+H(r)\left[\bm{d\phi}+W(r)\bm{dt}\right]^2.
\end{eqnarray}
On the other hand, if one chooses the stationary symmetry
$\bm{k=\partial_{t}}$ as the fundamental Killing field, the
stationary cyclic symmetric $2+1$ metric can be written as
\begin{eqnarray}\label{metricsc-ii}
&&{\it g}=-\frac{F(r)}{h(r)}\left[\bm{dt}
-{\omega(r)}\,\bm{d\phi}\right]^2+h(r)\,{\bm{d\phi}}^2
+\frac{{\bm{dr}}^2}{{F(r)}},\nonumber\\
&&{F}=F,\,
{H}={h}-\frac{F}{h}\omega^2,\,{W}\frac{H}{F}=\frac{\omega}{h},
h=\frac{H\, F}{F-W^2\,H^2}.
\end{eqnarray}
Mostly we will use the metric~(\ref{metricsc-i}) in the forthcoming
developments, but occasionally the metric
representation~(\ref{metricsc-ii}) will be used . When doing so, the
derived expressions will be given in terms of the set $\{F(r), h(r),
\omega(r)\}$ of structural functions. Omitting the dependence of the
structural functions on the variable $r$, the Maxwell
electromagnetic field contravariant tensor is given by
\begin{equation}\label{contfield}
\left(F^{\mu\nu} \right)= \left[
\begin {array}{ccc} 0&b&- \frac{c }{F}\\
\noalign{\medskip}-b&0&a\\
\noalign{\medskip}\frac{c }{F}&-a&0\end {array} \right],
\end{equation}
where $a$, $b$, and $c$ are constants related with the character of
the field. For instance, if only $b$ is different from zero, while
$a$ and $c$ vanish, the field is called (pure) electric field. When
$a\neq 0$, $b=0=c$, one deals with a pure magnetic field; other
possibilities do not receive a particular name. The covariant
components $F_{\mu\nu}$ of the field tensor are given by:
\begin{eqnarray}\label{covF}
&&F_{tr}=-b/H-W\,H(a-b\,W)/F,\, F_{t\phi}=c,\,F_{r\phi}=H(a-b\,W)/F.
\end{eqnarray}
\noindent The electromagnetic field quadratic invariant
$FF:=F_{\mu\nu}F^{\mu\nu}$ is given by
\begin{equation}\label{invFfield}
FF=-2\,{\frac {{c}^{2}}{F}}+2\,{\frac {{H} \left( a-Wb \right)
^{2}}{ F}}-2\,{\frac {{b}^{2}}{{H}}}.
\end{equation}
Notice that if one uses the vector--potential description of
the electromagnetic field
\begin{eqnarray}\label{apot}
\bm {F}:=\frac{1}{2}F_{\mu\nu}{\bm{dx}}^{\mu}\wedge{\bm{dx}}^{\nu}
=\bm{d}\left(A_{\mu}{\bm{dx}}^{\mu}\right)=:\bm{d\,A}\nonumber\\
\end{eqnarray}
one would have
\begin{eqnarray}\label{apota}
\bm{F}&=&\bm{d}\,{
\int^{r}[(\frac{1}{H}-\frac{H}{F}W^2)b+\frac{H}{F}W\,a]dr \times
\bm{dt}}+\bm{d}\,\int^{r}[-\frac{H}{F}W\,b
+\frac{H}{F}\,a]dr\times\bm{d\phi}\nonumber\\
&+&\bm{d}\left[\frac{1}{2}c(t\bm{d\phi}-\phi
\bm{dt})\right]=\bm{d\,A}.
\end{eqnarray}
The energy--momentum tensor associated with the
metric~(\ref{metricsc-i}) occurs to be
\begin{equation}\label{energyFfield}
 \left(T^{\mu}_{\nu}\right)= \left[
\begin {array}{ccc} \frac{{b}^{2}
({H}^{2}{W}^{2}-F)-{a}^{2}{H}^{2}-{c}^{
2}{H}}{8\pi\,F \,{H}}&{\frac {ac}{4\,\pi }}&-\frac{a\left[
b({H}^{2}{W}^{2}-F)-a{H}^{2}W \right]}{4\pi\,F
\,{H}}\\
\noalign{\medskip}{\frac {c{H} \left( W\,b-a \right)}{4\,\pi {F}^{2}
}}&\frac{-{b}^{2}F+{H}^{2}\left( W\,b-a
\right)^2+{c}^{2}{H}}{8\pi\,F \,{H}}&-\frac{c\left[
b({H}^{2}{W}^{2}-F)-a{H}^{2}W \right]}{4\pi\,F^2
\,{H}}\\
\noalign{\medskip}\frac {b{H} \left( W\,b-a \right)}{4\,\pi {F}
}&{\frac {bc}{4\,\pi
}}&-\frac{{b}^{2}({H}^{2}{W}^{2}-F)-{a}^{2}{H}^{2}+{c}^{
2}{H}}{8\pi\,F \,{H}}\end {array} \right]
\end{equation}
and possesses the trace $T:=T^{\mu}_{\mu}$ given by
\begin{eqnarray}\label{traceFfield}
T&&=-\frac{1}{8\,\pi}\,{\frac {{c}^{2}}{F }}
+\frac{1}{8\,\pi}\,{\frac {{H} \left( a-W\,b
 \right) ^{2}}{F }}-\frac{1}{8\,\pi}\,{\frac
 {{b}^{2}}{{H}}}=\frac{1}{16\pi}\,FF,
\end{eqnarray}
and the electromagnetic energy momentum quadratic invariant
$TT:=T_{\mu\nu}T^{\mu\nu}$
\begin{eqnarray}\label{quadTT}
TT&&=\frac{3}{64\pi^2}\frac{\left[H^2(a-b\,W)^2-b^2F-c^2H\right]^2}{F^2H^2}=\,\frac{3}{256\pi^2}\,FF^2.
\end{eqnarray}
The Einstein--Maxwell equations
\begin{equation}
E_{\mu\nu}:= R_{\mu\nu}-\frac{R}{2}g_{\mu\nu}+\Lambda
g_{\mu\nu}-8\pi\,T_{\mu\nu}=0
\end{equation}
for a negative cosmological constant $\Lambda=-1/{l^2}$ explicitly
read:
\begin{subequations}\label{Einstein}
\begin{eqnarray}\label{Eins11}
{E_{t}}^{t}&&=\frac{1}{2}\,F\frac {H_{,r,r}}{H}+\frac{1}{4}\,\frac
{H_{,r}F_{,r}}{H}-\frac{1}{4}\frac{F}{H^2}{H_{,r}}^2
+\frac{1}{2}\,{H}^{2}W\,W_{,r,r}
+{H}WW_{,r}H_{,r}+\frac{1}{4}\,{H}^{2}{W_{,r}}^{2}\nonumber\\
&& +{b}^{2}\frac
{F-{H}^{2}{W}^{2}}{F\,H}+\frac{{c}^{2}}{F}+\frac{{a}^{2}H}{F}
-\frac{1}{{l}^2},
\end{eqnarray}
\begin{eqnarray}\label{Eins12}
{E_{t}}^{r}&&=-2\,ca,
\end{eqnarray}
\begin{eqnarray}\label{Eins13}
{E_{t}}^{\Phi}&&=\frac{1}{2}\,W\,F_{,r,r}-F\,W\,\frac
{H_{,r,r}}{H}-W\,{\frac {H_{,r}F_{,r}}{H}} -2\,a^2\,\frac{{H}\,W}{F}
-2\,a\,b\,\frac{ F-{H}^{2}{W}^{2}}{F{H}}\nonumber\\&&-\frac{1}{2}
\left( F+{H}^{2}{W}^{2}
 \right) \left(W_{,r,r}+2 W_{,r}\frac{H_{,r}}{H}\right)
 +F\,W\,{\frac {{H_{,r}}^{2}}{{H}^{2}}} -{H}^{2}W{W_{,r}}^{2},
\end{eqnarray}
\begin{eqnarray}\label{Eins21}
{E_{r}}^{t}&&=-2c\,\frac {{H} }{{F}^{2}}\left( W b-a \right),
\end{eqnarray}
\begin{eqnarray}\label{Eins22}
{E_{r}}^{r}&&=\frac{1}{4}\,\frac
{H_{,r}F_{,r}}{H}-\frac{1}{4}\,F\frac
{{H_{,r}}^{2}}{{H}^{2}}+\frac{1}{4} \,{H}^{2}{W_{,r}}^{2}+
\frac{b^{2}}{H}-\frac{c^{2}}{F}-\frac{H}{F}(b{W}-{a})^{2}
-\frac{1}{l^{2}},
\end{eqnarray}
\begin{eqnarray}\label{Eins23}
{E_{r}}^{\Phi}&&= -2\,c\,b\,\frac { F-{H}^{2}{W}^{2}}{{F}^{2}{H}
}-2\,c\,a\frac{H\,W}{F^2},
\end{eqnarray}
\begin{eqnarray}\label{Eins31}
{E_{\Phi}}^{t}&&=\frac{1}{2}\,{H}^{2}W_{,r,r}+{H}W_{,r}H_{,r}-2\,b\,{\frac
{{H}}{F}}\left( W b-a \right),
\end{eqnarray}
\begin{eqnarray}\label{Eins32}
{E_{\Phi}}^{r}&&=-2\,bc,
\end{eqnarray}
\begin{eqnarray}\label{Eins33}
 {E_{\Phi}}^{\Phi}&&=\frac{1}{2}\,F_{,r,r}-\frac{1}{2}\,F\frac
{H_{,r,r}}{H}-\frac{3}{4}\, \frac {H_{,r}F_{,r}}{H}
+\frac{3}{4}\,F\,\frac
{{H_{,r}}^{2}}{{H}^{2}}-\frac{1}{2}\,{H}^{2}W\,W_{,r,r}-{H}
\,W\,W_{,r}H_{,r}\nonumber\\&&
-\frac{3}{4}\,{H}^{2}{W_{,r}}^{2}-{b}^{2}\frac
{F-{H}^{2}{W}^{2}}{F\,H}+\frac{{c}^{2}}{F}
-\frac{{a}^{2}H}{F}-\frac{1}{l^2}.
\end{eqnarray}
\end{subequations}
The vanishing of ${E_{t}}^{r}$ and ${E_{\Phi}}^{r}$ yields
respectively $ac=0=bc$. Therefore one can distinguish the branches:\\
$c=0$ with $a$ and $b$,
not vanishing simultaneously; and\\
$c\neq0$ with $a$ and $b$ vanishing simultaneously.

\noindent In the forthcoming sections we shall deal with the
integration and characterization of each branch starting from the
simplest static solutions.

\subsection{Complex extension and real cuts}\label{complex}

It would be of some interest to add some lines about the complex
extension of the metric under consideration. Accomplishing in the
metric~(\ref{metricsc-i}) the complex transformations
\begin{equation}
t\rightarrow{i\,\Phi},\,\,{\phi}\rightarrow{-i\,T},\,\,
\end{equation}
one arrives at
\begin{eqnarray}\label{cmetricsc1}
\bm{g}_{c}&&=(\frac{F}{H}-HW^2)\bm{d\Phi^2}+\frac{\bm{dr^2}}{F}
-H\bm{d\,T}^2+2H\,W\,\bm{d\,T}\,\bm{d\Phi},
\end{eqnarray}
which can be brought to the form
\begin{equation}\label{commetric}
\bm{g}_{c}=-\frac{\mathcal{F}}{\mathcal{H}}\bm{dT^2}
+\frac{\bm{dr^2}}{\mathcal{F}}
+\mathcal{H}\left(\bm{d\Phi}+\mathcal{W}\,\bm{dT}\right)^2,
\end{equation}
accompanied by the identification
\begin{eqnarray}\label{compfun}
\mathcal{F}=F,\, \mathcal{H}=\frac{F}{H}-HW^2,\,
\mathcal{W}=\frac{H\,W}{\mathcal{H}}.
\end{eqnarray}
At the level of the field tensor $F^{\mu\nu}$ one has
\begin{equation}\label{comfield}
\mathcal{F}^{\mu\nu}= \left[
\begin {array}{ccc} 0&\mathcal{B}&-\frac{\mathcal{C}}{\mathcal{F}}\\
\noalign{\medskip}-\mathcal{B}&0&\mathcal{A}\\
\noalign{\medskip}\frac{\mathcal{C}}{\mathcal{F}}&-\mathcal{A}&0\end
{array} \right]=\left[
\begin {array}{ccc} 0&-i\,a\,& {\frac{c}{F} }\\
\noalign{\medskip}i\,a&0&i{b}\\
\noalign{\medskip}- \frac{c}{F}&-i{b}&0\end {array} \right]
\end{equation}
thus the following correspondence for the field constants arises
\begin{equation}\label{comfielda}
{-i\,\it{a}}\rightarrow\mathcal{B},\,{i\,\it{b}}\rightarrow\mathcal{A},
\,{-\it{c}}\rightarrow\mathcal{C}.
\end{equation}
Summarizing, one may say that the role of the Killingian coordinates
has been interchanged: the time-like coordinate $t$ becomes the
space-like $\Phi$--coordinate, while the cyclic $\phi$--coordinate
becomes the new time--coordinate $T$. Correspondingly, one has to
think of the tensor components of the participating quantities from
this perspective. This procedure can be used to determine new
classes of solutions from known ones. For instance, one can generate
magnetic solutions from electric ones. The relations arising  from
this kind of complex transformations have been called ``duality
mapping'' by  Cataldo \cite{Cataldo96}, although strictly there is
no electric-magnetic duality in 2+1 dimensions.

\subsection{Positive $\Lambda$ solutions}\label{de--Sitter}

For completeness and to avoid duplication of works it is worthwhile
to notice that solutions for positive cosmological constant are
easily obtainable from the anti--de Sitter $(\Lambda=-1/l^2)$ ones;
First, notice that the Einstein equations for any sign of a
cosmological constant $\Lambda$--positive or negative--are recovered
from Eqs.~(\ref{Einstein}) simply by replacing
$-1/l^2\rightarrow\Lambda$. Next, having at disposal a concrete
solution of the Einstein equations mentioned above~(\ref{Einstein}),
by replacing there $l^2$ by $-\,l^2$, one determines the
corresponding metric structure for positive cosmological constant
$\Lambda=1/l^2$. This replacement is equivalent to accomplishing the
complex change $l\rightarrow{i\,l}$ in the $\Lambda<0$ solution,
nevertheless one ought to take care of possible additional
arrangements of constants, if any, and also of possible changes in
the signature.

{\it In the publication appears (3.15a) instead of the correct set
of equations (3.15).}

\subsection{Quasilocal mass, energy and angular momentum}\label{Quasilocal}

To evaluate quasilocal mass, energy, and angular momentum of
spacetimes with asymptotic different from the flat one, in
particular the anti-de Sitter, one uses the quasilocal formalism
developed in \cite{BrownY-prd93, BrownCM-prd94}.

For a stationary cyclic symmetric metric of the form
\begin{eqnarray}\label{BrownCAG}
\bm{g}&=&-N(\rho)^2\,\bm{d\,T}^2+{L(\rho)^{-2}}{\bm{d\rho}^2}
+K(\rho)^2\left[\bm{d\Phi}+W(\rho)\bm{d\,T}\right]^2,
\end{eqnarray}
the surface energy density is given by
\begin{eqnarray}\label{enerden1}
\epsilon(\rho)=-\frac{L(\rho)}{\pi\,K(\rho)}\frac{d\,}{d\rho}K(\rho)-\epsilon_{0},
\end{eqnarray}
where $\epsilon_{0}$ is the reference energy density, which in the
case of solutions with negative cosmological constant
$\Lambda=-1/l^2$ corresponds to the density of the anti--de Sitter
spacetime, namely
$\epsilon_{0}=-\frac{1}{\pi\rho}\sqrt{1+\frac{\rho^2}{l^2}}$. The
momentum density is determined from
\begin{eqnarray}\label{momentden}
j(\rho)=\frac{K(\rho)^2\,L(\rho)}{2\,\pi\,N(\rho)}\frac{d\,}{d\rho}W(\rho).
\end{eqnarray}
The integral momentum $J(\rho)$, global energy $E(\rho)$, and the
integral mass $M(\rho)$ are correspondingly given by
\begin{eqnarray}\label{globalq}
J(\rho)&=&2\pi\,K(\rho)\,j(\rho),\nonumber\\
E(\rho)&=&2\pi\,K(\rho)\,\epsilon(\rho),\nonumber\\
M(\rho)&=&E(\rho)\,N(\rho)-W(\rho)\,J(\rho).
\end{eqnarray}
As far as the evaluation of these physical quantities for the
studied classes of solutions is concerned, a work on these lines is
being developed and will be published elsewhere.

{\it In the publication there is a misprint in $M(\rho)$; instead
 of the correct $N(\rho)$ was typed $K(\rho)$.}

\section{Static cyclic symmetric solutions for Maxwell fields}\label{static}

In this section, we derive all the static solutions of the
Einstein--Maxwell equations (\ref{Einstein}); there are only three
families within this class. In the static solution $W(r)=0$,
consequently the metric~(\ref{metricsc-i}) becomes
\begin{equation}\label{metricST}
\bm{g}=-\frac{F(r)}{H(r)}\bm{dt}^2+\frac{\bm{dr}^2}{F(r)}
+H(r){\bm{d\phi}}^2,
\end{equation}
and the Einstein--Maxwell equations simplify drastically:
\begin{eqnarray}\label{EinsteinSt}
{E_{t}}^{t}&&=\frac{1}{2}\,F\frac {H_{,r,r}}{H}+\frac{1}{4}\,\frac
{H_{,r}F_{,r}}{H}-\frac{1}{4}\,F\frac
{{H_{,r}}^{2}}{{H}^{2}}+\frac{{b}^{2}}{{H}}+\frac{{c}^{2}}{F}+\frac{{a}^{2}{H}}{F}
-\frac{1}{{l}^2},\nonumber\\
{E_{r}}^{r}&&=\frac{1}{4}\,\frac
{H_{,r}F_{,r}}{H}-\frac{1}{4}\,F\frac {{H_{,r}}^{2}}{{H}^{2}}
+\frac{{b}^{2}}{{H}}-\frac{{c}^{2}}{F} -\frac{{a}^{2}{H}}{F}
-\frac{1}{l^{2}},\nonumber\\
{E_{\Phi}}^{\Phi}&&=\frac{1}{2}\,F_{,r,r}-\frac{1}{2}\,F\frac
{H_{,r,r}}{H}-\frac{3}{4}\, \frac {H_{,r}F_{,r}}{H}
+\frac{3}{4}\,F\,\frac {{H_{,r}}^{2}}{{H}^{2}}
-\frac{{b}^{2}}{{H}}+\frac{{c}^{2}}{F}-\frac{{a}^{2}{H}}{F}-\frac{1}{l^2},\nonumber\\
{E_{t}}^{r}&&=-2\,ca,\, {E_{t}}^{\Phi}=-2a\,b\frac{ 1 }{{H}},\nonumber\\
{E_{r}}^{t}&&=2a\,c\,\frac {{H} }{{F}^{2}},\,
{E_{r}}^{\Phi}=-2b\,c\frac {1}{{F}{H} },\nonumber\\
{E_{\Phi}}^{t}&&=2a\,b\,\frac {{H}}{F},\,{E_{\Phi}}^{r}=-2\,bc,
\end{eqnarray}
Each of these ${E_{\mu}}^{\nu}$--equations has to be equated to
zero, therefore one can distinguish the following three families of
static
solutions:\\
the electric class: $b\neq0, a=0, c=0$,\\
the magnetic class: $a\neq0, b=0, c=0$,\\
the hybrid class: $c\neq0, a=0, b=0$.

\noindent In the next subsections we proceed to integrate each class
separately.

\subsection{\label{Statelectric} Electrostatic
solutions; $b\neq0, a=0 $ }

The substraction ${E_{t}}^{t}(a=0=c)-{E_{r}}^{r}(a=0=c)$ yields
\begin{eqnarray}\label{eqec0W0a0H}
{\frac {d^2}{dr^2}}H \left( r \right)=0\Rightarrow
H(r)=C_{0}+C_{1}\,r,
\end{eqnarray}
where $C_{0}$, and $C_{1}$ are constants of integration; $C_{1}$ at
this stage is assumed to be different from zero, the zero case
deserves special attention and will be treated separately.
Substituting this structural function $H$ into the equation
${E_{t}}^{t}(a=0=c)$ one arrives at a first--order differential
equation for $F$
\begin{eqnarray}
\frac{H_{,r}}{H}F_{,r}-\left(\frac{H_{,r}}{H}\right)^2\,F
+4\frac{b^2}{H}-\frac{4}{l^2}=0,
\end{eqnarray}
which by introducing an auxiliary function $f(r)$ through
$$F(r)= H(r)\,f(r)=\left( C_{0}+C_{1}\,r \right) { f}
(r),$$ reduces to the simple equation
\begin{equation}
\frac{df(r)}{dr}=\frac{4}{C_{1}\,{l}^{2}}\,{\frac
{C_{1}\,r+C_{0}-{b}^{2}{l}^{2}}{  C_{0}+C_{1}\,r }}
\end{equation}
with general integral
\begin{equation}
f=\frac{4}{C_{1}^2\,{l}^{2}}\left( K_{0}+C_{1}\,r-{b}^{2}{l}^{2}\ln
\left( C_{0}+C_{1}\,r \right) \right),
\end{equation}
where $K_{0}$ is a new integration constant, into which of course
one has incorporated $C_{0}$.\\
Summarizing, one arrives at the metric
\begin{eqnarray}\label{staticq}
\bm{g}&=&-\frac{F(r)}{h(r)}\bm{dt}^2+\frac{\bm{dr}^2}{F(r)}
+h(r){\bm{d\phi}}^2,\nonumber\\
F(r)&=&\frac{4}{
C_{1}^2\,{l}^{2}}\left[K_{0}+h(r)-{b}^{2}{l}^{2}\ln{h(r)}\right]h(r),\nonumber\\
{h(r)}&=&C_{1}\,r+C_{0}.
\end{eqnarray}
This solution is characterized by:\\
the vector field
\begin{eqnarray}\label{staticqA}
&&\bm{A}=A_{t}{\bm{d t}}=\frac{b}{C_{1}}\ln{h}\,{\bm{d t}},
\end{eqnarray}
the electromagnetic field tensors
\begin{eqnarray}
&&F^{\mu\nu}=2b{\delta_{[t}}^{\mu}{\delta_{r]}}^{\nu},
\,\,F_{\mu\nu}=-2\frac{b}{h(r)}{\delta_{[\mu}}^{t}{\delta_{\nu]}}^{r}\nonumber\\
\end{eqnarray}
with field invariant
\begin{equation}
F_{\mu\nu}F^{\mu\nu}=-2\,\frac {{b}^{ 2}}{h},
\end{equation}
the energy momentum tensor
\begin{equation}
 {T_{\mu}}^{\nu}=  -\frac{{b}^{2}}{8\,\pi}\frac{1}{h}\left(
{\delta_{\mu}}^{t}{\delta_{t}}^{\nu}+{\delta_{\mu}}^{r}{\delta_{r}}^{\nu}
-{\delta_{\mu}}^{\phi}{\delta_{\phi}}^{\nu} \right),
\end{equation}
with quadratic energy momentum invariant and trace
\begin{equation}
T_{\mu\nu}T^{\mu\nu}=\frac{3}{64{\pi}^2}\frac{b^4}{h^2},\,
T_{\mu}^{\mu}=-\frac{1}{8{\pi}}\frac{b^2}{h}.
\end{equation}
A familiar representation of the above--mentioned solution is achieved for the
choice $C_{0}=0,\,C_{1}=2,\,K_{0}=b^2\,l^2\,\ln{2r_{0}}$, which
yields
\begin{eqnarray}\label{staticqcl1}
\bm{g}&=&-\left[\frac{2\,r}{l^2}-{b}^{2}\ln{\frac{r}{r_{0}}}\right]{\bm{dt}}^2
+2\,r{\bm{d\phi}}^2 +\frac{{\bm{dr}}^2}{2\,r\,\left[\frac{2\,r}{l^2}
-{b}^{2}\ln{\frac{r}{r_{0}}}\right]},\nonumber\\
\bm{A}&=&\frac{b}{2}\ln{\frac{r}{r_{0}}}\,{\bm{d t}}.
\end{eqnarray}
This solution, endowed with mass, electric charge, and radial
parameters, allows for a charged black hole interpretation. The mass
may assume positive as well as negative values, whereas the charge
is not upper bound.

\subsubsection{Gott--Simon--Alpern,
Deser--Mazur, and Melvin electrostatic solution $\Lambda=0$}

According to the existing references Gott, Simon, and Alpern were
the first to derive solutions within Maxwell theory in 2+1
gravity~\cite{GottSA1, GottSA2}; they found, among other things, the
electrostatic solution without cosmological constant. Introducing in
the above expressions, (\ref{staticq}) and (\ref{staticqA}), the
radial coordinate $\rho$ through $C_{0}+C_{1}\,r\rightarrow{\rho^2}$
together with
$t\rightarrow{C_{1}\,t/2},K_{0}\rightarrow{l^2\,k_{0}}$, and by letting
$1/l^2\rightarrow 0$ one arrives at the electrostatic solution in
the form
\begin{eqnarray}\label{metricSTlambz}
&&\bm{g}=-F\bm{dt}^2+\frac{\bm{d\rho}^2}{F} +\rho^2{\bm{d\phi}}^2,\nonumber\\
&&F(\rho)=k_{0}-2b^2\ln{\rho}=\frac{\kappa}{2\pi}
\,Q^2\ln{\frac{{\rho}_{c}}{\rho}},\nonumber\\
&&\bm{A}=A_{t}{\bm{d t}}={b}\,\ln{\rho}\,{\bm{d t}}.
\end{eqnarray}
Some authors refer to the coordinate system $\{t, \rho, \phi\}$, in
which the perimeter of the circle equates
$2\pi\,\rho=\int_{0}^{2\pi}\rho\,\bm{d\phi}$, as to the
"Schwarzschild" coordinates.\\
Practically at the same time Deser and Mazur \cite{DeserM85}
published their version of the electrostatic solution for
$\Lambda=0$. Moreover, by then, the work by Melvin
\cite{Melvin86} was published with the derivation of the
electrostatic as well as the magnetostatic solutions for vanishing
$\Lambda$. Later, Kogan \cite{Kogan92} reported and analyzed the
(electro and magneto) static solutions of the (2+1)--dimensional
Einstein--Maxwell equations for both positive and negative signs of
the gravitational constant $\kappa$; recall that in the three dimensions
there is not restriction on its sign. The $r$--coordinate used there
was such that $g_{rr}=1$ for the signature used in the present
report.

\subsubsection{Peldan electrostatic solution with $\Lambda$}
The electrostatic solution with cosmological constant in polar
coordinates arises from the general expressions above,
(\ref{staticq}) and (\ref{staticqA}), by means of the coordinate and
parameter changes $C_{0}+C_{1}\,r\rightarrow{\rho^2},
t\rightarrow{C_{1}\,t/2},K_{0}\rightarrow{-l^2\,m}$. In this way one
obtains
\begin{eqnarray}\label{metricSTP}
\bm{g}&=&-F\bm{dt}^2+\frac{\bm{d\rho}^2}{F}
+\rho^2{\bm{d\phi}}^2,\nonumber\\
F(\rho)&=&\frac{\rho^2}{l^2}-m-2b^2\ln{\rho},\nonumber\\
\bm{A}&=&{b}\,\ln{\rho}\,{\bm{d t}}.
\end{eqnarray}
The corresponding field tensors are given by
\begin{eqnarray}
&&{T_{\mu}}^{\nu}= -\frac{{b}^{2}}{8\,\pi}\frac{1}{\rho^2}\left(
{\delta_{\mu}}^{t}{\delta_{t}}^{\nu}+{\delta_{\mu}}^{\rho}{\delta_{\rho}}^{\nu}
-{\delta_{\mu}}^{\phi}{\delta_{\phi}}^{\nu} \right),\nonumber\\
&&F_{\mu\nu}=-2\frac{b}{\rho}{\delta_{[\mu}}^{t}{\delta_{\nu]}}^{r},
\end{eqnarray}
To achieve the specific Peldan \cite{Peldan93} writing, one has to
accomplish the additional identifications $m\rightarrow -C_{1},
1/l^2\rightarrow-\lambda/2, b^2\rightarrow{q^2/4},
t\rightarrow{C_{2}t},\rho\rightarrow r$.\\

\subsection{\label{Statemagnetic}
Magnetostatic solutions; $a\neq0, b=0 $}

To derive the magnetostatic solution, one starts from the addition
${E_{t}}^{t}+{E_{r}}^{r}$ which yields
\begin{eqnarray}
\frac{d}{dr}\left[\frac{F}{H}\,\frac{d}{dr}H-4\,r\,/l^2\right]=0,
\end{eqnarray}
with integral
\begin{eqnarray}\label{eqta2}
\frac{F}{H}\,\frac{d}{dr}H=4\,\frac{r}{l^2}+C_{1}.
\end{eqnarray}
The substraction ${E_{t}}^{t}-{E_{r}}^{r}$ gives
\begin{eqnarray}\label{eqta1}
F^2\frac{d^2}{dr^2}H+4a^2H^2=0.
\end{eqnarray}
Substituting $F(r)$ from Eq.~(\ref{eqta2}) into Eq.~(\ref{eqta1})
one arrives at a first--order equation for $\frac{d}{dr}H$
\begin{eqnarray}\label{eqta4}
 \left(
4\,{r}+C_{1}{l}^{2}
 \right)^{2} \frac {d^{2}H}{d{r}^{2}}
 +4\,{a}^{2}{l}^{4} \left( \frac {dH}{dr} \right)^{2}=0,\nonumber\\
\end{eqnarray}
which is rewritten as
\begin{eqnarray}\label{eqta5}
{d}(\frac{d}{dr}H)^{-1}=- {{a}^{2}{l}^{4}}d{\left(
4\,{r}+C_{1}{l}^{2}\right)^{-1}}
\end{eqnarray}
with first integral
\begin{eqnarray}\label{eqta6}
(\frac{d}{dr}H)^{-1}=\frac{C_{2}(4\,r +C_{1}\,{l}^{2})
-{a}^{2}{l}^{4}}{4r+C_{1}\,{l}^{2}}.
\end{eqnarray}
A subsequent integration gives
\begin{eqnarray}\label{eqta7}
&&H(r)=\frac {\left[R(r) l^2+{a}^{2} {l}^{4}\ln
\left(R(r) l^2\right)+C_{3}\right]}{4C_{2}^2},\nonumber\\
&&F(r)=R(r)H(r),\nonumber\\
&&R(r)\,l^2:=C_{2}\,(4\,r+C_{1}\,{l}^{2})-{a}^{2}{l}^{4},
\end{eqnarray}
where Eq.~(\ref{eqta2}) it has been used to evaluate $F(r)$. These
structural functions completely determine the magnetostatic
solution; without any loss of generality, by letting
$C_{2}\rightarrow{C_{1}}l^2/4,$ $C_{1}\rightarrow
4\,(a^2\,l^2+C_{0})/(C_{1}\,l^2),$ $C_{3}\rightarrow
K_{0}\,l^2-a^2\,l^4\,\ln{l^2}$, the magnetostatic metric can be
given as
\begin{eqnarray}\label{eqta7a}
\bm{g}&=&-h(r)\bm{dt}^2+\frac{\bm{dr}^2}{H(r)h(r)}
+H(r){\bm{d\phi}}^2,\nonumber\\
H(r)&=&\frac {4}{{C_{1}}^2\,l^2}\left[K_{0}+h(r)+{a}^{2}
{l}^{2}\ln{h(r)}\right],\nonumber\\
F(r)&=&\,H(r)\,h(r),\, h(r):=C_{1}\,r+C_{0}.
\end{eqnarray}
This solution is characterized by:\\
the electromagnetic field vector
\begin{equation}
\bm{A}=\frac{a}{C_{1}}\ln{h}\,{\bm{d\phi}},
\end{equation}
the electromagnetic field tensors
\begin{eqnarray}
&&F^{\mu\nu}=-2a{\delta_{[\phi}}^{\mu}{\delta_{r]}}^{\nu},
\,\,F_{\mu\nu}=-2\frac{a}{h(r)}{\delta_{[\mu}}^{\phi}{\delta_{\nu]}}^{r}\nonumber\\
\end{eqnarray}
with field invariant
\begin{equation}
F_{\mu\nu}F^{\mu\nu}=2\,\frac {{a}^{ 2}}{h(r)},
\end{equation}
the energy--momentum tensor
\begin{equation}
 {T_{\mu}}^{\nu}=  \frac{{a}^{2}}{8\,\pi}\frac{1}{h(r)}\left(
-{\delta_{\mu}}^{t}{\delta_{t}}^{\nu}+{\delta_{\mu}}^{r}{\delta_{r}}^{\nu}
+{\delta_{\mu}}^{\phi}{\delta_{\phi}}^{\nu} \right),
\end{equation}
with energy field invariants
\begin{equation}
T_{\mu\nu}T^{\mu\nu}={\frac {3}{64}}\, {\frac {{a}^{4}}{{\pi }^{2}
 }}\,\frac{1}{h(r)^2},\,
T_{\mu}^{\mu}=a^2\frac{1}{8\pi}\,\frac{1}{h(r)}.
\end{equation}
This class of solutions allows for a hydrodynamics interpretation in
terms of a perfect fluid energy--momentum tensor for a stiff fluid, $
\rho=p$, where $\rho$ and $p$ are, respectively, the fluid energy
density and the fluid pressure. In fact, the energy momentum tensor
for a perfect fluid is given by
$$T_{\mu\nu}=(\rho+p)u_{\mu}u_{\nu}+p\,g_{\mu\nu}.$$
Therefore choosing the fluid $4$--velocity along the time direction
$u^{\mu}=\delta_{t}^{\mu}/\sqrt{-g_{tt}}$ one establishes that
$\rho=\frac{{a}^{2}}{8\pi}\frac {H}{F^{2}}=p$.

\subsubsection{Peldan magnetostatic solution with $\Lambda$}

Introducing in the metric~(\ref{eqta7a}) new coordinates according
to $ h=C_{1}r+C_{0}\rightarrow{\rho^2}, t\rightarrow t,
\phi\rightarrow \phi\,C_{1}/2, K_{0}\rightarrow{k_{0}l^2}$ one gets
\begin{eqnarray}\label{eqta7apel}
\bm{g}&=&-\rho^2\bm{dt}^2+\frac{\bm{d\rho}^2}{F(\rho)}
+F(\rho){\bm{d\phi}}^2,\nonumber\\
F(\rho)&=&k_{0}+\frac{\rho^2}{l^2}+2{a}^{2}\ln{\rho},\nonumber\\
\bm{A}&=&a\ln{\rho}\,{\bm{d \phi}},
\end{eqnarray}
characterized by the field tensors
\begin{eqnarray}
&&{T_{\mu}}^{\nu}=  \frac{{a}^{2}}{8\,\pi}\frac{1}{\rho^2}\left(
-{\delta_{\mu}}^{t}{\delta_{t}}^{\nu}+{\delta_{\mu}}^{\rho}{\delta_{\rho}}^{\nu}
+{\delta_{\mu}}^{\phi}{\delta_{\phi}}^{\nu} \right),\nonumber\\&&
F_{\mu\nu}=-2\frac{a}{\rho}{\delta_{[\mu}}^{\phi}{\delta_{\nu]}}^{\rho}.
\end{eqnarray}
This solution has been derived and analyzed in \cite{Peldan93}.

\subsubsection{Hirschmann--Welch solution with $\Lambda$}

Accomplishing in the general magnetic static metric~(\ref{eqta7a})
the transformations
\begin{eqnarray}
C_{1}r+C_{0}&&\rightarrow({\rho^2+r_{+}^2}-ml^2)/l^2=:{h(\rho)},
\nonumber\\
2\phi/(C_{1}\,l^2)&{}&\rightarrow{\phi},\, a^2\,l^4=\chi^2,K_{0}=m,
\end{eqnarray}
one ends with the Hirmanch--Welch \cite{Hirschmann96} representation
of the magnetic solution
\begin{eqnarray}\label{metricHW}
\bm{g}&&=-\frac{1}{l^2}(\rho^2+r_{+}^2-ml^2)\bm{dt^}2
+[\rho^2+r_{+}^2+\chi^2\ln(|h(\rho)|)]{\bm {d\phi}}^2\nonumber\\
+&&\frac{l^2\,\rho^2\,\bm{d\rho}^2}{(\rho^2+r_{+}^2-ml^2)
[\rho^2+r_{+}^2+\chi^2\ln(|h(\rho)|)]},\nonumber\\
&&{h(\rho)}=({\rho^2+r_{+}^2}-ml^2)/l^2,
\end{eqnarray}
with vector potential
\begin{equation}
\bm{A}=\frac{1}{2}\chi\ln|(\rho^2+r_{+}^2)/l^2-m|\bm {d\Phi},
\end{equation}
For $\rho=0$, one determines the constant $r_{+}$ fulfilling
\begin{equation}
r_{+}^2+\chi^2\ln|r_{+}^2/l^2-m|=0.
\end{equation}
This solution is endowed with mass, magnetic charge and radial
parameters. The coordinate $\rho$ ranges from zero to infinity. This
magnetic solution does not allow the existence of an event horizon
since time--like geodesics can reach the origin at finite proper time,
while null geodesics approach the origin at finite affine parameter;
hence it does not describe a magnetic black hole. Moreover the Ricci
tensor, and consequently the curvature tensor, as well as the
electromagnetic field, are well behaved in this spacetime.

Cataldo {\it et al.} \cite{CataldoCCdC04} commented on this static
circular magnetic solution of the 2+1 Einstein-Maxwell equations,
derived previously by other authors, and came to the conclusion that
this solution, considered up to that moment as a two-parameter one,
is in fact a one-parameter solution, which describes a distribution
of a radial magnetic field in a 2+1 anti-de Sitter background
spacetime, and that the mass--parameter is just a pure gauge and can
be re--scaled to minus one.

\subsubsection{Melvin, and Barrow--Burd--Lancaster
magnetostatic solution $\Lambda=0$}

Melvin \cite{Melvin86} derived the electric and the magnetic
static solutions for vanishing cosmological constant $\Lambda=0$.
The corresponding solution can be obtained from (\ref{eqta7apel})
and written as
\begin{eqnarray}\label{Melvin-mag}
\bm{g}&=&-\rho^2\bm{dt}^2+\frac{\bm{d\rho}^2}{F(\rho)}
+F(\rho){\bm{d\phi}}^2,\nonumber\\
F(\rho)&=&k_{0}+2{a}^{2}\ln{\rho},
\end{eqnarray}
or by introducing $a^4\,r^2\,{\rm
e}^{k_{0}/a^2}=k_{0}+a^2\ln{\rho^2}$, and scaling the variables $t$
and $\phi$ one brings it to the form
\begin{eqnarray}\label{metric_st_mag_BBL}
\bm{g}&=&e^{r^2}(-\bm{dt^}2+\bm{dr}^2)+r^2\bm{d\phi}^2.
\end{eqnarray}
In the paragraph
devoted to stiff perfect fluid, Barrow, Burd, and Lancaster, see \cite{Barrow86}, pointed out that for a fluid aligned
along the time--coordinate, " in $(2+1)$ dimensions the stiff fluid
has an energy-momentum tensor identical to that of a static magnetic
field", and they continued with a statement very close to the
following: if one sets the electric field components $F_{0i}=0$ and
magnetic components ${F_{i}}^{j}={\epsilon_{i}}^{j}\sqrt{2\rho}$ in
the electromagnetic energy-momentum tensor
$T_{\mu\nu}=F_{\mu\lambda}{F_{\nu}}^{\lambda}
-g_{\mu\nu}F_{\alpha\lambda}{F}^{\lambda\alpha}/4$ reduces to the
perfect fluid energy-momentum tensor
$T_{\mu\nu}=(\rho+p)u_{\mu}u_{\nu}+p\,g_{\mu\nu}$, with energy
density  $\rho$ equalling the pressure $p$, $\rho=p$.

\subsection{Static hybrid $\bm{A}=\frac{c}{2}(t\bm{d\phi}-\phi \bm{dt})$ solution}

In this case, the integration starts from the combination
${E_{t}}^{t}(a=0=b)+2{E_{r}}^{r}(a=0=b)+{E_{\Phi}}^{\Phi}(a=0=b)$
which yields
\begin{eqnarray}\label{metricCat4}
{\frac {d^{2}}{d{r}^{2}}}F -\frac{8}{l^2}=0,
\end{eqnarray}
with integral
\begin{eqnarray}\label{metricCat5}
F(r)=4\,{\frac { \left( r-r_{1} \right)  \left( r- r_{2} \right)
}{{l} ^{2}}}.
\end{eqnarray}
As the equation for $H(r)$ one may consider the first--order equation
${E_{r}}^{r}(a=0, b=0)$, which can be written as
\begin{eqnarray}\label{metricCat6}
\left( \frac {H _{,r}}{2\,H}-\frac {F _{,r}}{4\,F} \right)
^{2}=\left(\frac {F _{,r}}{4\,F} \right)^2- {\frac {{c}^{2}}{F^2
}}-\frac{1}{l^2\,F}.\nonumber\\
\end{eqnarray}
Evaluating the right--hand side of this equation, one arrives at
\begin{eqnarray*}\label{metricCat6a}
\left[\frac {d}{dr}\ln\left(\frac{H}{F^{1/2}}\right) \right]
^{2}=4\frac{\left( ({r_{2}}-r_{1})^2- {c}^{2}l^4\right)}{l^4\,F^2}.
\end{eqnarray*}
For definiteness we assume ${r_{2}}>r_{1}$. Accomplishing the
integration one obtains
\begin{eqnarray*}\label{metricCat7}
\ln {\left( {\frac {H }{\sqrt{F}}} \right)}&&=\mp\,\frac{\sqrt
{({r_{2}}-r_{1})^2- {c}^{2}l^4}}{2\left( {r_{2}}-r_{1} \right)}
\times\ln \left( {\frac {r-{ r_{1}}}{r-{r_{2}}}} \right).
\end{eqnarray*}
Introducing the constant $\alpha$ through
\begin{eqnarray}\label{metricCat8}
&& \alpha=1-{\frac {{l}^{4}{c}^{2}}{ \left( {r_{2}}-r_{1} \right)
^{2 }}},\, {c}^{2}={\frac { \left( {r_{2}}-r_{1} \right) ^{2} \left(
1-\alpha
 \right) }{{l}^{4}}},
\end{eqnarray}
one obtains $H(r)$ in the form
\begin{eqnarray}\label{metricCat9}
 H \left( r \right)={K_{0}}^2\,\sqrt{F
\left( r \right) } \left( { \frac {r-r_{1}}{r-{r_{2}}}} \right)
^{\pm\,\frac{\sqrt {\alpha}}{2}}.
\end{eqnarray}
Summarizing, this class of solutions is given by the metric
\begin{eqnarray}\label{hybridstatic}
\bm{g}&=&-\frac{F}{H}{\bm{dt}}^2+
\frac{1}{F}{\bm{d{r}}}^2+H\,{\bm{d\phi}}^2,\nonumber\\
F&=&\frac{4}{l^2}(r-r_{1})(r-r_{2}),\nonumber\\
H&=&\frac{2\,K_{0}^2}{l}\,
(r-r_{1})^{(1\pm\sqrt{\alpha})/2}(r-r_{2})^{(1\mp\sqrt{\alpha})/2},\nonumber\\
\bm{A}&=&\frac{c}{2}(t\bm{d\phi}-\phi \bm{dt}).
\end{eqnarray}
The field tensor characterization of this solution is given by
\begin{eqnarray}
&&F^{\mu\nu}= -2\frac
{c}{F}{\delta^{[\mu}}_{t}{\delta^{\nu]}}_{\phi},\,F_{\mu\nu}= 2
{c}{\delta_{[\mu}}^{t}{\delta_{\nu]}}^{\phi}, \nonumber\\
&&{T^{\mu}}_{\nu}={ \frac {{c}^{2}}{8\,\pi \,F
}}\,\left(-{\delta^{\mu}}_{t}{\delta^{t}}_{\nu}
+{\delta^{\mu}}_{r}{\delta^{r}}_{\nu}
-{\delta^{\mu}}_{\phi}{\delta^{\phi}}_{\nu}\right),\nonumber\\
&&FF=-2\,{\frac {{c}^{2}}{F }}, \,TT=\frac {3}{64}\,\frac
{{c}^{4}}{{ \pi }^{2} F
 ^{2}},\nonumber\\
&&T_{\mu}^{\mu}=-\frac{1}{8}\,{\frac {{c}^{2}}{\pi \,F }}.
\end{eqnarray}
By scaling transformations of the Killingian coordinates $\phi$ and
$t$, the arbitrary constant $K_{0}$ can be equated to $1$.

\subsubsection{Cataldo azimuthal electromagnetic solution}

Subjecting the metric~(\ref{hybridstatic}) to the coordinate
transformations
\begin{eqnarray}
t&=&\frac{1}{\sqrt{2}}K_{0}l^{\mp\sqrt{\alpha}/2}t^{\prime},
\phi=\frac{1}{\sqrt{2}K_{0}}l^{\pm\sqrt{\alpha}/2}\phi^{\prime},\,
r={\rho}^2+r_{1}, M:=\frac{1}{l^2}(r_{2}-r_{1}),
\end{eqnarray}
dropping primes, one brings the static hybrid metric to the form
\begin{eqnarray}\label{metricsc3}
\bm{g}&=&-\,{\rho}^{1\mp\sqrt{\alpha}}
\left(\frac{{\rho}^2}{l^2}-M\right)^{(1\pm\sqrt{\alpha})/2}{\bm{dt}}^2+{\rho}^{1\pm\sqrt{\alpha}}
\left(\frac{{\rho}^2}{l^2}-M\right)^{(1\mp\sqrt{\alpha})/2}{\bm{d\phi}}^2\nonumber\\
&&+ \left(\frac{{\rho}^2}{l^2}-M\right)^{-1}{\bm{d{\rho}}}^2.
\end{eqnarray}
The electromagnetic field tensor under the above--mentioned transformations
becomes
\begin{eqnarray}
&&F_{\mu\nu}=
M\sqrt{1-\alpha}{\delta_{[\mu}}^{t}{\delta_{\nu]}}^{\phi},\nonumber\\
&{}&{T_{\nu}}^{\mu}=\frac{M^2}{32\pi\,\rho^2}\frac{(1-\alpha)}{(\rho^2/l^2-M)}
\times\left(-{\delta_{\nu}}^{t}{\delta_{t}}^{\mu}
+{\delta_{\nu}}^{r}{\delta_{r}}^{\mu}
-{\delta_{\nu}}^{\phi}{\delta_{\phi}}^{\mu}\right).
\end{eqnarray}
This solution corresponds to the static charged solution reported in
~\cite{Cataldo02}, where the name of azimuthal static solution was
coined.

\section{Uniform electromagnetic solutions}\label{uniform}

To determine all uniform electromagnetic solutions, i.e., those
possessing vanishing covariant derivatives of $F_{\mu\nu}$,
$F_{\mu\nu\,;\sigma}=0$, one has to start the integration process
from the differential relations arising from these conditions. The
hybrid class $c\neq 0$ does not allow for such kind of solutions.
The other families with $a\neq0$ and (or) $b\neq0$ give rise to
non--trivial solutions.

\subsection{General uniform electromagnetic
 solution for $a\neq0\neq\,b$}\label{uniformab}

A class of uniform electromagnetic stationary solutions, for
$a\neq0,\,b\neq0$ and $c=0$, can be constructed by demanding the
vanishing of the covariant derivatives of the electromagnetic tensor
field, $F_{\mu\nu\,;\sigma}=0$, which yields two independent
equations: $F_{t\phi;t}=0$ and $F_{tr;r}=0$. From the last one, one
isolates ${dW}/{dr}$
\begin{eqnarray}\label{Newunix1}
\frac{d W}{dr}=-b\frac{F}{H^3(a-b\,W)}\frac{d H}{dr},
\end{eqnarray}
which when used in the first equation $F_{t\phi;t}=0$ allows us to write
\begin{eqnarray}\label{Newunix2}
\frac{d F}{dr}=\frac{F}{H}\frac{d
H}{dr}-b^2\frac{F^2}{H^3(a-b\,W)^2}\frac{d H}{dr}.
\end{eqnarray}
As the next step, one substitutes recursively these first
derivatives into the Einstein equations. One gets, among other
relations, a simple expression for the equation ${E_{\phi}}^t$
\begin{eqnarray}\label{Newunix3}
\frac{d^2H}{dr^2}=-4\frac{H^2}{F^2}(a-b\,W)^2,
\end{eqnarray}
which when substituted back into the Einstein equations reduces them to a
single relation
\begin{eqnarray}\label{Newunix4}
F(b^2l^2-H)=l^2H^2(a-b\,W)^2,
\end{eqnarray}
from which one has
\begin{eqnarray}\label{Newunix5}
W(r)=\frac{a}{b}\mp\frac{F}{l b H}\sqrt{b^2l^2-H}.
\end{eqnarray}
Using the relation~(\ref{Newunix4}) in Eq.~(\ref{Newunix3}) one
obtains
\begin{eqnarray}\label{Newunix6}
\frac{d^2H}{dr^2}=-\frac{4}{l^2F}(b^2l^2-H).
\end{eqnarray}
On the other hand, substituting $W(r)$ from Eq.~(\ref{Newunix5})
into Eq.~(\ref{Newunix2}) one arrives at the relation
\begin{eqnarray}\label{Newunix7}
\frac{d F}{dr}(b^2l^2-H)+F\frac{d H}{dr}=0,
\end{eqnarray}
with integral
\begin{eqnarray}\label{Newunix8}
F(r)=\frac{b^2l^2-H(r)}{l^2\,\beta^2},\,\beta=\rm{constant}.
\end{eqnarray}
The substitution of $F(r)$ from~(\ref{Newunix8}) into
Eq.~(\ref{Newunix6}) yields
\begin{eqnarray}\label{Newunix9}
\frac{d^2H}{dr^2}=-{4}{\beta^2},
\end{eqnarray}
hence
\begin{eqnarray}\label{Newuniz}
H(r)=-{2}{\beta^2}r^2+c_{1}r+c_{0}.
\end{eqnarray}
Consequently the function $W(r)$ becomes
\begin{eqnarray}\label{Newunix5a}
W(r)=\frac{a}{b}\mp\frac{1}{l^2\, b\,\beta}\frac{b^2l^2-H}{H}.
\end{eqnarray}
No restriction arises from the remaining Eq.~(\ref{Newunix1}).

Thus, we have determined the general uniform electromagnetic
stationary cyclic symmetric solution given by the metric and the
field vector
\begin{eqnarray}\label{Newunif5}
&&\bm{g}=-\frac{b^2\,l^2-H(r)}{l^2\,\beta^2\,H(r)}\bm{dt}^2
+\frac{l^2\,\beta^2\,\bm{dr}^2}{b^2\,l^2-H(r)}+H(r)\left[\bm{d\phi}+\left(\frac{a}{b}\mp
\frac{1}{l^2\, b\,\beta}\frac{b^2l^2-H}{H}\right)\bm{dt}\right]^2,\nonumber\\
&&H(r)=-{2}{\beta^2}r^2+C_{1}r+C_{0},      \nonumber\\
&&\bm{A}=-\beta\,r\,[\bm{d\phi}-\frac{1\pm\,
al^2\beta}{l^2\,b\,\beta}\bm{dt}],
\end{eqnarray}
characterized by the uniform electromagnetic field tensors
\begin{eqnarray}\label{Newunif6}
&&F_{\mu\nu}=-2\frac{1\pm
al^2\beta}{l^2b}{\delta_{[\mu}}^{t}{\delta_{\nu]}}^{r}\,
\pm 2\beta {\delta_{[\mu}}^{r}{\delta_{\nu]}}^{\phi},\nonumber\\
&& {8\pi\,l^2}{T_{\nu}}^{\mu}=-{(1\pm\,2al^2\beta)}
\left({\delta_{\nu}}^{t}{\delta_{t}}^{\mu}
-{\delta_{\nu}}^{\phi}{\delta_{\phi}}^{\mu}\right)
-{\delta_{\nu}}^{r}{\delta_{r}}^{\mu}\mp2\,\beta\,bl^2{\delta_{\nu}}^{\phi}{\delta_{t}}^{\mu}
+\frac{2a}{b}(1\pm\,al^2\beta){\delta_{\nu}}^{t}{\delta_{\phi}}^{\mu}\nonumber\\
\end{eqnarray}
with $FF$ invariant $FF=-2/l^2.$

Although the solution above has been derived for $\Lambda=-1/l^2$,
the branch corresponding to $\Lambda=1/l^2$ is achieved from the
above expressions by changing $l^2\rightarrow{-l^2}$.

\subsection{Uniform ``stationary'' electromagnetic
$\bm{A}={r}/({b\,l^2})(\bm{dt}-\omega_{0}\bm{d\phi})$
solutions}\label{uniformb}

Consider now the case $a=0=c$ for the metric~(\ref{metricsc-ii})
\begin{eqnarray}\label{commetricCLu}
\bm{g}&=&-\frac{F}{h}\left(\bm{dt}
-{\omega}\,\bm{d\phi}\right)^2+h\,{\bm{d\phi}}^2
+\frac{{\bm{dr}}^2}{{F}}
 =-\frac{{F}}{{H}}\bm{dt}^2
+\frac{\bm{dr}^2}{{F}}
+{H}\left(\bm{d\phi}+{W}\,\bm{dt}\right)^2,\nonumber\\
{F}&=&F,\, {H}={h}-\frac{F}{h}\omega^2,\,
{W}=\frac{\omega}{H}\frac{F}{h}.
\end{eqnarray} The electromagnetic
tensor amounts to
$F^{{\mu}{\nu}}=2\,b\,{\delta^{\mu}}_{[t}{\delta^{\nu}}_{r]}$, and
\begin{eqnarray*}\label{ClementCLmg2xu}
F_{{\mu}{\nu}}=-2\,\frac{b}{H
F}(F-H^2\,W^2)\,{\delta_{\mu}}^{[t}{\delta_{\nu}}^{r]}+2\,b\,\frac{H
W}{F}\,{\delta_{\mu}}^{[\phi}{\delta_{\nu}}^{r]} =-2\,\frac{b}{h(r)}
\,{\delta_{\mu}}^{[t}{\delta_{\nu}}^{r]}
+2\,b\,\frac{\omega(r)}{h(r)}
\,{\delta_{\mu}}^{[\phi}{\delta_{\nu}}^{r]}.
\end{eqnarray*}
The covariant derivatives $F_{{\phi}{r;r}}$ and $F_{{t}{r;r}}$ of
the field $F_{{\mu}{\nu}}$ are equal to zero if
$$\omega(r)=\omega_{0},\,h(r)=h_{0}.$$
Therefore, the structural functions $\omega(r)=\omega_{0}$ and
$h(r)=h_{0}$ are constants. The Einstein--Maxwell equations require
the fulfillment of the equations
\begin{eqnarray}\label{fmnCL3u}
\frac{d^2}{d\,r^2}F(r)&=&\frac{4}{l^2}\rightarrow\,
F(r)=\frac{2}{l^2}r^2+c_{1}\,r+c_{0},\, h_{0}=b^2\,l^2.
\end{eqnarray}
Consequently the derived solution can be given as
\begin{eqnarray}\label{ClementCL4u}
\bm{g}&=&-\frac{F}{h_{0}}(\bm{dt}-\omega_{0}\bm{d\phi})^2
+h_{0}\,{\bm{d\phi}}^2+\frac{\bm{dr}^2}{F(r)}
,\nonumber\\
F(r)&=&\frac{2r^2}{l^2}+c_{1}r+c_{0}, h_{0}=b^2\,l^2,\nonumber\\
\bm{A}&=&\frac{r}{b\,l^2}(\bm{dt}-\omega_{0}\bm{d\phi}),
\end{eqnarray}
and hence by a shifting transformation of the $t$--coordinate the
derived metric becomes a static one. The electromagnetic tensors
characterizing this uniform ``stationary'' cyclic symmetric solution
are given by
\begin{eqnarray}
F_{\mu\,\nu}&&= -2\frac{1}{b\,l^2}{\delta_{[\mu}}^{t}{\delta_{\nu]}}^{r}
+2\frac{\omega_{0}}{b\,l^2}{\delta_{[\mu}}^{\phi}{\delta_{\nu]}}^{r},\nonumber\\
{T_{\mu}}^{\nu}&&=
\frac{1}{8\,\pi\,l^2}(-{\delta_{\mu}}^{t}{\delta_{t}}^{\nu}
-{\delta_{\mu}}^{r}{\delta_{r}}^{\nu}+{\delta_{\mu}}^{\phi}{\delta_{\phi}}^{\nu})
+\frac{\omega_{0}}{4\,\pi\,l^2}{\delta_{\mu}}^{t}{\delta_{\phi}}^{\nu}.
\end{eqnarray}
This solution can be generated from the static solution, which is given
in subsection \ref{Matyja} by the metric (\ref{metricBRZ}), via the
transformation $t\rightarrow\,t-\omega_{0}\phi,\,\phi\rightarrow
\phi.$

\subsubsection{Clement uniform ``stationary'' electromagnetic
solution}

Clement \cite{Clement93} reported the uniform ``stationary''
generalization of the electrostatic solution in the form of
\begin{eqnarray}\label{Clement25u}
&&\bm{g}=-\frac{F}{H_{0}}\left(\bm{dt}
-\omega_{0}\bm{d\,\phi}\right)^2+\frac{{\bm{dr}}^2}{F(r)}
+H_{0}{\bm{d\phi}}^2,\nonumber\\
&&F(r)=\frac {2r^2}{l^2}+c_{1}r+c_{0},\,
\nonumber\\
&&\bm{A}=\frac{1}{\sqrt{H_{0}}\,l}\,r\,\left(\bm{dt}
-\omega_{0}\bm{d\,\phi}\right),
\end{eqnarray}
which is equivalent to the metric~(Cl.25, $\Lambda=-1/l^2$), where
we have changed the signature. In Clement's parametrization
$H_{0}=\frac{\pi_{0}^2\,l^2}{4m}$ , with $m=1/(2\,\kappa)$.

\subsubsection{No uniform generalization of the electrostatic
solution for $\Lambda=1/l^2$}

On the other hand for $b\neq 0$ and positive cosmological constant
$\Lambda=1/l^2$ there is no a uniform electromagnetic stationary
cyclic symmetric solution; the reason is hidden in the resulting
erroneous signature. In the considered case, for the
metric~(\ref{metricsc-ii}) with structural functions $F(r),
h(r),\omega(r)$ the electromagnetic tensor amounts, for $a=0$, to
\begin{eqnarray*}\label{ClementCLmg2xua}
F^{{\mu}{\nu}}&&=2\,b\,{\delta^{\mu}}_{[t}{\delta^{\nu}}_{r]},
\nonumber\\
F_{{\mu}{\nu}}&&=-2\,b/h(r)\,{\delta_{\mu}}^{[t}{\delta_{\nu}}^{r]}
+2\,b\,\omega(r)/h(r)\,{\delta_{\mu}}^{[\phi}{\delta_{\nu}}^{r]}.
\end{eqnarray*}
Therefore $F_{{\mu}{\nu}; \lambda}=0$ is achieved for $
h(r)=h_{0},\,\omega(r)=\omega_{0}$. The Einstein equations requires
$ {E_{r}}^{r}={b^2}/{h_{0}}+{1}/{l^2}
=0\rightarrow{h_{0}=-b^2\,l^2}$. The covariant tensor components
$g_{tt}$ and $g_{rr}$ explicitly amount to $
g_{tt}=F(r)/(b^2\,l^2)>0,\,g_{rr}=1/F(r)>0, $ which contradicts the
adopted signature $\{-,+,+\}$, therefore this case does not
represent a compatible solution.

\subsection{Matyjasek--Zaslavskii uniform electrostatic
$\bm{A}={r}/({b\,l^2})\,{\bm{d t}}$ solution}\label{Matyja}

The sub--branch $\{a=0, b\neq\,0, W(r)=0\}$ of uniform electrostatic
solutions arises for constant $H(r)$, $H(r)=H_{0}=\rm{constant.}$
The equation $E_{t}^{t}$ implies that the constant $H_{0}$ has to be
$H_{0}=b^2\,l^2$. The remaining equation ${E_{t}}^{\phi}$ amounts to
\begin{eqnarray}
\frac{d^2\,F}{d\,r^2}-\frac{4}{l^2}=0,\,\rightarrow
F(r)=2\frac{r^2}{l^2}+4\,c_{1}r+c_{0};
\end{eqnarray}
consequently the metric and the field vector become
\begin{equation}\label{metricBRZ}
\bm{g}=-\frac{F(r)}{b^2\,l^2}{{\bm{dt}}^2}+\frac{{\bm{dr}}^2}{F(r)}
+{b^2\,l^2}\bm{d\phi}^2,\,\bm{A}=\frac{r}{b\,l^2}\,{\bm{d t}}.
\end{equation}
The electromagnetic field tensors of this solution possess constant
eigenvalues and also exhibit the uniform character; explicitly they
are given by
\begin{eqnarray}
&&F_{\mu\,\nu}= \left[ \begin {array}{ccc}
0&-\frac{1}{b\,l^2}&0\\\noalign{\medskip}\frac{1}{b\,l^2}
&0&0\\\noalign{\medskip}0&0&{0}\end {array}
\right],\,{T_{\mu}}^{\nu}= \left[
\begin {array}{ccc} -\,{ \frac {1}{8\pi
\,{l}^{2}}}&0&0\\\noalign{\medskip}0&-\,{\frac {1}{ 8\pi
\,{l}^{2}}}&0\\\noalign{\medskip}0&0&{\frac {1}{8\pi \,{l}^{2}
}}\end {array} \right] ,
\end{eqnarray}
with constant field invariants given by
$$FF=-\frac{2}{l^2},
\,TT=\frac{3}{64}\frac{1}{\pi^2\,l^4}.$$ Incorporating the constant
$H_{0}=b^2\,l^2$ in the new definitions of $t$ and $\phi$,
$t/b\,l\rightarrow t,b\,l\phi\rightarrow \phi$, one can set $b\,l=1$
in the metric~(\ref{metricBRZ}).

For the sake of comparison with the previous reports let us introduce
hyperbolic functions:
\begin{eqnarray*}
F(r)&&=2\frac{r^2}{l^2}+4\,c_{1}r+c_{0}=|c_{0}-2\,
{l^2}c_{1}^2|\left(\frac{2(r+l^2c_{1})^2}{l^2|c_{0}-2\,{l^2}c_{1}^2|}\pm
1\right),
\end{eqnarray*}
\begin{eqnarray}\label{funcBRZ}
&&+:r+l^2c_{1}=\sqrt{l^2|c_{0}-2\,{l^2}c_{1}^2|/2}\,\sinh(\alpha x
),\,F(r)\rightarrow{|c_{0}-2\,{l^2}c_{1}^2|\cosh^2(\alpha x
)},\nonumber\\
&&-:r+l^2c_{1}=\sqrt{l^2|c_{0}-2\,{l^2}c_{1}^2|/2}\,\cosh(\alpha x
),\,F(r)\rightarrow{|c_{0}-2\,{l^2}c_{1}^2|\sinh^2(\alpha x )},\nonumber\\
&&c_{0}=2\,{l^2}c_{1}^2:
r+l^2c_{1}=\exp{\sqrt{2}x/l},\,F(r)\rightarrow\frac{2}{l^2}{\exp{(2\sqrt{2}x/l)}}.
\end{eqnarray}
The above--mentioned quantities, i.e. metric~(\ref{metricBRZ}) and structural
functions~(\ref{funcBRZ}), determine the solution derived in
\cite{Matyjasek-cqg04}. \noindent Expressions (\ref{metricBRZ}) and
(\ref{funcBRZ}) are equivalent to the Bertotti \cite{Bertotti59} and
Robinson \cite{Robinson59} uniform electromagnetic--gravitational
field solution,  for a constant slice of one of the spatial
coordinates; the BR $3+1$ solution allows for a product of two
surfaces of constant curvature as manifold. Moreover, the
two--dimensional BR--metric sector $ds_{-}^2$ reduces to
$r_{-}^2\bm{d\phi}^2$ in the $2+1$ case, therefore the
(2+1)--dimensional uniform electrostatic field can be considered as
a dimensional reduction of the $3+1$ Bertotti--Robinson solution.

With the aim of demonstrating the uniqueness of this class of
uniform solutions with $H(r)=h_{0}=\rm{constant}$, even in the
framework of stationary fields, let us consider the general metric
with $W(r)$:\\
\noindent In the case $\{a=0=c, b\neq\,0, H(r)=h_{0}\}$, the
combination of equations
${E_{t}}^{t}-{E_{r}}^{r}-W(r){E_{t}}^{\phi}=2\,{\frac {  W \left( r
\right)^{2}{h_{0}}\,{b}^{2}}{ F \left( r \right) }}$ implies
$W(r)=0$ and consequently the gravitational field is static;
further integration gives rise to the above--mentioned uniform electrostatic fields.\\
\noindent Case $\{a\neq\,0, b=0=c, H(r)=h_{0}\}$:
from${E_{t}}^{t}-{E_{r}}^{r}-W(r){E_{t}}^{\phi}=2\,a^2h_{0}/F(r)$
therefore there is no solution here.\\
\noindent Case $\{a=0=b, c\neq\,0, H(r)=h_{0}\}$: the combination
${E_{t}}^{t}-{E_{r}}^{r}-W(r){E_{\phi}}^{t}=2c^2/F(r)$, hence there
is no solution.

\subsection{Uniform ``stationary'' electromagnetic
$\bm{A}={r}/(a\,l^2)(\bm{d\phi}+W_{0}\bm{dt})$
solutions}\label{uniforma}

In the case of positive cosmological constant $\Lambda=1/l^2$ there
exists a uniform ``stationary'' magnetic solution with constant
$W(r)=W_{0}$.

The electromagnetic tensor possesses the structure
$$F_{{\mu}{\nu}}=-2\,a \frac{H}{F}W\,{\delta_{\mu}}^{[t}{\delta_{\nu}}^{r]}
-2\,a\,\frac{H}{F}\,{\delta_{\mu}}^{[\phi}{\delta_{\nu}}^{r]},$$ and
its covariant derivatives occurs to be zero if
$$F_{{r}{t};r}=-\frac{a\,H^2}{2F}\frac{dW}{dr},$$
$$F_{{r}{\phi};r}=-\frac{a}{4\,F\,H}(-2H^2\frac{d}{dr}\frac{F}{H}+H^3\frac{dW}{dr})$$
vanish. Hence
\begin{eqnarray}\label{ClementCLmg2}
F(r)=\beta\,H(r),\,W(r)=W_{0}=\rm{constant},\nonumber\\
\end{eqnarray}
and consequently
\begin{eqnarray}\label{ClementCLmg3}
F_{t\,r}=-W_{0}\frac{a}{\beta},\,F_{\phi\,r}=-\frac{a}{\beta}.
\end{eqnarray}
The Einstein equations reduce to
\begin{eqnarray}\label{ClementCLmg4}
{E_{r}}^{r}&&=-\frac{a^2}{\beta}+\frac{1}{l^2}
=0\rightarrow{\beta=a^2\,l^2},\,
2\,{E_{t}}^{t}=\frac{d^2}{dr^2}F(r)+\frac{4}{l^2}=0\nonumber\\&&
\rightarrow{F(r)=-\frac{2}{l^2}r^2+c_{1}r+c_{0}}.
\end{eqnarray}
The metric and fields for the derived solution can be expressed as
\begin{eqnarray}\label{Clementcospos}
\bm{g}&&=-a^2\,l^2\bm{dt}^2+\frac{\bm{dr}^2}{F(r)}
+\frac{F(r)}{a^2l^2}(\bm{d\phi}+W_{0}\bm{dt})^2,\nonumber\\
&&F(r)=-\frac{2r^2}{l^2}+c_{1}r+c_{0},\,\nonumber\\
&&\bm{A}=\frac{r}{a\,l^2}(\bm{d\phi}+W_{0}\bm{dt}),
\end{eqnarray}
with uniform electromagnetic field tensor is
$$F_{\mu\nu}=-2\frac{W_{0}}{a\,l^2}\delta_{[\mu}^{t}\delta_{\nu]}^{r}
-2\frac{1}{a\,l^2}\delta_{[\mu}^{\phi}\delta_{\nu]}^{r},$$ and
energy--momentum tensor
$${8\pi\,T_{\nu}}^{\mu}=\frac{1}{l^2}[-\delta_{\nu}^{t}\delta_{t}^{\mu}
+\delta_{\nu}^{r}\delta_{r}^{\mu}+\delta_{\nu}^{\phi}\delta_{\phi}^{\mu}
+2\,W_{0}\delta_{\nu}^{t}\delta_{\phi}^{\mu}].$$

\noindent This solution is equivalent to the Clement's solution
given by Eq.~(Cl.26, $\Lambda=1/l^2$) of ~\cite{Clement93}.

\subsubsection{No uniform stationary generalization
of the magnetostatic solution for $\Lambda=-1/l^2$}

On the contrary, as far as to the stationary uniform electromagnetic
branch with $a\neq 0$ and negative cosmological constant
$\Lambda=-1/l^2$ is concerned, one establishes that there is no
solution at all. Following a similar procedure as the one used in
the previous case, where now $b=0,\,
F(r)=\beta\,H(r),\,W(r)=W_{0}=\rm{constant.}$ The Einstein equation
${E_{r}}^{r}=-\frac{a^2}{\beta}-\frac{1}{l^2} =0$ yields
${\beta=-a^2\,l^2}$. The covariant tensor components $g_{rr}$ and
$g_{\phi\,\phi}$ explicitly amount to $
g_{rr}=1/F(r)>0,\,g_{\phi\,\phi}=-{a^2\,l^2}F(r)<0, $ which yields a
contradiction with the adopted signature $\{-,+,+\}$. Hence this
case does not represent a solution compatible with the 2+1 metric
signature.

\section{Constant electromagnetic invariants'
solutions}\label{constant}

This section is devoted to the derivation of the electromagnetic
fields coupled to stationary (static) cyclic symmetric 2+1
gravitational fields such that their electromagnetic invariants
$FF$, $T$ and $TT$ are constants;  because of the proportionality of
$T$ and $TT$ to $FF$, it is enough to establish under which
conditions
$$F_{\mu\nu}F^{\mu\nu}=-2\,{\frac {{c}^{2}}{F}}+2\,{\frac {{H}
\left( a-Wb \right) ^{2}}{ F}}-2\,{\frac {{b}^{2}}{{H}}}$$
vanishes.\\
It should be pointed out that  the hybrid $c\neq 0$ class of
spacetimes does not allow for constant invariants' solution when a
cosmological constant is present. Therefore it is sufficient to
restrict oneself to the case $a\neq 0$ or $b\neq 0$.

In what follows we shall search for sub--classes of solutions with
constant electromagnetic invariant, namely those with $\{a\neq 0,
b\neq\,0, W(r)=W(r)\}$, $\{a= 0, \,b\neq\,0,
\omega(r)=\omega_{0}\}$,
and $\{a\neq 0, b=0\,, W(r)=W_{0}\}$ families of solutions.\\
At this stage it is worthwhile to point out that constant invariant
electromagnetic fields contain, as sub--classes, the covariantly
constant electromagnetic field solutions, while the inverse
statement does not hold.

\subsection{General constant electromagnetic
invariant $F_{\mu\nu}F^{\mu\nu}=2\gamma$ case for $a\neq0\neq b$}

Restricting the present section to the study of the cases $a\neq0$
and $b\neq0$, the constancy of $FF=2\gamma$ is guaranteed by
\begin{eqnarray}\label{homog1}
\frac{F}{H^2}(b^2+\gamma\,H)=(a-b\,W)^2,
\end{eqnarray}
which yields
\begin{eqnarray}\label{homog2}
W(r)=\frac{a}{b}\pm\frac{\sqrt{F(r)}}{b\,H(r)}\sqrt{b^2+\gamma\,H(r)}.
\end{eqnarray}
In general, the following combinations of the Einstein equations
give
\begin{eqnarray}\label{homog3}
&&{E_{t}}^{t}+{E_{\phi}}^{\phi}+2{E_{r}}^{r}=
\frac{d^2F}{dr^2}-\frac{8}{l^2}+4\frac{b^2F-H^2(a-b\,W)^2}{F\,H}=0,\nonumber\\
&&2HF({E_{t}}^{t}-{E_{r}}^{r}-W\,{E_{\phi}}^{t})=
F^2\frac{d^2H}{dr^2}+4H^2(a-b\,W)^2=0.
\end{eqnarray}
Using the relation~(\ref{homog1}) one brings Eq.~(\ref{homog3}) to
the form
\begin{eqnarray}\label{homog4}
\frac{d^2F}{dr^2}=\frac{8}{l^2}+4\gamma,\,\frac{d^2H}{dr^2}
+\frac{4\gamma}{F}H=-\frac{4\,b^2}{F},\nonumber\\
\end{eqnarray}
with solution for $F(r)$
\begin{eqnarray}\label{homog4x}
F(r)&&=(\frac{4}{l^2}+2\gamma)r^2+C_{1}r+C_{0}
 =(\frac{4}{l^2}
+2\gamma)(r-r_{1})(r-r_{2}),
\end{eqnarray}
and the function $H$, as solution of its second--order equation
(\ref{homog4}), is expressed in terms of hypergeometric functions.
Nevertheless there exists a shortcut for the integration of the
function $H(r)$ by noticing that ${E_{r}}^r$ contains only first
derivatives of the structural functions: replacing W(r)
from~(\ref{homog2}) in the quoted equation, after extracting square
root, one arrives at
\begin{eqnarray*}\label{homogHx}
&&\gamma\,F\,\frac{d H}{d r}-(b^2+\gamma\,H)\,\frac{d F}{d r} =\pm
4\frac{b}{l}\,\sqrt{1+l^2\gamma}\,\sqrt{F}\,\sqrt{b^2+\gamma\,H},
\end{eqnarray*}
which, when introducing the auxiliary function
$Q(r)^2:=b^2+\gamma\,H(r)$, becomes
\begin{eqnarray*}\label{homogH2}
\frac{d}{dr}\frac{Q}{\sqrt{F}}
\mp\,2\frac{b}{l\,F}\sqrt{1+l^2\gamma}=0.
\end{eqnarray*}
Integrating this equation, using $F(r)$ (\ref{homog4x}), one obtains
\begin{eqnarray}\label{homogQ}
Q&=&\pm\,{b}{l}\frac{\sqrt{1+l^2\gamma}}
{2+l^2\gamma}\frac{\ln{(r-r_{2})}-\ln{(r-r_{1})}}{r_{2}-r_{1}}
+\beta\sqrt{F(r)}.
\end{eqnarray}
Fulfilling a single constraint still remains; using the
expressions of the function $W(r)$ and its derivatives from the
${E_\mu}^\nu$--equations as well as the derivatives of F(r),
together with $H(r)$ in terms of $Q(r)$ and the derivative of the
latter from (\ref{homogH2}), one gets a single equation
\begin{eqnarray}\label{conQ}
&&2(1+l^2\gamma)\left(Q(r)^2 +b^2\right)\sqrt{F}+l\,b\,Q(r)
\sqrt{1+l^2\gamma}\frac{dF}{dr}=0,
\end{eqnarray}
which is incompatible with $Q(r)$ determined in (\ref{homogQ})
except for $\gamma=-1/l^2$.

Hence, we conclude that there are no solutions for arbitrary
constant electromagnetic invariant $FF=2\,\gamma$.

\subsubsection{Constant electromagnetic invariant $FF=\mp2/l^2$
solution}

A class of constant electromagnetic invariants' stationary solutions
with $a\neq0$, $b\neq0$ and $\gamma=-1/l^2$, $FF=-2/l^2$, arises for
\begin{eqnarray}\label{Newunif1c}
W(r)=\frac{a}{b}\mp\frac{F}{b\,l\,H(r)}\sqrt{b^2\,l^2-H}\,.
\end{eqnarray}
As in the previous case, $F(r)$ and $H(r)$ fulfill the
Eq.~(\ref{homog4}) for $\gamma=-1/l^2$ and correspondingly their
integrals are given by
\begin{eqnarray*}\label{Newunif5c}
F(r)&&=\frac{2}{l^2}r^2+c_{1}r+c_{0},\,\, H(r)=b^2l^2-\beta^2
l^2\,F(r).
\end{eqnarray*}
There are no further constraints from the field equations.
Consequently the final result can be written as
\begin{eqnarray*}\label{Newunif7c}
&&\bm{g}=-\frac{b^2\,l^2-H(r)}{l^2\,\beta^2\,H(r)}\bm{dt}^2
+l^2\,\beta^2\,\frac{\bm{dr}^2}{b^2\,l^2-H(r)}
+H(r)\left[\bm{d\phi}+\left(\frac{a}{b}\mp
\frac{1}{l^2\, b\,\beta}\frac{b^2l^2-H}{H}\right)\bm{dt}\right]^2,\nonumber\\
&&H(r)=-{2}{\beta^2}r^2+C_{1}r+C_{0},\nonumber\\
&&\bm{A}=-\beta\,r\,[\bm{d\phi}-\frac{1\pm\,
al^2\beta}{l^2\,b\,\beta}\bm{dt}],
\end{eqnarray*}
which coincides with the uniform solution (\ref{Newunif5}).
Therefore we have determined a class of uniform constant
electromagnetic invariant stationary solutions for both non--vanishing
constants $a\neq0\neq b$. Although the solution mentioned above has been
derived for $\Lambda=-1/l^2$, the branch with positive
$\Lambda=1/l^2$ is achieved from the above--mentioned expressions by changing
$l^2\rightarrow{-l^2}$.

\subsubsection{Vanishing invariant
$F_{\mu\nu}F^{\mu\nu}=0$ solution}

The next in simplicity class of solutions corresponds to the family
with vanishing invariant $FF=0$, $\gamma=0$. In such a case
\begin{eqnarray}\label{homog4kk1}
W(r)&=&\frac{a}{b}\pm\frac{\sqrt{F(r)}}{H(r)}.
\end{eqnarray}
The first--order equations (\ref{homogHx}) and (\ref{conQ}) give rise
to a single first order equation for $F(r)$, namely
\begin{eqnarray}\label{homog4kk1a}
\frac{d F}{d
r}=\mp\frac{4}{l}\sqrt{F}\rightarrow\,F(r)=\frac{4}{l^2}\,(r-C)^2.\nonumber\\
\end{eqnarray}
The integration of the equation (\ref{homog4}) for $H(r)$
$$\frac{d^2 H}{d r^2}=-\frac{4b^2}{F} $$
yields
\begin{eqnarray}\label{homog4kk1b}
H(r)=C{0}+C_{1}r+{b^2\,l^2}\ln{(r-C)}.
\end{eqnarray}
Therefore, the vector field potential occurs to be
\begin{eqnarray}
\bm{A}
=\mp\frac{b\,l}{2}\ln{(r-C)}\left(\bm{d\phi}+\frac{a}{b}\bm{dt}\right).
\end{eqnarray}
This solution corresponds to a possible representation of the
Kamata--Koikawa \cite{KamataK95} solution
to be treated in detail in Section~\ref{AsdMf}.\\
It should be pointed out that this solution does not belong to the
family of uniform solutions, i.e., the fields possessing
vanishing covariant derivatives.

\subsection{Constant electromagnetic invariant $FF=-2/l^2$ solution
for $b\neq0$}

The constant electromagnetic invariant solution with $b\neq0$ can be
determined as solution of the Einstein--Maxwell equations by
considering the stationary metric in the form
\begin{eqnarray*}\label{commetricCLc1}
\bm{g}&=&-\frac{F}{h}\left(\bm{dt}
-{\omega}\,\bm{d\phi}\right)^2+h\,{\bm{d\phi}}^2
+\frac{{\bm{dr}}^2}{{F}} =-\frac{{F}}{{H}}\bm{dt}^2
+\frac{\bm{dr}^2}{{F}}
+{H}\left(\bm{d\phi}+{W}\,\bm{dt}\right)^2,\nonumber\\
{F}&=&F,\, {H}={h}-\frac{F}{h}\omega^2,\,
{W}=\frac{\omega}{H}\frac{F}{h}.
\end{eqnarray*}
Demanding the electromagnetic invariant $FF$ in the case $b\neq0=a$
to be constant, one establishes
\begin{eqnarray}\label{fmnCLc1}
FF=-\frac{2b^2}{h(r)}\rightarrow\,h(r)=h_{0}=\rm
{constant.}\nonumber\\
\end{eqnarray}
Substituting $h(r)=h_{0}$ into the Einstein equations one obtains
from ${E_{\phi}}^{t}$ that
\begin{eqnarray}\label{fmnCLc2}
\frac{d^2}{dr^2}\omega=-\frac{2}{F}\frac{d
F}{dr}\frac{d\omega}{dr},
\end{eqnarray}
which when used in ${E_{t}}^{t}$-${E_{r}}^{r}$ yields
\begin{eqnarray}\label{fmnCLc3}
\frac{F}{h_{0}^2}(\frac{d\omega}{dr})^2=0\rightarrow\,\omega(r)=\omega_{0}.
\end{eqnarray}
Replacing $\omega=\omega_{0}$ and $h=h_{0}$ in the remaining
equations one establishes
\begin{eqnarray}\label{fmnCL3y}
\frac{d^2}{d\,r^2}F(r)&=&\frac{4}{l^2}\rightarrow\,
F(r)=\frac{2}{l^2}r^2+c_{1}\,r+c_{0},\, h_{0}=b^2\,l^2
\end{eqnarray}
Therefore we arrive at a constant electromagnetic invariant
solution in the form
\begin{eqnarray}\label{ClementCL4y}
\bm{g}&=&-\frac{F}{h_{0}}(\bm{dt}-\omega_{0}\bm{d\phi})^2
+h_{0}\,{\bm{d\phi}}^2+\frac{\bm{dr}^2}{F(r)}
,\nonumber\\
F(r)&=&\frac{2r^2}{l^2}+c_{1}r+c_{0}, h_{0}=b^2\,l^2,\nonumber\\
\bm{A}&=&\frac{r}{b\,l^2}(\bm{dt}-\omega_{0}\bm{d\phi}),
\end{eqnarray}
which in all respects is identical to the uniform electromagnetic
solution (\ref{ClementCL4u}) derived in the previous section. Notice
that this solution exists only for negative cosmological constant,
$\Lambda=-1/l^2$, there is no extension to $\Lambda=1/l^2$. It is
evident that this solution can be generated from the static one,
(\ref{metricBRZ}), via the transformations
$t\rightarrow\,t-\omega_{0}\phi,\,\phi\rightarrow \phi.$\\

For $\omega_{0}=0$, the above--mentioned metric and field reduce to
the Matyjasek--Zaslavskii solutions, see Section~(\ref{Matyja}),
thus this class of constant electromagnetic invariants' static
solutions occurs to be unique with the additional property of being
a uniform static solution.

\subsection{Constant electromagnetic invariant
$FF=2/l^2$ stationary solution for $a\neq0$}

In the case of positive cosmological constant $\Lambda=1/l^2$
there exists a constant electromagnetic invariant stationary
$a\neq0$ solution. Requiring the constancy of the electromagnetic
invariant $FF=2a^2\frac{H}{F}$, one gets
\begin{eqnarray}\label{fmnCLc1x}
FF=2a^2\frac{H}{F}\rightarrow\,H(r)=\beta^2F(r).
\end{eqnarray}
From ${E_{\phi}}^{t}$ one establishes
\begin{eqnarray}\label{fmnCLc2x}
\frac{d^2}{dr^2}W=-\frac{2}{F}\frac{dF}{dr}\frac{dW}{dr},
\end{eqnarray}
which when used in ${E_{t}}^{t}$-${E_{r}}^{r}$ yields
\begin{eqnarray}\label{fmnCLc3x}
\frac{F}{\beta^4}(\frac{dW}{dr})^2=0\rightarrow\,W(r)=W_{0}
\end{eqnarray}
Using $W=W_{0}$ and $H(r)=\beta^2F(r)$ in the remaining Einstein
equations, one gets
\begin{eqnarray}\label{fmnCL3x}
&&\frac{d^2}{d\,r^2}F(r)+\frac{4}{l^2}=0 \rightarrow\,
F(r)=-\frac{2}{l^2}r^2+c_{1}\,r+c_{0}.
\end{eqnarray}
Therefore we have established that there is a unique constant
electromagnetic invariants' solution given by
\begin{eqnarray}\label{Clementcosposx}
\bm{g}&=&-a^2l^2\bm{dt}^2+\frac{\bm{dr}^2}{F(r)}
+\frac{F(r)}{a^2l^2}(\bm{d\phi}+W_{0}\bm{dt})^2,\nonumber\\
F(r)&=&-\frac{2r^2}{l^2}+c_{1}r+c_{0},\,\nonumber\\
\bm{A}&=&\frac{r}{a\,l^2}(\bm{d\phi}+W_{0}\bm{dt}),
\end{eqnarray}
which is identical to the uniform electromagnetic solution
(\ref{Clementcospos}) derived in the previous Section~\ref{uniform}
dealing with uniform electromagnetic solutions. Notice that this
solution exists only for positive cosmological constant,
$\Lambda=1/l^2$, there is no $\Lambda=-1/l^2$ solution within this
class.

\section{(Anti--) Self--dual Maxwell fields; $FF=0$ }\label{AsdMf}

A particular family of stationary cyclic symmetric solutions arises
by demanding the vanishing of the electromagnetic invariant
$F_{\mu\nu}F^{\mu\nu}$,
\begin{eqnarray}
F_{\mu\nu}F^{\mu\nu}&&=2\,\frac { H  \left( a-b\,W  \right) ^{2}}{F
}-2\,{\frac {b^{2}}{H }}=0\rightarrow\,W(r)=\frac{a}{b}\pm \frac {
\sqrt {F }}{  H }.
\end{eqnarray}
For the above--mentioned $W(r)$, the equation ${E_{r}}^{r}$ gives
\begin{eqnarray}
&&{l}^{2}\left( {\frac {d\,F}{dr}} \right) ^{2}-16\,F
=0\rightarrow\,F(r)=4\,{\frac { \left( r-C \right) ^{2}}{{l}^{2}}}.
\end{eqnarray}
After the substitution of $W(r)$ and F(r) into the Einstein
equations, the remaining equation to be solved amounts to
\begin{eqnarray*}
&&\left( r-C \right) ^{2}\frac {d^{2}}{ d{r}^{2}}H+{b}^{2}{l}^{2} =0
 \rightarrow\,H(r)=C_{0}+C_{1}\,r+{b}^{2}{l}^{2} \ln
\left(r-C\right).
\end{eqnarray*}
The gravitational and electromagnetic fields of this solution can be
given as
\begin{eqnarray}\label{kkmetric}
&&\bm{g}=-\frac{F}{H}\bm{dt}^2+\frac{\bm{dr}^2}{F}
+H(\bm{d\phi}+W\bm{dt})^2,\nonumber\\
&&H(r)=C_{0}+C_{1}\,r+{b}^{2}{l}^{2} \ln \left(r-C \right) ,\,
F(r)=4\,{\frac { \left( r-C \right) ^{2}}{{l}^{2}}},\,
W(r)=\frac{a}{b}\pm \frac { \sqrt {F }}{  H },\nonumber\\
&&\bm{A}=\mp\frac{\,l}{2}\ln{(r-C)}\left(a\bm{dt}+b\bm{d\phi}\right).
\end{eqnarray}
\noindent This solution is characterized by the field tensor
\begin{eqnarray}
F_{\mu\nu}&&=\pm\,\frac{l}{\,(r-C)}\left(a{\delta_{[\mu}}^{t}{\delta_{\nu]}}^{r}
 +\,b{\delta_{[\mu}}^{\phi}{\delta_{\nu]}}^{r}\right),\nonumber\\
F^{\mu\nu}&&= 2b{\delta_{t}}^{[\mu}{\delta_{r}}^{\nu]}
 -2a{\delta_{\phi}}^{[\mu}{\delta_{r}}^{\nu]},
\end{eqnarray}
with energy--momentum tensor
\begin{eqnarray}
{ T_{\mu}}^{\nu}&&=
\frac{l}{4\,\pi\,(r-C)}\left(-{a\,b}{\delta_{\mu}}^{t}{\delta_{t}}^{\nu}
+{a\,b}{\delta_{\mu}}^{\phi}{\delta_{\phi}}^{\nu}
+a^2{\delta_{\mu}}^{t}{\delta_{\phi}}^{\nu}-b^2
{\delta_{\mu}}^{\phi}{\delta_{t}}^{\nu}\right).
\end{eqnarray}
Notice that the three invariants $F_{\mu\nu}F^{\mu\nu}$,
$T_{\mu}^{\mu}$ and $T_{\mu\nu}T^{\mu\nu}$ are equal to zero.
Without any loss of generality one can always set $C=0$.

\subsection{Kamata--Koikawa solution}

Kamata and Koikawa \cite{KamataK95} reported their electrically
charged BTZ black hole with negative cosmological constant
such that the Maxwell field is self (anti-self) dual, condition
which is imposed on the orthonormal basis components of the electric
field and the magnetic field. This solution describes an
electrically charged extreme black hole with mass $M$, angular
momentum $J$, and electric charge $Q$. To achieve their
representation one accomplishes in metric~(\ref{kkmetric}) the
substitutions
\begin{eqnarray}
&&r=\rho^2,\,t\rightarrow{t\sqrt{Q}/2},
\,\phi\rightarrow{\phi/\sqrt{Q}},\,C_{1}\rightarrow{Q},l\rightarrow
{|\Lambda|^{-1/2}}\,{C_{0}}\rightarrow{-b^2/\Lambda\,\ln{\rho_{0}^2}},
\,C=\rho_{0}^2,\nonumber\\
&&H/Q\rightarrow{K^2},W\,Q/2
\rightarrow{\frac{a}{b}\frac{Q}{2}+N^{\phi}},\, F\rightarrow
{4\,\rho^2\,L^2},
\end{eqnarray}
arriving at the solution
\begin{eqnarray}\label{kkmetricp}
&&\bm{g}=-\rho^2\frac{L^2}{K^2}\bm{dt}^2+\frac{\bm{d\rho}^2}{L^2}
+K^2[\bm{d\phi}+(\frac{a}{b}\frac{Q}{2}+N^{\phi})\bm{dt}]^2,\nonumber\\
&&L^2=|\Lambda|(\rho-\rho_{0}^2/\rho)^2,\,K^2=\rho^2+\frac{b^2}{Q\,\Lambda}\ln
\left(\frac{\rho^2-\rho_{0}^2 }{\rho_{0}^2}\right),\,N^{\phi}=\pm
\frac { \rho\,L}{ K^2},\nonumber\\ &&\bm{A}=
\frac{b}{2\sqrt{Q\,|\Lambda|}}\ln{(\frac{\rho^2-\rho_{0}^2}{\rho_{0}^2})}
\left(\bm{d\phi}+\frac{a}{b}\frac{Q}{2}\bm{dt}\right).
\end{eqnarray}
The electromagnetic field tensors occur to be
\begin{eqnarray}
&&F_{\mu\nu}=\frac{\rho}{\sqrt{|\Lambda|}(\rho^2-\rho_{0}^2)}
\left({a}\sqrt{Q}{\delta_{[\mu}}^{t}{\delta_{\nu]}}^{r}
+2\frac{b}{\sqrt{Q}}{\delta_{[\mu}}^{\phi}{\delta_{\nu]}}^{r}
\right),\nonumber\\
&&{T_{\mu}}^{\nu}=\frac{1}{8\pi}\frac{1}{\sqrt{|\Lambda|}(\rho^2-\rho_{0}^2)}
\left[a\,b({\delta_{\mu}}^{t}{\delta_{t}}^{\nu}-{\delta_{\mu}}^{\phi}{\delta_{\phi}}^{\nu})
-\frac{Q\,a^2}{2}{\delta_{\mu}}^{\phi}{\delta_{t}}^{\nu}
-\frac{2b^2}{Q}{\delta_{\mu}}^{t}{\delta_{\phi}}^{\nu} \right].
\end{eqnarray}
Next, one restores the factor $\pi\,G$ in the above--mentioned
solution through the identifications of the physical parameters:
\begin{eqnarray*}
&&\rho_{0}^2=4\pi\,G\,\,Q^2/|\Lambda|
=\frac{\epsilon}{2|\Lambda|^{1/2}}J,\,
b=2\sqrt{\pi\,G}{Q^{3/2}},\,a=\pm\,4\sqrt{\pi\,G}|\Lambda|^{1/2}Q^{1/2},
\end{eqnarray*}
arriving at the metric~(\ref{kkmetricp}) with structural functions
\begin{eqnarray}
&&L^2=|\Lambda|(\rho-\rho_{0}^2/\rho)^2,\, \,N^{\phi}=\pm \frac {
\rho\,L}{ K^2},\, K^2=\rho^2+\rho_{0}^2\,\ln
\left(\frac{\rho^2-\rho_{0}^2 }{\rho_{0}^2}\right)
\end{eqnarray}
and electromagnetic field tensors
\begin{eqnarray}
&&\bm{A}=
Q\sqrt{\pi\,|\Lambda|}\ln{(\frac{\rho^2-\rho_{0}^2}{\rho_{0}^2})}
\times\left(\frac{1}{\sqrt{|\Lambda|}}\bm{d\phi}+\bm{dt}\right),\nonumber\\
&&F_{\mu\nu}=-4Q\sqrt{\pi\,G}\frac{\rho}{\rho^2-\rho_{0}^2}
\left({\delta_{[\mu}}^{t}{\delta_{\nu]}}^{r}
+\frac{1}{\sqrt{|\Lambda|}}{\delta_{[\mu}}^{\phi}{\delta_{\nu]}}^{r}
\right),\nonumber\\
&&{T_{\mu}}^{\nu}=\frac{Q^2\,G}{\rho^2-\rho_{0}^2}
\left(-{\delta_{\mu}}^{t}{\delta_{t}}^{\nu}+{\delta_{\mu}}^{\phi}{\delta_{\phi}}^{\nu}
-\frac{1}{\sqrt{|\Lambda|}}{\delta_{\mu}}^{\phi}{\delta_{t}}^{\nu}
+{\sqrt{|\Lambda|}}{\delta_{\mu}}^{t}{\delta_{\phi}}^{\nu} \right).
\end{eqnarray}

It should be pointed out that Clement \cite{Clement93} also reported
a metric expression and electromagnetic vector field describing a
solution with  vanishing electromagnetic invariants. Comments
concerning the mass content of this solution can be found
in~\cite{Chan96}. This solution is horizonless and consequently does
not permit a black hole interpretation.

\section
{General stationary cyclic symmetric solution for
 electromagnetic field ${\ast {\bm{F}}}
=c\frac{g_{rr}}{\sqrt{-g}}\bm{dr}$}\label{hybrid}

The main goal of this section is to derive the stationary
cyclic symmetric spacetime corresponding to the case $c\neq0$, i.e.,
for the vector potential
\begin{eqnarray}
\bm{A}=\frac{c}{2}(t\bm{d\phi}-\phi \bm{dt}).
\end{eqnarray}
It is worthwhile to point out that this case has no analog in
stationary axial symmetric spacetimes of the standard 3+1
Einstein--Maxwell theory. \noindent The set of field equations is
given by: $\{{E_{t}}^{t}, {E_{t}}^{\phi}, {E_{r}}^{r},
{E_{\phi}}^{t}, {E_{\phi}}^{\phi}\}$. In the forthcoming subsections
two main families of solutions exhibiting the
hybrid feature of the vector potential are derived.

\subsection{Ayon--Cataldo--Garcia hybrid electromagnetic stationary
solution}

The starting point in the integration process of the system of field
equations is ${E_{\phi}}^t(a=0=b)=0$, (\ref{Eins31}), which
possesses a first integral of the form
\begin{eqnarray}\label{eeq10}
W_{,r}= \frac{J}{H^2},
\end{eqnarray}
where $J$ is an integration constant. \noindent The combination
$E{_{t}}^{t}+2{E_{r}}^{r}+{E_{\phi}}^{\phi}$, for $a=0=b$, yields
\begin{eqnarray}\label{eeq11}
{F_{,r,r}}-\frac{8}{l^2}=0,
\end{eqnarray}
which possesses the general solution
\begin{eqnarray}\label{eeq12}
F=\frac{4}{l^2}(r-r_{1})(r-r_{2}),
\end{eqnarray}
where $r_{1}$ and $r_{2}$ are constant of integration. \noindent
Next, using $W_{,r}$ from Eq.~(\ref{eeq10}) in ${E_{r}}^{r}(a=0=b)$,
 (\ref{Eins22}), one arrives at
\begin{eqnarray}\label{eeq13}
\frac{1}{4}&&\left(\frac{H_{,r}}{H}-\frac{1}{2}\frac{F_{,r}}{F}\right)^2-
\frac{{J}^2}{4H^2\,F}=\frac{{F_{,r}}^2}{{16}F^2} -\frac{{c}^2}{F^2}-
\frac{1}{l^2\,F}.
\end{eqnarray}
The evaluation the right--hand side of this equation gives the same
result as in the static case, thus one gets
\begin{eqnarray}\label{metricCat6ax}
\left[ \frac {d}{dr}\ln\left(\frac{H}{F^{1/2}}\right) \right] ^{2}-
\frac{{J}^2}{H^2\,F}=4\frac{ ({r_{2}}-r_{1})^2\alpha}{l^4\,F^2},
\end{eqnarray}
where $\alpha$ is defined through
$$\nonumber\\
c^2=\frac{(r_{2}-r_{1})^2}{l^4}(1-\alpha).$$
 From the above equation it becomes
apparent that $H$ can be sought in the form of
\begin{eqnarray}\label{eeq13M}
H(r)={h\left( r \right)}\,\sqrt {F \left( r \right) }.
\end{eqnarray}
Replacing $H(r)$ from above into (\ref{metricCat6ax}) one obtains an
equation for $h(r)$  which can be given as
\begin{eqnarray}
&&\frac {d\,h  }{\sqrt {{\alpha_{0}}^2\, h ^{2}+{J}^{2}}}=
\mp\,\frac {d\,r}{F },\,
{\alpha_{0}}:=2({r_{2}}-r_{1})\sqrt{\alpha}/l^2,
\end{eqnarray}
with integral
\begin{eqnarray}
&&\ln{\left( \alpha_{0}\,h  + \sqrt {{\alpha_{0}}^2\,  h ^{2}
+{J}^{2}} \right)} =\ln{\left[k_{1}\left( {\frac {r-r_{1}}
{r-{r_{2}}}} \right)^{\pm\,\sqrt{\alpha}/2}\right]},
\end{eqnarray}
where $k_{1}$ is an integration constant. Therefore $h(r)$ can be
expressed as
\begin{eqnarray}
h(r)=&&\frac{l^2\,k_{1}}{4({r_{2}}-r_{1})\sqrt{\alpha}}\,
\left[\left(\frac {r-r_{1}}{r-{r_{2}}}\right)^{\pm\,\sqrt{\alpha}/2}
-\frac{J^{2}}{k_{1}^2} \left(\frac
{r-r_{1}}{r-{r_{2}}}\right)^{\mp\,\sqrt{\alpha}/2}\right].
\end{eqnarray}
The integration of the Eq.~(\ref{eeq10}) for $W$ does not present
problem.\\

Summarizing the derived above results, one has that this family of
solutions can be given by
\begin{eqnarray*}\label{metricscH}
\bm{g}&=&-\frac{F}{H}\bm{dt}^2+\frac{\bm{dr}^2}{F}
+H(\bm{d\phi}+W\bm{dt})^2,
\end{eqnarray*}
\begin{eqnarray}
&&F=\frac{4}{l^2}(r-r_{1})(r-r_{2}),\nonumber\\
&&H(r)=l\frac{\sqrt{(r-r_{1})(r-r_{2})}}{2K_{1}
(r_{2}-r_{1})\sqrt{\alpha}}
\times\left[\left(\frac{r-r_{1}}{r-r_{2}}\right)^{\pm\frac{\sqrt{\alpha}}{2}}
-K_{1}^2\,J^2\left(\frac{r-r_{1}}{r-r_{2}}\right)^{\mp\frac{\sqrt{\alpha}}{2}}\right],
\nonumber \\
&&W(r)=W_{0}\pm\frac{4}{l^2}J\,K_{1}^2\sqrt{\alpha}(r_{2}-r_{1})
\times\left[\left(\frac{r-r_{1}}{r-r_{2}}\right)^{\pm\,\sqrt{\alpha}}
-K_{1}^2\,J^2\right]^{-1},\nonumber\\
&&\bm{A}=\frac{c}{2}(t\bm{d\phi}-\phi \bm{dt}),
\end{eqnarray}
where the constant $K_{1}$ stands for $1/k_{1}$, $K_{1}=1/k_{1}$,
and $W_{0}$ is an integration constant. Recall that the parameter
$\alpha$ is related to $c$, $r_{1}$ and $r_{2}$ through
$c^2=\frac{(r_{2}-r_{1})^2}{l^4}(1-\alpha)$.

\noindent Correspondingly, the electromagnetic field tensors are
\begin{eqnarray}
F_{\mu\nu}&&=2c{\delta_{[\mu}}^{t}{\delta_{\nu]}}^{\phi},
\nonumber\\
{T_{\mu}}^{\nu}&&= { \frac {{c}^{2}}{8\pi \,F
}}(-{\delta_{\mu}}^{t}{\delta_{t}}^{\nu}
+{\delta_{\mu}}^{r}{\delta_{r}}^{\nu}
-{\delta_{\mu}}^{\phi}{\delta_{\phi}}^{\nu}),
\end{eqnarray}
with invariants
\begin{eqnarray*}\label{field}
F_{\mu\nu}F^{\mu\nu}&&= -2\frac{c^2}{F},\,T_{\mu\nu}T^{\mu\nu}=\frac
{3}{64}\,\frac {{c}^{4}}{{ \pi }^{2} F
 ^{2}},\,
T_{\mu}^{\mu}=-\frac{1}{8}\,{\frac {{c}^{2}}{\pi \,F }}.
\end{eqnarray*}
This solution has been reported, for the first time to our
knowledge, in~\cite{AyonCG04}. The static hybrid solution
~(\ref{hybridstatic}) arises from the stationary one above by
setting $J=0=W_{0}$ and identifying
${4(r_{2}-r_{1})\sqrt{\alpha}\,K_{0}^2}=l^2/K_{1}$.

\subsubsection{The ACG hybrid solution allowing for BTZ limit}

To achieve a representation of this hybrid solution in terms of the
radial coordinate $\rho$, such that at the limit of vanishing
electromagnetic parameter $c=0 \rightarrow \alpha=1$, the stationary
BTZ solution would arise, one has to accomplish the coordinate
transformations
\begin{subequations}
\begin{equation}
t=\frac{l}{4K_{1}}\frac{1-K_{1}^2J^2}{r_{2}-r_{1}}\,T,
\end{equation}
\begin{equation}
\phi=\Phi-\left(W_{0}\frac{l}{4K_{1}}\frac{1-K_{1}^2J^2}{r_{2}-r_{1}}
+\frac{K_{1}}{l}J\right)\,T,
\end{equation}
\begin{eqnarray}
r&&=\frac{1}{1-K_{1}^2\,J^2}\left(r_{1}-r_{2}K_{1}^2\,J^2
-2\frac{K_{1}}{l}(r_{2}-r_{1})\,{\rho}^2\right),
\end{eqnarray}
\end{subequations}
where with $\{T,\rho,\Phi\}$ are denoted the corresponding BTZ
coordinates, which ought to be accompanied with the identification
\begin{eqnarray}
J^2\,K_{1}&&=-R_{(-)}:=-\left(M\,l-\sqrt{M^2l^2-J^2}\right),\,
M=-\frac{1+K_{1}^2\,J^2}{2\,l\,K_{1}}.
\end{eqnarray}
In this way this solution can be given in the standard
representation as
\begin{widetext}
\begin{eqnarray}\label{metriCAG2}
\bm{g}&=&-\frac{\rho^2\,f(\rho)}{H(\rho)}\bm{d\,T}^2+\frac{\bm{d\rho}^2}{f(\rho)}
+H(\rho)\left[\bm{d\Phi}+W(\rho)\bm{d\,T}\right]^2,\nonumber\\
f(\rho)&=&\frac{\rho^2}{l^2}-M+\frac{J^2}{4\rho^2},\nonumber\\
H(\rho)&=&\frac{\sqrt{2\rho^2-lR_{-}}\sqrt{2\rho^2-lR_{+}}}{4\sqrt{\alpha}K_{1}\sqrt{M^2\,l^2-J^2}}
\left[J^2K_{1}^2({2\rho^2-lR_{-}})^{-\sqrt{\alpha}/2}
({2\rho^2-lR_{+}})^{\sqrt{\alpha}/2}
\right.\nonumber
\\&&
\left.-({2\rho^2-lR_{-}})^{\sqrt{\alpha}/2}
({2\rho^2-lR_{+}})^{-\sqrt{\alpha}/2}\right],
\nonumber\\
W(\rho)&=&\frac{R_{-}}{Jl}\left[({2\rho^2-lR_{+}})^{\sqrt{\alpha}}
(2\sqrt{\alpha}\sqrt{M^2\,l^2-J^2}+R_{-})R_{-}
-({2\rho^2-lR_{-}})^{\sqrt{\alpha}}J^2\right]\times\nonumber\\
&&\left[({2\rho^2-lR_{+}})^{\sqrt{\alpha}}R_{-}^2
-({2\rho^2-lR_{-}})^{\sqrt{\alpha}}J^2\right]^{-1}.
\end{eqnarray}
\end{widetext}
When the electromagnetic field is turned off, $c=0\rightarrow
\alpha=1$, the above metric components reduce to
\begin{eqnarray}
g_{TT}&&=M-\frac{\rho^2}{l^2}, g_{T\Phi}=\frac{J}{2},
g_{\Phi\Phi}=\rho^2,\,
g_{\rho\rho}=\left(\frac{\rho^2}{l^2}-M+\frac{J^2}{4\rho^2}\right)^{-1},
\end{eqnarray}
which correspond to the BTZ ones.

This solution possesses mass $M$, angular momentum $J$,
electromagnetic parameter $\alpha$, and negative cosmological
constant, and describes a black hole.

\subsection{Constant electromagnetic invariants' hybrid
solution for $\Lambda=0$ }\label{Humf}

This section is devoted to the studied of the hybrid electromagnetic
stationary solution with constant electromagnetic invariant $FF$ and
by virtue of the field structure, constant $T$ and $TT$. The
constant character of $FF= -\frac{2c^2}{F(r)}$ is achieved by
requiring $F(r)=F_{0}$, and consequently all electromagnetic
invariants equal to constants
\begin{eqnarray}\label{unifhy2}
FF&=& -\frac{2c^2}{F_{0}},\,
TT=\frac{3}{64}\frac{c^4}{\pi^2F_{0}^2},\,
{T_{\mu}}^{\mu}=-\frac{c^2}{8\pi\,F_{0}}.
\end{eqnarray}

Again the integration of the Einstein equations start from
${E_{\phi}}^{t}(a=0=b)=0$, which gives the relation
\begin{eqnarray}\label{unifhy3}
\frac{d}{dr}W(r)=\frac{J}{H(r)^2}.
\end{eqnarray}
for the integration of the function $H(r)$ the substitution of
$\frac{d}{dr}W(r)$ and $F(r)=F_{0}$ into the remaining Einstein
equations requires the cosmological constant to vanish, $\Lambda=0.$
Under such condition, the equation for $H(r)$ becomes
\begin{eqnarray*}\label{unifhy5}
F_{0}\left(\frac{d\,H(r)}{dr}\right)^2-J^2+4\,c^2\,F_{0}{H(r)^2}=0
\end{eqnarray*}
with solution
\begin{eqnarray}\label{unifhy6}
&&H(r)=\epsilon_{H}\,\frac{J}{2\,c}\,\sqrt{F_{0}}
\,\sin{\frac{2\,c\,}{F_{0}}(r-C_{0})},\,\epsilon_{H}=\pm 1,
\end{eqnarray}
which, used in (\ref{unifhy3}), after integration yields
\begin{eqnarray}\label{unifhy7}
&&W(r)=W_{0}+\epsilon_{W}\,\frac{2\,c}{J}
\,\cot{\frac{2\,c\,}{F_{0}}(r-C_{0})},\, \epsilon_{W}=\pm 1,
\end{eqnarray}
where $\epsilon_{H}$ and $\epsilon_{W}$ assume their signs
independently; one has to take care on the ranges of the variable
$r$ to guarantee a correct signature. Moreover, notice that the
integration constant $C_{0}$ can be always equated to zero.
Therefore the corresponding metric and electromagnetic field vector
amount to
\begin{eqnarray}\label{hymetricscH}
&&\bm{g}=-\frac{F_{0}}{H}\bm{dt}^2+\frac{\bm{dr}^2}{F_{0}}
+H(\bm{d\phi}+W\bm{dt})^2,\nonumber\\
&&H=\epsilon_{H}\,\frac{J}{2\,c}\,\sqrt{F_{0}}\,\sin{\frac{2\,c\,}{F_{0}}\,r},\,
W=W_{0}+\epsilon_{W}\,\frac{2\,c}{J}\,\cot{\frac{2\,c\,}{F_{0}}\,r},\nonumber\\
&&\bm{A}=\frac{c}{2}(t\bm{d\phi}-\phi \bm{dt}).
\end{eqnarray}
The electromagnetic field tensors are
\begin{eqnarray}\label{unifhy1}
F_{\mu\nu}&=& 2c{\delta_{\mu}}^{[t}{\delta_{\nu}}^{\phi]},\nonumber\\
{T_{\mu}}^{\nu}&=&
\frac{c^2}{8\pi\,F_{0}}[-{\delta_{\mu}}^{t}{\delta_{t}}^{\nu}
+{\delta_{\mu}}^{r}{\delta_{r}}^{\nu}-{\delta_{\mu}}^{\phi}{\delta_{\phi}}^{\nu}].
\end{eqnarray}
By means of scaling transformations $F_{0}$ can be set always equal
to unit, $F_{0}=1$, hence this solution is endowed with two
effective parameters $c$ and $J$.

\section{Stationary cyclic symmetric solutions for $a\neq\,0$ or
$b\neq\,0$}\label{StaMagab}

This section deals with the search of stationary solutions for
the branches where one of the electromagnetic constants is zero,
$a\neq0= b$ or $b \neq 0= a$. It occurs that for these families the
integration problem reduces to find the solution of a master four
order (reducible to a third order) nonlinear equation for $F(r)$,
and to fit a differential constraint on the found structural
functions $F(r)$ and $H(r)$\, $(\mathcal{H}(r))$. The integration of
$W(r)$ \,$(\mathcal{W}(r))$ is trivial.

\subsection{Stationary magneto--electric solution for $a\neq0=b$,
}\label{StaMag-a}

If the structural function $W(r)$ is different from a constant (the
constant case will be treated at the end of this paragraph) then
${E_{\phi}}^{t}$ reads
\begin{eqnarray}\label{Stanza1}
{E_{t}}^{\phi}={\frac {d}{dr}} \left(  H^{2}{\frac {d}{dr}}W
\right)=0,
\end{eqnarray}
which yields
\begin{eqnarray}\label{Stanza2}
{\frac {d}{dr}}W  ={\frac {J}{  H ^{2}}}.
\end{eqnarray}
The remaining independent Einstein-Maxwell equations arise
respectively from combinations
$(4\,{E_{r}}^{r}+2{E_{t}}^{t}+2{E_{\phi}}^{\phi})$,
$(-2{H}({E_{r}}^{r}-{E_{t}}^{t}+W\,{E_{t}}^{\phi})/{F})$, and
${E_{r}}^{r}$:
\begin{eqnarray}\label{Stanza3}
EQ_{F}&=&
\frac{d^2\,F}{d\,r^2}-4\,a^2\frac{H}{F}-8\frac{1}{l^2}=0,\nonumber\\
EQ_{H}&=&\frac{d^2}{dr^2}H+4a^2\frac{H^2}{F^2}=0,\nonumber\\
{E_{r}}^{r}&=& \frac{1}{4H}\frac{d\,H}{d\,r}\frac{d\,F}{d\,r}
-\frac{F}{4H^2}(\frac{d\,H}{d\,r})^2 +\frac{J^2}{4\,H^2}
-a^2\frac{H}{F}-\frac{1}{l^2}=0.
\end{eqnarray}
The equation ${E_{r}}^{r}$ can be written in the form
\begin{eqnarray}\label{Stanza4aH}
EQ_{H1}=\left( \frac{1}{2H}{\frac {d H}{dr}}-\frac{1}{4F}\, \frac{d
F}{dr} \right) ^{2}  -\frac{1}{16\,F^2}\,\left( {\frac {d
F}{dr}}\right) ^{2} +{\frac {{a}^{2}H}{F ^{2}}}-\frac{1}{4}\,\frac
{{J}^{2}}{F  H ^{2}}+\frac {1}{{l}^{2}F}=0.
\end{eqnarray}
On the other hand using $EQ_{F}$ one expresses $H$ in terms of $F$
and its derivative
\begin{eqnarray}\label{Stanza7}
H(r)=\frac{1}{4\,a^2}\, \left( {\frac {d^{2}F}{d{r}^{2}}}
-\frac{8}{l^2}\right) F.
\end{eqnarray}
Substituting the above $H(r)$ into $EQ_{H}$~(\ref{Stanza3}) one gets
\begin{eqnarray}\label{Stanza8}
&{}&F\,\frac {d^{4}F}{d{r}^{4}}+2\,\frac {d^{3}F}{d{r}^{3}}\,
  \frac {d F}{dr}
  +2\,\left( {\frac {d^{2}F}{d{r}^{2}}}\right)
 ^{2}
 -\frac{24}{l^2}
\,\frac {d^{2}F}{d{r}^{2}}+\frac{64}{l^4}=0.
\end{eqnarray}
Therefore, integrating, if possible, Eq.~(\ref{Stanza8}) for $F(r)$,
substituting the solution $F(r)$ into Eq.~(\ref{Stanza7}) one
determines $H(r)$. The resulting functions $F(r)$ and $H(r)$ ought
to fulfil the Eq.~(\ref{Stanza4aH}) or ${E_{r}}^{r}$ equation from
Eq.~(\ref{Stanza3}). By integrating the linear first order
Eq.~(\ref{Stanza2}) one determines $W(r)$.

The contravariant components of electromagnetic tensor are
\begin{eqnarray}\label{EFuu}
F^{{\mu}{\nu}}=-2\,a\,{\delta^{\mu}}_{[\phi}{\delta^{\nu}}_{r]}.
\end{eqnarray}
The Eq.~(\ref{Stanza8}) for $F(r)$ can be reduced to a third--order
non-linear equation. In this equation, the problem for deriving
solutions in this branch actually resides.

Another possibility arises with the introduction of the auxiliary
function $h(r)$ by means of
\begin{eqnarray}\label{Stanza5}
H(r)=F(r)^{1/2}\,h(r),
\end{eqnarray}
the $EQ_{H1}$ acquires the form
\begin{eqnarray}\label{Stanza6}
EQ_{h}=- {l}^{2} \,h^{2}\,\left( {\frac {d F}{dr} }\right)^{2}
+4\,{l}^{2} F ^{2} \left( {\frac {d h}{dr}}\right) ^{2}
+16\,{l}^{2}\, {a}^{2}\, h ^{3}\sqrt {F} -4\,{l}^{2}\,{J}^{2}+16\,
h^{2}\,F=0,
\end{eqnarray}
and one could try to determine solutions for this variant.

\subsubsection{``Stationary'' magneto--electric
$\bm{A}=A(r)(\bm{d\phi}-J_{0}\bm{dt})$ solution}\label{consWmagsol}

A particular solution to Eq.~(\ref{Stanza8}) is given by $F(r)$ from
~(\ref{eqta7a}), namely
\begin{eqnarray}\label{solw1}
&&F(r)=\frac {4\,h(r)}{{C_{1}}^2\,l^2}\left[K_{0}+h(r)+{a}^{2}
{l}^{2}\ln{h(r)}\right], \,\, h(r):=C_{1}\,r+C_{0}.
\end{eqnarray}
which, being substituted into Eq.~(\ref{Stanza7}), leads to
\begin{eqnarray}\label{solw2}
H(r)&=&\frac {4}{{C_{1}}^2\,l^2}\left[K_{0}+h(r)+{a}^{2}
{l}^{2}\ln{h(r)}\right].
\end{eqnarray}
Entering with these particular solutions $F(r)$ and $H(r)$ in the
constraint Eq.~(\ref{Stanza4aH}) one arrives at
\begin{eqnarray}\label{solw3}
\frac{J^2}{F(r)}=0 \rightarrow{W(r)}=-J_{0}=\rm{constant.}
\end{eqnarray}
Summarizing, this solution is given by the same structural
functions~(\ref{eqta7a}) of the  magnetostatic solution except that
in the present case the function $W(r)$ is a constant. The
corresponding metric line element and field vector can be written as
\begin{eqnarray}\label{metricscrho}
\bm{g}&=&-h(r)\bm{dt}^2+\frac{\bm{dr}^2}{H(r)\,h(r)}
+H(r)(\bm{d\phi}-J_{0}\bm{d\,t})^2,\nonumber\\
\bm{A}&=&\frac{a}{C_{1}}\ln{h(r)}(\bm{d\phi}-J_{0}\bm{d\,t}).
\end{eqnarray}
The electromagnetic field tensors and their invariants are given by
\begin{eqnarray}
F^{{\mu}{\nu}}=2\,a\,{\delta^{\mu}}_{[r}{\delta^{\nu}}_{\phi]},\,
F_{{\mu}{\nu}}=2\,a\,J_{0}/h(r)\,{\delta_{\mu}}^{[t}{\delta_{\nu}}^{r]}
-2\,a/h(r)\,{\delta_{\mu}}^{[\phi}{\delta_{\nu}}^{r]},\,
FF=2\frac{a^2}{h},
\end{eqnarray}
and
\begin{eqnarray}
{T_{\nu}}^{\mu}=
\,\frac{a^2}{8\pi\,h}\left[-\delta_{\nu}^{t}\delta_{t}^{\mu}
+\delta_{\nu}^{r}\delta_{r}^{\mu}
+\delta_{\nu}^{\phi}\delta_{\phi}^{\mu}
-2\,J_{0}\delta_{\nu}^{t}\delta_{\phi}^{\mu}\right],\,
T=\frac{3}{64\pi^2}\frac{a^4}{h^2}.
\end{eqnarray}
Because of the structure of the energy--momentum tensor above, this
solution can be interpreted as a rigidly rotating perfect fluid
\begin{eqnarray}
{T_{\mu}\nu}&&=(\rho+p)u_{\mu} u_{\nu}+p\,g_{\mu\nu}, \,u^{\mu}=
\frac{1}{\sqrt{F/H}}({\delta^{\mu}}_{t}+J_{0}{\delta^{\mu}}_{\phi}),
\end{eqnarray}
with energy density $\rho$ and pressure $p$ given by
$$\rho=\frac{1}{8\pi}\,{\frac { {a}^{2}}{h }}=p.$$

This solution can be generated via transformations
$t\rightarrow{t},\,\phi\rightarrow{\phi-J_{0}t}$ from the
magnetostatic solution~(\ref{eqta7a}).

\subsubsection{Clement ``rotating'' electromagnetic
$\bm{A}=A(r)(\bm{d\phi}+\omega_{0}\bm{dt})$
solution}{\label{Clement93mag}}

Clement \cite{Clement93} published the dual family of
electromagnetic ``stationary'' cyclic symmetric solutions,
Eq.~(Cl.24), changing signature and $V_{Cl}\rightarrow{-V}$, given
by
\begin{eqnarray}\label{Clement23i}
\bm{g}&=&V(\bm{d\phi}+\omega_{0}\bm{dt})^2
+\frac{1}{\xi_{0}^2}\frac{\bm{d\rho}^2}{2\rho V}
-2\rho{\bm{dt}}^2,\nonumber\\
V&=&-2\Lambda\rho+\frac{\pi_{1}^2}{4m}\ln(\frac{\rho}{\rho_{0}}),\nonumber\\
\bm{A}&=&-\frac{\pi_{1}}{2}\ln(\frac{\rho}{\rho_{0}})(\bm{d\phi}+\omega_{0}\bm{dt}),
\end{eqnarray}
where $m, \pi_{1}, \xi_{0}$ and $\rho_{0}$ are constants,
$\Lambda=\pm 1/l^2$ stands for the cosmological constant of both
signs; for anti--de Sitter $\Lambda=-1/l^2$. The parameter
$\omega_{0}$ is related to the angular momentum constant.

\noindent It is worthwhile to notice that the Clement expressions
(\ref{Clement23i}) satisfy the 2+1 Einstein--Maxwell equations if
$\xi_{0}^2=1$ and for $2\,m=1/\kappa$; for the adopted in the
Clement's convention, $\kappa\neq 1$,
$G_{\mu\nu}+\Lambda\,g{\mu\nu}=4\pi\kappa\,T_{\mu\nu}$, the
evaluation of the right hand side of the Einstein equations for the
structural functions~(\ref{Clement23i}), for $\xi_{0}^2=1$, yields
$${ G_{\nu}}^{\mu}=\frac{\pi_{1}^2}{8m\rho}[-\delta_{\nu}^{t}\delta_{t}^{\mu}
+\delta_{\nu}^{r}\delta_{r}^{\mu}+\delta_{\nu}^{\phi}\delta_{\phi}^{\mu}
+2\,\omega_{0}\delta_{\nu}^{t}\delta_{\phi}^{\mu}],$$ while the
right hand side amounts to
$${4\pi\,\kappa\,T_{\nu}}^{\mu}=
\frac{\kappa\,\pi_{1}^2}{4\rho}[-\delta_{\nu}^{t}\delta_{t}^{\mu}
+\delta_{\nu}^{r}\delta_{r}^{\mu}+\delta_{\nu}^{\phi}\delta_{\phi}^{\mu}
+2\,\omega_{0}\delta_{\nu}^{t}\delta_{\phi}^{\mu}]$$ hence
$2m=1/\kappa$.\\
If one were adopting $\kappa=1$, then modifying the electromagnetic
vector $\bm{A}$ to be
$\bm{A}_{mod}=-\frac{\pi_{1}}{2\sqrt{2\,m}}\ln(\frac{\rho}{\rho_{0}})
(\bm{d\phi}+\omega_{0}\bm{dt})$, one would arrive at the solution in
our convention.

It is apparent that this Clement's solutions correspond to a variant
of the solution derived in the previous Section~(\ref{consWmagsol}),
with the identification $r\rightarrow\rho$ followed by minor scaling
transformations of $t$ and $\phi$. Notice also that the above
generalization with $W(r)=\omega_{0}\neq0$ of the magnetostatic
solution~(\ref{eqta7a}) can be determined applying to it $SL(2,R)$
transformations of the form $\phi\rightarrow
\phi+\omega_{0}\,t,\,t\rightarrow t$.

\subsection{Stationary electro--magnetic solution for
$b\neq0=a$}\label{StaElec-b}

A straightforward way to derive the equations and solutions of this
class of fields is just by using the complex extension of the
stationary magnetic field we derived in the previous subsection
taking into account the specific structure of the
functions~(\ref{compfun}) of the extended metric~(\ref{commetric})
and the metric components from~(\ref{metricscrho}) of the magnetic
solution together with the formal change $a^2\rightarrow{-b^2}$.

Another close possibility is to accomplish the substitution
\begin{eqnarray}\label{compfunr}
F=\mathcal{F},\,
H=\frac{\mathcal{F}}{\mathcal{H}}-\mathcal{H}\mathcal{W}^2,
W=\frac{\mathcal{H}\,\mathcal{W}}{{H}},
\end{eqnarray}
in the corresponding Einstein equations for this case $b\neq 0= a$,
arriving at the following set of independent field equations
\begin{eqnarray}\label{Stanzb1x}
EQ_{\mathcal{F}}&=&
\frac{d^2\,\mathcal{F}}{d\,r^2}+4\,b^2\frac{\mathcal{H}}{\mathcal{F}}
-8\frac{1}{l^2}=0,\nonumber\\
EQ_{\mathcal{H}}&=&\frac{d^2}{dr^2}\mathcal{H}^2
-4b^2\frac{\mathcal{H}^4}{\mathcal{F}^2}=0,\nonumber\\
E^{r}_{r}&=&
\frac{1}{4\mathcal{H}}\frac{d\,\mathcal{H}}{d\,r}\frac{d\,\mathcal{F}}{d\,r}
-\frac{\mathcal{F}}{4\mathcal{H}^2}(\frac{d\,\mathcal{H}}{d\,r})^2
+\frac{1}{4\,\mathcal{H}^2}\,J^2+b^2\frac{\mathcal{H}}{\mathcal{F}}
-\frac{1}{l^2}=0,\nonumber\\
{\frac {d}{dr}}\mathcal{W} &=&{\frac {J}{\mathcal{H}^{2}}}.
\end{eqnarray}
Continuing with the parallelism, isolating $\mathcal{H}$ from
$EQ_{\mathcal{F}}$ and replacing it into $EQ_{\mathcal{H}}$ one
obtains
\begin{eqnarray}\label{Stanzb8}
&{}&\mathcal{F}\,\frac {d^{4}\mathcal{F}}{d{r}^{4}}+2\,\frac
{d^{3}\mathcal{F}}{d{r}^{3}}\,
  \frac {d \mathcal{F}}{dr}
  +2\,\left( {\frac {d^{2}\mathcal{F}}{d{r}^{2}}}\right)
 ^{2}-\frac{24}{l^2} \,\frac
{d^{2}\mathcal{F}}{d{r}^{2}}+\frac{64}{l^4}=0.
\end{eqnarray}
Thus, as before, the first step in the integration of the problem
depends upon the Eq.~(\ref{Stanzb8}) for $\mathcal{F}(r)$,
structurally identical to Eq.~(\ref{Stanza8}). Substituting the
solution $\mathcal{F}(r)$ into $EQ_{\mathcal{H}}$ from
Eq.~(\ref{Stanzb1x}) one determines $\mathcal{H}$. The resulting
functions $\mathcal{F}(r)$ and $\mathcal{H}$ have to fulfill
${E_{r}}^{r}$ from Eq.~(\ref{Stanzb1x}). By integrating the linear
first order equation for $\mathcal{W}$ one determines
$\mathcal{W}(r)$.

\subsubsection{``Stationary'' electro--magnetic
$\bm{A}=A(r)(\bm{dt}+J_{0}\bm{d\phi})$ solution} \label{StaElec-bp}

The only known until now particular solution for $\mathcal{F}(r)$ of
Eq.~(\ref{Stanzb8}) and its corresponding solutions for
$\mathcal{H}$ and $\mathcal{W}$ are
\begin{eqnarray}\label{eqta7aWx}
&&\mathcal{F}=\frac{4}{{C_{1}}^2\,l^2}\left[K_{0}+h(r)
-{b}^{2}{l}^{2}\ln{h(r)}\right]\,h(r),\,h(r):=C_{1}\,r+C_{0},\nonumber\\
&&\mathcal{H}=\frac{\mathcal{F}}{h},\,\mathcal{W}=-J_{0}=\rm
constant.
\end{eqnarray}
Substituting these expressions into Eq.~(\ref{compfunr}), one gets
\begin{eqnarray}\label{compfunrs}
&&F=\mathcal{F}=\mathcal{H}h,\, H=h-\mathcal{H}{J_{0}}^2,\nonumber\\
&&W=-J_{0}\frac{\mathcal{H}}{h-\mathcal{H}{J_{0}}^2},\,
\mathcal{H}:=\frac{4}{{C_{1}}^2\,l^2}\left[K_{0}+h(r)
-{b}^{2}{l}^{2}\ln{h(r)}\right],
\end{eqnarray}
therefore, the corresponding metric and field vector can be written
as
\begin{eqnarray}\label{electrica}
\bm{g}&=&-{\mathcal{H}}(\bm{dt}+J_{0}\bm{d\phi})^2+\frac{\bm{dr}^2}{{\mathcal{H}}\,h(r)}
+h(r)\bm{d\phi}^2,\nonumber\\\
\bm{A}&=&\frac{b}{C_{1}}\ln{h(r)}(\bm{dt}+J_{0}\bm{d\phi}).
\end{eqnarray}
The electromagnetic field tensors and their invariants are given by
\begin{eqnarray*}
&&F^{{\mu}{\nu}}=2\,b\,{\delta^{\mu}}_{[t}{\delta^{\nu}}_{r]},\,
F_{{\mu}{\nu}}=-2\,\frac{b}{h(r)}\,{\delta_{\mu}}^{[t}{\delta_{\nu}}^{r]}
-2\,b\,\frac{J_{0}}{h(r)}\,{\delta_{\mu}}^{[\phi}{\delta_{\nu}}^{r]},\,
FF=-2\frac{b^2}{h}, \nonumber\\&&
{T_{\nu}}^{\mu}=
\,\frac{b^2}{8\pi\,h}[-\delta_{\nu}^{t}\delta_{t}^{\mu}
-\delta_{\nu}^{r}\delta_{r}^{\mu}+\delta_{\nu}^{\phi}\delta_{\phi}^{\mu}
-2\,J_{0}\delta_{\nu}^{t}\delta_{\phi}^{\mu}],\,
TT=\frac{3}{64\pi^2}\frac{b^4}{h^2}.
\end{eqnarray*}

As we shall see in the forthcoming section, this stationary
electromagnetic solution can be generated via transformations
$t\rightarrow{t+J_0\phi},\,\phi\rightarrow{\phi}$ from the
electrostatic solution~(\ref{staticq}).

\subsubsection{Clement ``rotating'' electromagnetic
$\bm{A}=A(r)(\bm{dt}-\omega_{0}\bm{d\phi})$
solution}{\label{Clement93elec}}

Also Clement \cite{Clement93} published a class of electromagnetic
``stationary'' cyclic symmetric metrics, Eq.~(Cl.23), changing
signature, given by
\begin{eqnarray}\label{Clement23x}
\bm{g}&=&-U(\bm{dt}-\omega_{0}\bm{d\phi})^2+\frac{1}{\xi_{0}^2}\frac{\bm{d\rho}^2}{2\rho
U}
+2\rho{\bm{d\phi}}^2,\nonumber\\
U&=&-2\Lambda\rho-\frac{\pi_{0}^2}{4m}\ln(\frac{\rho}{\rho_{0}}),\nonumber\\
\bm{A}&=&\frac{\pi_{0}}{2}\ln(\frac{\rho}{\rho_{0}})
(\bm{dt}-\omega_{0}\bm{d\phi}),
\end{eqnarray}
where $m, \pi_{0}, \xi_{0}$ and $\rho_{0}$ are constant parameters,
$\Lambda=\pm 1/l^2$ stands for the cosmological constant of both
signs; for anti--de Sitter $\Lambda=-1/l^2$. The parameter
$\omega_{0}$ is a constant related to the angular momentum. The
evaluation of the right hand side of the Einstein equations for the
structural functions~(\ref{Clement23x}), for $\xi_{0}^2=1$, yields
$${ G_{\nu}}^{\mu}=-\frac{\pi_{0}^2}{8mr}[\delta_{\nu}^{t}\delta_{t}^{\mu}
+\delta_{\nu}^{r}\delta_{r}^{\mu}-\delta_{\nu}^{\phi}\delta_{\phi}^{\mu}
-2\,\omega_{0}\delta_{\nu}^{\phi}\delta_{t}^{\mu}],$$ while the
energy momentum tensor in the left hand side, for the vector
$\bm{A}$, amounts to
$$4\pi\,{T_{\nu}}^{\mu}=-\frac{\pi_{0}^2}{4r}
[\delta_{\nu}^{t}\delta_{t}^{\mu}
+\delta_{\nu}^{r}\delta_{r}^{\mu}-\delta_{\nu}^{\phi}\delta_{\phi}^{\mu}
-2\,\omega_{0}\delta_{\nu}^{\phi}\delta_{t}^{\mu}],$$ therefore
Einstein--Maxwell equations are fulfilled if $2m=1/\kappa$ or, for
$\kappa=1$, modifying the electromagnetic vector $\bm{A}$ to be
$\bm{A}_{mod}=\frac{\pi_{0}}{2\sqrt{2\,m}}\ln(\frac{\rho}{\rho_{0}})
(\bm{dt}-\omega_{0}\bm{d\phi})$. Recall that additionally one has to
set $\xi_{0}^2=1$.

\noindent It is clear that this solution is equivalent to the one
treated in the previous Section ~\ref{StaElec-bp} for the
identification $C_{1}\,r+C_{0}\rightarrow \rho$ accompanied with
minor scaling transformations of $t$ and $\phi$.

\noindent Notice that this branch of rotating solutions with
$W(r)=\omega_{0}$ can be determined from the static electric field
solution, i.e., metric~(\ref{staticq}) and vector $ \bm{A}$
~(\ref{staticqA}), via $SL(2,R)$ transformations: $t\rightarrow
t-\omega_{0}\phi,\,\phi\rightarrow \phi$.

\subsubsection{Constant $W$ electric solution}

In the case $W(r)=-J=\rm{constant}$ the equation ${E_{\phi}}^{t}$,
from~(\ref{Einstein}), reduces to $b^2\,H\,J/F=0$, then
$J=0\rightarrow W=0$. Hence, the set of equations reduces to the
corresponding one of the static case.

\section{Stationary cyclic symmetric solutions for $a\neq\,0$ and
$b\neq\,0$}\label{GENSTATIONARY}

It is clear that the derivation of a general solution to the whole
system of Einstein--Maxwell equations (\ref{Einstein}) is far from
being an easy task. Nevertheless, it occurs that some
simplifications of the system of equations can be achieved by a
useful change of the structural functions and combinations of the
Einstein equations; the integration problem on the whole for the
three structural functions is constrained to three differential
equations without any further restrictions. Although we could not
find sufficiently general classes of solutions, we were able to
determine new families of solutions within particular combinations
of elementary functions.

\subsection{Alternative representation of the Einstein equations}
Having in mind the derivation of other possible families of
Einstein--Maxwell solutions with $a$ and $b$ different from zero, it
is desirable to have at hand the most simple set of equations. For
this purpose, introducing $W(r)=\Omega(r)/H(r)$, the independent
Einstein equations can be written as;
\begin{eqnarray}\label{EEMom1}
EQ_{H2}=F^2\frac{d^2}{dr^2}H+4(a\,H-b\Omega)^2=0,
\end{eqnarray}
\begin{eqnarray}\label{EEMom2}
EQ_{\Omega2}=H\,F^2\frac{d^2}{dr^2}\Omega
+4\,\Omega(a\,H-b\Omega)^2+ 4b\,F(a\,H-b\Omega)=0,
\end{eqnarray}
\begin{eqnarray}\label{EEMom3}
EQ_{F1}&&=F\frac{dF}{dr}\,H\frac{dH}{dr}
-F^2\,\left(\frac{dH}{dr}\right)^2
-4\,H\left(a\,H-b\Omega\right)^2+F\,\left( H\frac{d\Omega}{dr}-
\Omega\frac{dH}{dr}\right)^2\nonumber\\
&&+4F\,H(b^2 -H/l^2)=0,\text{ \it correct: equate to zero},
\end{eqnarray}
\begin{eqnarray}\label{EEMom4}
EQ_{F2}=&&\Omega\,H^2\,\frac{d^2F}{dr^2}
-2\,\Omega\,H\frac{dH}{dr}\,\frac{dF}{dr}
+2\Omega\left(\frac{dH}{dr}\right)^2\,F -2\,\Omega\,\left(
H\frac{d\Omega}{dr}- \Omega\frac{dH}{dr}\right)^2=0.\nonumber\\&&
\text{ \it correct: equate to zero},
\end{eqnarray}

It is worthwhile to point out that the equation $EQ_{F2}$ can be
considered as an integrability condition of the system of equations;
differentiating the $EQ_{F1}$ one obtains the second derivative
$\frac{d^2F}{dr^2}$ together with second derivatives of $H$ and
$\Omega$, which can be replaced through $EQ_{H2}$ and
$EQ_{\Omega2}$, next substituting $\frac{d^2F}{dr^2}$ into $EQ_{F2}$
one arrives at an equation of the form
$\left(H\,\frac{dF}{dr}+F\,\frac{dH}{dr}\right)\times \,EQ_{F1}$,
equal to zero by virtue of the same $EQ_{F1}$. Although one can
adopt a different point of view; the $EQ_{F2}$--equation arises as
the differentiation of $EQ_{F1}$ together with the use of $EQ_{H2}$
and $EQ_{\Omega2}$, and therefore it is not an independent equation.

Using the experience gathered until now, we shall search for
particular solutions of the form
\begin{eqnarray}\label{EEMom1search}
F(r)&=&P(r)+Q(r)\ln(r),\nonumber\\
H(r)&=&A(r)+B(r)\ln(r),\nonumber\\
W(r)&=&\Omega(r)/H(r), \nonumber\\
\Omega(r)&=&V(r)+Z(r)\ln(r),
\end{eqnarray}
where it is assumed the explicit dependence on $\ln(r)$.
Substituting these guessed functions into the quoted system of
equations and equating to zero the coefficients of different powers
of $\ln(r)$ one arrives at a very large non--linear system of
equations; since the non--trivial Einstein equations are five, then
one may expect 40 secondary equations. For instance, from equations
arising from the coefficients of $\ln(r)$ to the seventh power, one
has
$$
{E_{t}^{t}}_{ln^7}=B^2\,Q\,l^2(Z\frac{dB}{dr}-B\frac{dZ}{dr})^2
-2\,B^3\,Z\,Q\,l^2(Z\frac{d^2B}{dr^2}-B\frac{d^2Z}{dr^2}),$$
$$
{E_{\phi}^{\phi}}_{ln^7}+3{E_{r}^{r}}_{ln^7}
=2\,l^2B^3Q(Z\frac{d^2B}{dr^2}-B\frac{d^2Z}{dr^2}),
$$
hence
$$Z\frac{d^2B}{dr^2}-B\frac{d^2Z}{dr^2}=0,\,\,Z\frac{dB}{dr}-B\frac{dZ}{dr}=0,$$
therefore
$$Z(r)=c_{1}B(r).$$
After a very lengthy and time--consuming integration process we
succeeded in getting two branches of stationary electromagnetic
solutions of the Einstein--Maxwell equations. The structural
functions $H$ and $W$ possess a multiplicative factor $a/b$ which
can be absorbed by re--scaling of the Killingian coordinates
according to: $\sqrt{|a/b|}t\rightarrow t$ and
$\sqrt{|b/a|}\phi\rightarrow \phi$, $\sqrt{|a|\,|b|}=\pm\alpha$.

\subsection{Stationary electromagnetic solution  with BTZ--limit
 }\label{stationary-elec}

This class of solutions, depending on three parameters, is given by
\begin{widetext}
\begin{eqnarray*}\label{EEMom1searcha}
\bm{g}&=&-\frac{F(r)}{H(r)}\bm{d\,t}^2 +\frac{\bm{dr}^2}{F(r)}
+H(r)\left[\bm{d\phi}+W(r)\bm{dt}\right]^2,\nonumber\\
F(r)&=&4\frac{r^2}{l^2}
+2\frac{r}{l}\left(l\,w_{1}+\sqrt{l^2w_{1}^2-4}\right)
\left[w_{0}+W_{0}\ln(r)\right],\nonumber\\
H(r)&=&\frac{r}{l}\sqrt{l^2w_{1}^2-4}-\left[w_{0}+W_{0}\ln(r)\right],\nonumber\\
W(r)&=&\Omega(r)/H(r),\,\,\Omega(r):=w_{0}+w_{1}\,r+W_{0}\ln(r),\nonumber\\
W_{0}&:=&-\frac{1}{2}\,l^2\alpha^2\left(l^2w_{1}^2-2-l\,w_{1}\sqrt{l^2w_{1}^2-4}\right),\nonumber\\
\bm{A}&=&\frac{1}{4}\,\alpha\,\left(l\,w_{1}-\sqrt{l^2w_{1}^2-4}\right)(\bm{dt}-\bm{d\phi})\ln(r).
\end{eqnarray*}
\begin{eqnarray}\label{EFuubtz}
F^{{\mu}{\nu}}&=&2\,\alpha\,\left({\delta^{\mu}}_{[t}{\delta^{\nu}}_{r]}
-{\delta^{\mu}}_{[\phi}{\delta^{\nu}}_{r]}\right),
\,F_{{\mu}{\nu}}=-\frac{\alpha\,l}{4\,r}\,\left(l\,w_{1}-\sqrt{l^2w_{1}^2-4}
\right)\left({\delta_{\mu}}^{[t}{\delta_{\nu}}^{r]}
-{\delta_{\mu}}^{[\phi}{\delta_{\nu}}^{r]}\right),\nonumber\\
{T^{\mu}}_{\nu}&=& \frac{\alpha^2}{8\pi\,F\,H}\left[
\begin {array}{ccc} -[F+H^2(1-W^2)]&0&{2[F+H^2W(1-W)]}
\\\noalign{\medskip}0&-[F-H^2(1-W)^2]&0\\
\noalign{\medskip}{-2H^2(1-W)}&0&[F+H^2(1-W^2)]\end {array}
\right],\nonumber\\
\end{eqnarray}
\end{widetext}
with electromagnetic invariants
$FF=-2\frac{\alpha^2}{H}+2\frac{\alpha^2}{F}\,H\,(1-W)^2$, and
$TT=\frac{3}{64}\,\frac{\alpha^4}{\pi^2}\frac{1}{F^2\,H^2}[-{F}+H^2\,(1-W)^2]^2$.

\subsubsection{Transformation to BTZ-like coordinates}

Since this solution possesses as a limit for $\alpha=0$ the BTZ
solution, it is natural to search for new coordinates in which it
will become apparent the BTZ standard structure. First one
determines the radial transformation
$r=\beta_{0}(\rho^2+\gamma_{0})$; since $g_{rr}\rightarrow
{g_{\rho\rho}}$, then
\begin{eqnarray*}
F(r)&=&4\frac{r^2}{l^2}
+2\frac{r}{l}\left(l\,w_{1}+\sqrt{l^2w_{1}^2-4}\right)w_{0}
\rightarrow{{F(\rho)}=\frac{\rho^4}{l^2}-M\rho^2+\frac{J^2}{4}},
\end{eqnarray*}
hence
\begin{eqnarray*}
&&\gamma_{0}^2/l^2+\gamma_{0}M+J^2/4=0 \rightarrow{\gamma_{0}/l
=-\frac{Ml}{2}\mp\frac{1}{2}\sqrt{l^2M^2-J^2}},\nonumber\\
&&{w_{0}}\left(l\,w_{1}+\sqrt{l^2w_{1}^2-4}\right)
=\pm\,2\,\beta_{0}\sqrt{l^2M^2-J^2}.
\end{eqnarray*}
 Next, the structure of the
Killingian transformations is of the form
$$ t=\alpha_{t}\, T +\beta_{t}\, \Phi,\,\,\phi=\delta_{t}\,\Phi.$$
Substituting these relations into the metric and comparing the
metric components with the corresponding ones of the BTZ--metric one
establishes that
\begin{eqnarray}\label{EEMom1vp}
w_{0}&=&\frac{1}{J} {\sqrt {{l}^{2}{M}^{2}-{J}^{2}}\left(
-l\,M+\sqrt {{l}^{2}{M}^{2 }-{J}^{2}} \right) },\,
\beta_{0}=-1,\,w_{1}=2\frac{M}{J}.
\end{eqnarray}
Therefore, the coordinate transformations to be used in the
electromagnetic solution in order to get the proper BTZ limit when
the electromagnetic $\alpha$--parameter is switch off is given by
\begin{eqnarray}\label{EEMom1v}
r&=&-\rho^2+\frac{M\,l^2}{2}+\frac{l}{2}\sqrt{l^2M^2-J^2},\nonumber\\
t&=&\frac{1}{\sqrt{2}}\left({l}^{2}{M}^{2}-{J}^{2}\right)^{-1/4}\left(\sqrt{\frac{J}{l}}\,T
-l\,M\,\sqrt{\frac{l}{J}}\,\Phi\right),\nonumber\\
\phi&=&\frac{1}{\sqrt{2}}\sqrt{\frac{l}{J}}\left({l}^{2}{M}^{2}
-{J}^{2}\right)^{1/4}\,\Phi.
\end{eqnarray}
Under these transformations the metric becomes
\begin{widetext}
\begin{eqnarray*}\label{metricscrhoa}
\bm{g}&=&-\frac{\rho^2\,F(\rho)}{H(\rho)}\bm{d\,T}^2
+\frac{\bm{d\rho}^2}{F(\rho)}
+H(\rho)(\bm{d\Phi}+W(\rho)\bm{d\,T})^2,\nonumber\\
\end{eqnarray*}
\begin{eqnarray*}\label{metricscrhofx}
&F(\rho)&=\frac{\rho^2}{l^2}-M+\frac{J^2}{4\rho^2} -\frac
{l\alpha^{2}}{2J\rho^2}\,\left[ {J}^{2}l
-2\,{\rho}^{2}R_{(-)}\right] \ln|r|,\nonumber\\
\end{eqnarray*}
\begin{eqnarray*}
H(\rho)&=&\frac{H_{n}}{H_{d}},\nonumber\\
H_{n}&:=&-{l}^{6}{J}^{2} {\alpha}^{4}R_{(-)}^2 \left(\ln|r|\right)^{2}\nonumber\\
&{}&-2\,{l}^{3}J \sqrt {{l}^{2}{M} ^{2}-{J}^{2}}
\left[-2\,\sqrt {{l}^{2}{M} ^{2}-{J}^{2}}\,R_{(-)}{\rho}^{2}
+M{J}^{2}{l}^{2} \right] {\alpha}^{2}\ln|r| \nonumber\\
&&+4\,{\rho}^{2}{J}^{2} \left( {\rho}^{2}-{l}^{2}M \right) \left(
{l}^{2}{M}^{2}-{J}^{2} \right),\nonumber\\
H_{d}&:=&4\,J^2({l}^{2}{M}^{2}-{J}^{2})
\left({\rho}^{2}-{l}^{2}M\right) -2\,J\sqrt{{l}^{2}{M}^{2}-{J}^{2}}
R_{(-)}^2 {l}^{3}{
\alpha}^{2}\ln|r|, \nonumber\\
\end{eqnarray*}
\begin{eqnarray}
W(\rho)&=&\frac{\Omega(\rho)}{H_{n}},\nonumber\\
\Omega(\rho)&:= &{l}^{5}J\,R_{(-)}^3\,{\alpha}^{4}
\left(\ln|r|\right) ^{2} +{l^2}{J}^{2}{\alpha}^{2}\sqrt
{{l}^{2}{M}^{2}-{J}^{2}} \left[{J}^{2}l+2\,l\,R_{(-)}^2
-2R_{(-)}\,\rho^2 \right]\ln|r|\nonumber\\
&{}&-2\,{J}^{3} \left( {\rho}^{2}-{l}^{2}M \right)  \left(
{l}^{2}{M}^{2}- {J}^{2} \right),\nonumber\\
r&:=&-\rho^2+\frac{M\,l^2}{2}+\frac{l}{2}\sqrt{l^2M^2-J^2},\,R_{(\pm)}:=M\,l\pm\sqrt{l^2M^2-J^2}.
\end{eqnarray}
\end{widetext}
Notice that $ (g_{tt}g_{\phi\phi}-g_{t\phi}^2)=-\rho^2\,F(\rho)$.
\noindent The correspondence of this function
representation of this electromagnetic solution with the BTZ solution in
the limit of vanishing electromagnetic parameter $\alpha$ becomes apparent:
\begin{eqnarray*}
&&F(\rho)=\frac{\rho^2}{l^2}-M+\frac{J^2}{4\rho^2},
\,H(\rho)={\rho^2}, \,W(\rho)=-\frac{J}{2\,\rho^2}.
\end{eqnarray*}
Thus, this anti-de Sitter solution has three parameters: mass $M$,
angular momentum $J$, and electromagnetic parameter $\alpha$.
Because of its close similarity to the BTZ solution, it could be
represent a black hole; a research in this direction is in progress.

\subsection{Stationary electromagnetic solution with BTZ-counterpart limit}
\label{stationary-mag}

The second possible solution in the studied class is given by
\begin{widetext}
\begin{eqnarray*}
\bm{g}=-\frac{\mathcal{F}}{\mathcal{H}}\bm{dT}^2+\frac{\bm{dr}^2}{\mathcal{F}}
+\mathcal{H}\left(\bm{d\Phi}+\mathcal{W}\,\bm{dT}\right)^2,
\end{eqnarray*}
\begin{eqnarray*}\label{EEmag}
\mathcal{F}(r)&=&4\frac{r^2}{l^2}
+2\frac{r}{l}(l\,w_{1}+\sqrt{l^2w_{1}^2-4})
\left(w_{0}+W_{0}\ln(r)\right),\nonumber\\
\mathcal{H}(r)&=&\frac{\mathcal{H}_{n}}
{\mathcal{H}_{d}},\nonumber\\
\mathcal{H}_{n}&=&\mathcal{F}(r)-\Omega(r)^2,
\nonumber\\
\mathcal{H}_{d}&=&\frac{r}{l}\sqrt{l^2\,w_{1}^2-4}
-\left(w_{0}+W_{0}\ln(r)\right),\nonumber\\
\mathcal{W}(r)&=&\frac{\Omega(r)}{\mathcal{H}(r)},
\,\,\Omega(r)=w_{0}+W_{0}\,\ln(r)
+w_{1}\,r,\nonumber\\
W_{0}&:=&\frac{1}{2}\,l^2\alpha^2
\left(l^2w_{1}^2-2-l\,w_{1}\sqrt{l^2w_{1}^2-4}\right),
\end{eqnarray*}
\begin{eqnarray}\label{EFuubtza}
F^{{\mu}{\nu}}&=&2\,\alpha\,\left({\delta^{\mu}}_{[t}{\delta^{\nu}}_{r]}
+{\delta^{\mu}}_{[\phi}{\delta^{\nu}}_{r]}\right),
\nonumber\\
{T^{\mu}}_{\nu}&=&
\frac{\alpha^2}{8\pi\,\mathcal{F}\,\mathcal{H}}\left[
\begin {array}{ccc} -[\mathcal{F}+\mathcal{H}^2(1-\mathcal{W}^2)]
&0&{-2[\mathcal{F}-\mathcal{H}^2\mathcal{W}(1+\mathcal{W})]}
\\\noalign{\medskip}0&-[\mathcal{F}-\mathcal{H}^2(1+\mathcal{W})^2]
&0\\\noalign{\medskip}{2\mathcal{H}^2(1+\mathcal{W})}&
0&[\mathcal{F}+\mathcal{H}^2(1-\mathcal{W}^2)]\end {array}
\right],\nonumber\\
\end{eqnarray}
\end{widetext}
where $w_{0}$ and $w_{1}$ are parameters related to mass and angular
momentum, while $\alpha$ is an electromagnetic parameter; the
electromagnetic invariants are
\begin{eqnarray}
&&FF=2\frac{\alpha^2}{\mathcal{F}\mathcal{H}}
 \left[-{\mathcal{F}}+
\,\mathcal{H}^2\,(1+\mathcal{W})^2\right],\,
TT=\frac{3}{64}\,\frac{\alpha^4}{\pi^2}\frac{1}
{\mathcal{F}^2\,\mathcal{H}^2}
[-{\mathcal{F}}+\mathcal{H}^2\,(1+\mathcal{W})^2]^2.\nonumber\\
\end{eqnarray}
The calligraphic capital letters have been used above to make their relationship
evident to those structural functions arising as
real cuts of the complex extensions of the studied class of metric,
see~(\ref{commetric}). This solution of the Einstein--Maxwell
equations can be considered also as a real cut of the complex
version of the stationary electromagnetic solution  with BTZ--limit
given in the previous paragraph; the structural functions
$\mathcal{F}$, $\mathcal{H}$, and $\mathcal{W}$ can be constructed
according to Eq.~(\ref{compfun}) with $F$, $H$, and $W$ from
Eq.~(\ref{EEMom1search}) accompanied by the replacement of the sign
in front of $\alpha^2$,
${\alpha^2}_{el}\rightarrow{-{\alpha^2}_{mg}}$. If one
searches for the anti--de Sitter limit of this solution, one would
arrive at an alternative real cut of the BTZ--solution, namely to
the "BTZ--solution counterpart", for short BTZ--counterpart.
\begin{eqnarray}\label{antiBTZ}
&&\bm{g}_{c}=-\rho^2\frac{\mathcal{F}}{\mathcal{H}}\bm{d\,T}^2
+\frac{\bm{d\rho}^2}{\mathcal{F}}+\mathcal{H}\left(\bm{d\Phi}
+\mathcal{W}\,\bm{dT}\right)^2,\nonumber\\
&&\mathcal{F}=\frac{\rho^2}{l^2}-M+\frac{J^2}{4\,\rho^2},
\,\,\mathcal{H}=\frac{\rho^2}{l^2}-M,\,
\mathcal{W}=\frac{J}{2\mathcal{H}}.
\end{eqnarray}
Recall that in the above--mentioned metric one can again introduce the radial
coordinate by changing
\begin{eqnarray}
&&{\rho^2}\rightarrow{R^2+M\,l^2},\,\mathcal{H}\rightarrow{R^2},\mathcal{W}\rightarrow\frac{J}{2R^2},
\nonumber\\
&&\mathcal{ F}(\rho)\rightarrow{F(R)
=\frac{R^2}{l^2}+M+\frac{J^2}{4\,R^2}}.
\end{eqnarray}

Since this solution possesses the
BTZ--counterpart as a limit for $\alpha=0$ , it is pertinent to search for new coordinates in
which it will become apparent the BTZ--solution counterpart
structure~(\ref{antiBTZ}).

\subsubsection{Transformation to BTZ-counterpart coordinates}

The constants and the coordinate transformations to be used in this
case are given by
\begin{eqnarray}\label{EEMom1va}
w_{0}&=&-\frac{1}{J} {\sqrt {{l}^{2}{M}^{2}-{J}^{2}} \left(
l\,M-\sqrt {{l}^{2}{M}^{2 }-{J}^{2}} \right) },\,
w_{1}=2\frac{M}{J},\nonumber\\
r&=&-\rho^2+\frac{M\,l^2}{2}+\frac{l}{2}\sqrt{l^2M^2-J^2},\nonumber\\
\phi&=&\frac{1}{\sqrt{2}}\left({l}^{2}{M}^{2}-{J}^{2}\right)^{-1/4}
\left(M\,l\,\sqrt{\frac{l}{J}}\,T+\sqrt{\frac{J}{l}}\,\Phi\right),\nonumber\\
t&=&\frac{1}{\sqrt{2}}\sqrt{\frac{l}{J}}\left({l}^{2}{M}^{2}-{J}^{2}\right)^{1/4}\,T.
\end{eqnarray}
Under these transformations the solution amounts to
\begin{widetext}
\begin{eqnarray*}\label{metricscrhob}
\bm{g}&=&-\rho^2\,\frac{F(\rho)}{H(\rho)}\bm{dT^2}+\frac{\bm{d\rho^2}}{F(\rho)}
+H(\rho)(\bm{d\Phi}+W(\rho)\bm{d\,T})^2,\nonumber\\
\end{eqnarray*}
\begin{eqnarray*}\label{metricscrhoF-mag}
F&=&{\frac {{\rho}^{2}}{{l}^{2 }}}-M+{\frac {{J}^{2}}{4{\rho}^{2}}}
+\frac{l\,\alpha^2}{2\,J\,\rho^2}\left(J^2\,l-2\,R_{(-)}\,\rho^2\right)
\ln|r|,
\end{eqnarray*}
\begin{eqnarray*}
H(\rho)&=&\frac{H_{n}}{H_{d}},\nonumber\\
H_{n}&:=&-{l}^{6}{J}^{6}{{\alpha}}^{4} \left( \ln|r|\right) ^{2}
+4\,{J}^{3}{l}^{3}\left( {l}^{2}M-{\rho} ^{2}
\right)\sqrt{l^2M^2-J^2}R_{(+)}^2\,{
\alpha}^{2}\ln|r|\nonumber\\
&& -4\, \left( {l}^{2}M-{\rho}^{2}
\right) ^{2} (l^2M^2-J^2)R_{(+)}^4,\nonumber\\
H_{d}&:=&-2{l}^{5}{J}^{3}\sqrt{l^2M^2-J^2}R_{(+)}^2
{\alpha}^{2}\ln|r|+4\,{l}^{2} \left( {l}^{2}M-{\rho}^{2}
\right)(l^2M^2-J^2) R_{(+)}^4,
\end{eqnarray*}
\begin{eqnarray}
W(\rho)&=&\frac{l^2}{J}\,\frac{\Omega(\rho)}{H_{n}},\nonumber\\
\Omega(\rho)&=&-{l}^{5}{J}^{6}R_{(+)} {{\alpha}}^{4} \left( \ln|r|
\right) ^{2}\nonumber\\
&&+ {l}^{ 2}\,{J}^{3} \left[{l}\,\left( R_{(+)}^2+2{J}^{2}\right)
\sqrt{l^2M^2-J^2} +\left({J}^{2}-R_{(+)}^2\right){\rho}^{2}\right]
R_{(+)}^2{\alpha}^{2}\ln|r|
\nonumber\\
&& +2\,{J}^{2} \left( l^2\,M^2-J^2 \right)
 \left( {\rho}^{2}-{l}^{2}M\right) R_{(+)}^4,\nonumber\\
r&:=&-\rho^2+\frac{M\,l^2}{2}+\frac{l}{2}\sqrt{l^2M^2-J^2},\,R_{(\pm)}:=M\,l\pm\sqrt{l^2M^2-J^2}.
\end{eqnarray}
\end{widetext}
The correspondence of this representation with the
BTZ--solution counterpart in the limit of vanishing electromagnetic
parameter $\alpha$ is evident.

Because of the complexity of the system of equations, we have found
very hard to determine other branches, if any, of exact solutions in
the general case.

\section{Generating stationary solutions via $ SL(2,R)$--transformations from
the static solutions}\label{GENERATING}

This section deals with $ SL(2,R)$--transformations applied on
static solutions to construct stationary cyclic symmetric classes of
solutions, namely the electric and magnetic stationary families.

\subsection{$ SL(2,R)$--transformations}

Let us consider the general metric
\begin{eqnarray*}\label{genericmetra}
\bm{g}=g_{tt}{\bm{dt}}^2+2g_{t\phi}{\bm{dt}}{\bm{d\phi}}
+g_{\phi\phi}\,{\bm{d\phi}}^2+g_{rr}{\bm{dr}}^2,
\end{eqnarray*}
and accomplish here a $SL(2,R)$ transformations of the Killingian
coordinates $t$ and $\phi$
\begin{eqnarray}\label{gentrans}
{t}&=&\alpha \tilde{t}+\beta\tilde{\phi},\, {\phi}=\gamma
\tilde{t}+\delta \tilde{\phi},\,
\Delta:=\alpha\delta-\beta\gamma\neq0 .
\end{eqnarray}
The transformed metric components are given by
\begin{eqnarray}\label{metrans}
g_{\tilde{t}\tilde{t}}&=&\alpha^2\,g_{tt}+2\alpha\gamma\,
g_{t\phi}+\gamma^2\, g_{\phi\phi},\,
g_{\tilde{t}\tilde{\phi}}=\alpha\beta\,
g_{tt}+(\alpha\delta+\beta\gamma)\,g_{t\phi}
+\gamma\delta \,g_{\phi\phi},\nonumber\\
g_{\tilde{\phi}\tilde{\phi}}&=&\beta^2\,g_{tt}+2\beta\delta\,
g_{t\phi}+\delta^2\, g_{\phi\phi},\, g_{rr}= g_{rr},
\end{eqnarray}
while under the considered transformations the electromagnetic field
tensor~(\ref{elecmag1}) becomes
\begin{eqnarray}
&&F^{\tilde{\alpha}\tilde{\beta}}= \frac {1}{\sqrt{-
\tilde{g}}}\left[
\begin {array}{ccc} 0&\tilde{b}&-\frac {\tilde{c} g_{rr}}
{\sqrt{-\tilde{g}}}\\
\noalign{\medskip}-\tilde{b}&0&\tilde{a}\\
\noalign{\medskip}\frac {\tilde{c}
g_{rr}}{\sqrt{-\tilde{g}}}&-\tilde{a}&0\end {array}
\right],\nonumber\\&&
\tilde{g}=\det(g_{\tilde{\mu}\tilde{\nu}}),
\end{eqnarray}
where the new constant are given in terms of the original ones
through
\begin{eqnarray}\label{constrel}
\tilde{a}&=&\frac{\alpha\,a+\gamma\,b}{\alpha\delta-\beta\gamma},
\,\,a=\delta \tilde{a}-\gamma \tilde{b}\nonumber\\
\tilde{b}&=&\frac{\beta\,a+\delta\,b}{\alpha\delta-\beta\gamma},
\,\,b=-\beta \tilde{a}+\alpha \tilde{b},\nonumber\\
\tilde{c}&=&\frac{c}{\alpha\delta-\beta\gamma}.
\end{eqnarray}
Notice that $$
g_{\tilde{t}\tilde{t}}g_{\tilde{\phi}\tilde{\phi}}-g_{\tilde{t}\tilde{\phi}}^2
=(g_{tt}g_{{\phi}{\phi}}-g_{t{\phi}}^2)(\alpha\delta-\beta\gamma)^2=-F\Delta^2,$$
therefore, in concrete applications it is more useful to use
normalized transformations with
$\Delta=\alpha\,\delta-\beta\,\gamma=1$.

The electromagnetic tensor occurs to be form--invariant under the
above--mentioned $SL(2,R)$--transformations if the field constants $a, b$, and
$c$ are identified based on Eq~(\ref{constrel}). This
property, on its turn, yields to the form--invariance of the
electromagnetic energy--momentum tensor
${T_{\mu}}^{\nu}=1/(4\pi)(F_{\mu\sigma}\,F^{\nu\sigma}
-1/4\delta_{\mu}^{\nu}F_{\tau\sigma}F^{\tau\sigma}),$ and
consequently to the form--invariance of the Einstein-Maxwell
equations.\\
\noindent Therefore, starting with an electromagnetic solution in
which a single electric ($b\neq0$) or magnetic ($a\neq0$) field is
present, by accomplishing the  above--mentioned $SL(2,R)$--transformations, one
can generate solutions with both electric and magnetic fields
 $\tilde{b}\neq0, \tilde{a}\neq0$ present. Conversely, if one
originally has had a solution endowed with both constant parameters
$a$ and $b$ then, via transformations, one could achieve a branch of
solutions with one single parameter. At this level, one may argue
that one deals with one specific solution in its different
coordinate representations. But there exists a second point of view
in $2+1$--gravity: to end with a new solution one has to change the
variety, i.e., the topology, requiring the ranges of change of the
new variable be, for instance, the same as the ranges of the
original variables. This procedure can be considered as a generating
solution technique and it has been used to construct stationary
solutions starting from static solutions as we shall show in the forthcoming
sections.

For the metric~(\ref{metricsc-i}), subjected to the above--mentioned
$SL(2,R)$--transformations, one gets
\begin{eqnarray}\label{tranmetric}
g_{\tilde{t}\tilde{t}}&=&-\alpha^2\,\frac{F}{H}+H\left(\alpha\,W
+\gamma\right)^2,\nonumber\\
g_{\tilde{t}\tilde{\phi}}&=&-\alpha\,\beta\,\frac{F}{H}+
H\,(\delta+\beta\,W)\,(\gamma+\alpha\,W),\nonumber\\
g_{\tilde{\phi}\tilde{\phi}}&=&-\beta^2\,\frac{F}{H}+H\left(\beta\,W
+\delta\right)^2,\nonumber\\
g_{rr}&=& \frac{1}{F},
\end{eqnarray}
hence, the expressions of the new structural functions are given in
the form
\begin{eqnarray}\label{transtrf}
{\tilde{H}}=-\beta^2\frac{F}{{H}}+{H}(\delta+\beta\,W)^2,\,
\tilde{W}\,{\tilde{H}}=-\alpha\,\beta\frac{F}{{H}}
+{H}(\delta+\beta\,W)(\gamma+\alpha\,W),\, \tilde{F}=F.
\end{eqnarray}
The transformed electromagnetic field tensor
$F^{\tilde{\mu}\tilde{\nu}}$, as it should be, exhibits its
form--invariant property
\begin{equation}\label{elecmagf2}
F^{\tilde{\mu}\tilde{\nu}}= \left[
\begin {array}{ccc} 0&\tilde{b}&-\frac {\tilde{c} }{F}\\
\noalign{\medskip}-\tilde{b}&0&\tilde{a}\\
\noalign{\medskip}\frac {\tilde{c} }{F}&-\tilde{a}&0\end {array}
\right],
\end{equation}
where as before the new field constant parameters are related with
the old ones according to Eq~(\ref{constrel}).

\subsection{Transformed electrostatic $b\neq0$ solution}

Starting with the general electrostatic Maxwell
solution~(\ref{staticq}) with metric
\begin{eqnarray}\label{staticeleCt}
g&=&-\frac{F}{H}{\bm{dt}}^2+\frac{1}{F}{\bm{dr}}^2+H{\bm{d\phi}}^2,\nonumber\\
F(r)&=&4\,\frac{H(r)}{
C_{1}^2\,{l}^{2}}\left[K_{0}+H(r)-{b}^{2}{l}^{2}\ln{H(r)}\right],\nonumber\\
{H(r)}&=&C_{1}\,r+C_{0},
\end{eqnarray}
under normalized $SL(R,2)$--transformations
\begin{eqnarray}\label{normaltrans}
t&=&
\frac{\alpha}{\sqrt{\Delta}}\tilde{t}+\frac{\beta}{\sqrt{\Delta}}\tilde{\phi},
\, \phi=
\frac{\gamma}{\sqrt{\Delta}}\tilde{t}+\frac{\delta}{\sqrt{\Delta}}\tilde{\phi},\,
\Delta=\alpha\delta-\beta\gamma\neq0.
\end{eqnarray}
(in the general (non-normalized) case the same expressions hold
except for the absence of $\Delta$, set simply $\Delta=1$), the new
metric, the rotated one, acquires the form
\begin{equation}\label{normalmatrix}
g_{\tilde{\mu}\tilde{\nu}}= \left[
\begin {array}{ccc} {-\frac{\alpha^2}{\Delta}\frac{F}{H}+\frac{\gamma^2}{\Delta}H}&{0}&
{-\frac{\alpha\beta}{\Delta}\frac{F}{H}+\frac{\delta\gamma}{\Delta}
H}
\\\noalign{\medskip}{0}&{\frac{1}{F}}&{0}
\\\noalign{\medskip}
{-\frac{\alpha\beta}{\Delta}\frac{F}{H}+\frac{\delta\gamma}{\Delta}
H}&{0}&{-\frac{\beta^2}{\Delta}\frac{F}{H}+\frac{\delta^2}{\Delta}H}\end
{array} \right]
\end{equation}
the electromagnetic field tensor becomes
\begin{equation}\label{normalfield}
F^{\tilde{\mu}\tilde{\nu}}= \left[
\begin {array}{ccc} 0&{\frac {\delta\,b}{\sqrt{\Delta}}
}&0\\\noalign{\medskip}-{\frac {\delta\,b}{\sqrt{\Delta}}}&0&{\frac
{\gamma\,b}{\sqrt{\Delta}}}
\\\noalign{\medskip}0&-{\frac {\gamma\,b}{\sqrt{\Delta}}}&0\end {array} \right],
\end{equation}
while the electromagnetic energy--momentum tensor amounts to
\begin{equation}\label{normalenergy}
{T_{\tilde{\mu}}}^{\tilde{\nu}} = \left[
\begin {array}{ccc} -\frac{1}{8\pi}\,{\frac { \left( \alpha\,\delta+\beta\,\gamma
 \right) {b}^{2}}{H \,\Delta}}
 &0&\frac{1}{4\pi}\,{\frac {\gamma\,
\alpha\,{b}^{2}}{H \, \Delta}}
 \\\noalign{\medskip}0&-\frac{1}{8\pi}
\,{\frac {{b}^{2}}{ H }}&0
\\\noalign{\medskip}-\frac{1}{4\pi}\,{\frac {\delta\,\beta\,{b}^{2}}{H \,
 \Delta }}&0&\frac{1}{8\pi}\,{\frac { \left( \alpha\,\delta+\beta\,
\gamma \right) {b}^{2}}{H \, \Delta} } \end {array}
 \right].
\end{equation}
Explicitly, the new metric is given by the non-zero components
\begin{eqnarray}\label{normalmetric}
g_{\tilde{t}\tilde{t}}&&=-\frac{\alpha^2}{\Delta}\frac{1}{H(r)\,g_{rr}}
+\frac{\gamma^2}{\Delta}\,H(r),\,
g_{\tilde{t}\tilde{\phi}}=-\frac{\alpha\beta}{\Delta}\frac{1}{H(r)\,g_{rr}}
+\frac{\delta\gamma}{\Delta}\,H(r),\nonumber\\
g_{\tilde{\phi}\tilde{\phi}}&&=-\frac{\beta^2}{\Delta}\frac{1}{H(r)\,g_{rr}}
+\frac{\delta^2}{\Delta}\,H(r),\
g_{rr}=\frac{1}{4\,H(r)}\frac{C_{1}^2\,{l}^{2}}
{\left[K_{0}+H(r)-{b}^{2}{l}^{2}\ln{H(r)}\right]},\nonumber\\
H(r)&=&C_{1}r+C_{0}.
\end{eqnarray}
For general $SL(2,R)$--transformations, with non-vanishing entries,
the electromagnetic field tensor $F^{\mu\nu}$ allows for the
presence of both electric and magnetic fields, corresponding to new
$b$ and $a$ different from zero.\\

If one accomplishes the transformation of the dependent variable $r$
to the radial (polar) coordinate $\rho$, $\rm{arc}=
\rho\,\bm{d\phi}$, one chooses
\begin{eqnarray}\label{polar}
H(r)=C_{1}\,r+C_{0}=\rho^2,\, C_{1}=2.
\end{eqnarray}

\subsubsection{Stationary electromagnetic solution}

In particular, for the $SL(2,R)$--transformation
\begin{eqnarray}
t&=&\tilde{t}-\omega\tilde{\phi},\,\phi=\tilde{\phi},\,\alpha=1,\,\beta=-\omega,
\,\gamma=0,\,\delta=1,\,\,\Delta=1,
\end{eqnarray}
one obtains a new solution, the rotated one, with metric components
\begin{eqnarray}\label{clem-b}
g_{\tilde{t}\tilde{t}}&=&-\frac{4}{
C_{1}^2\,{l}^{2}}\left[K_{0}+H(r)-{b}^{2}{l}^{2}\ln{H(r)}\right],\,
g_{\tilde{t}\tilde{\phi}}=\omega\,\frac{4}{
C_{1}^2\,{l}^{2}}\left[K_{0}+H(r)-{b}^{2}{l}^{2}\ln{H(r)}\right]
,\nonumber\\
g_{\tilde{\phi}\tilde{\phi}}&=& H(r)-\omega^2\frac{1}{H(r)\,g_{rr}}
,\, g_{rr}=\frac{1}{4\,H(r)}\frac{C_{1}^2\,{l}^{2}}
{\left[K_{0}+H(r)-{b}^{2}{l}^{2}\ln{H(r)}\right]},\nonumber\\
H(r)&=&C_{1}r+C_{0}.
\end{eqnarray}
The electromagnetic field tensor is given by
$$ F^{\tilde{\mu}\tilde{\nu}}= \left[
\begin {array}{ccc} 0&b
&0\\\noalign{\medskip}-b&0&0
\\\noalign{\medskip}0&0&0\end {array} \right],
{T_{\tilde{\mu}}}^{\tilde{\nu}} = \left[
\begin {array}{ccc} -\frac{1}{8\pi}\,{\frac { {b}^{2}}{H }}
 &0&0
 \\\noalign{\medskip}0&-\frac{1}{8\pi}
\,{\frac {{b}^{2}}{ H }}&0
\\\noalign{\medskip}\frac{1}{4\pi}\,\frac {\omega {b}^{2}}{H}
&0&\frac{1}{8\pi}\,\frac { {b}^{2}}{H}\end {array}
 \right].$$
Therefore, by means of a $SL(2,R)$--transformation applied to the
static electric cyclic symmetric $2+1$ Einstein--Maxwell solution
one can generate a unique electromagnetic stationary cyclic
symmetric solution in the sense of the structure of the field tensor
$F^{\mu\nu}$, which is equal in all respects to the electro--magnetic
solution determined by the metric (\ref{electrica}). \noindent It is
worthwhile to mention that in 1993 Clement reported a solution
belonging to this class, see ~\cite{Clement93}, Eq.~(Cl.24).

\subsubsection{Clement spinning charged BTZ solution}{\label{Clement96}}

The so--called Clement's spinning charged BTZ solution, derived in
~\cite{Clement96} deserves special attention. It arises as a result
of a $SL(2,R)$--transformation of the electrostatic solution given
in terms of the radial coordinate $\rho\rightarrow{r}$. Here the
main Clement results are reproduced in a way quite close to the
cited work.

Setting $C_{1}=2$, which is equivalent to
$t\rightarrow{t\,C_{1}/2}$, accomplishing the coordinate
transformation $h(r)=C_{1}r+C_{0}\rightarrow{r^2}$, and introducing
the definitions $r_{0}=\exp(K_{0}/(2b^2l^2))$, and
$b^2=4\pi\,G\,Q^2$, the metric~(\ref{staticeleCt}) becomes
\begin{eqnarray}\label{metricClem93}
\bm{g}&&=-{F(r)}\bm{dt^2}+\frac{\bm{dr^2}}{F(r)}
+r^2\bm{d\phi}^2,\nonumber\\
F(r)&&=\frac{K_{0}}{l^2}+\frac{r^2}{l^2}-b^2\ln{r^2}
=\frac{r^2}{l^2}-8\pi
GQ^2\ln{\frac{r}{r_0}},\nonumber\\
\bm{A}&&=2\,Q\,\sqrt{\pi\,G}\,\ln{\frac{r}{r_{0}}}\,\bm{dt}.
\end{eqnarray}
To establish the range of values of $r_0$  allowing the existence of
a black hole, let us consider $F(r)$ in the form
\begin{eqnarray}\label{Clem931bh}
F(r)&=&\frac{r^2}{l^2}(1-\frac{k}{r^2}\ln{\frac{r^2}{r_0^2}}),
\,\,k=4\pi GQ^2l^2,\nonumber\\
\end{eqnarray}
the factor $(1-\frac{k}{r^2}\ln{\frac{r^2}{r_0^2}})$ vanishes in the
set of points $r_{h}$ determined through the LambertW function, $
{\rm LambertW(x)}\exp(\rm{LambertW(x)})= {x}$, namely
\begin{eqnarray}\label{Clem931bhroot}
r_{h}^2=-k\,\rm{LambertW}(-r_0^2/k),
\end{eqnarray}
which is positive for $r_0^2=k*{\rm exp}(-1) \epsilon$,
$0<\epsilon\leq1$, or explicitly
\begin{equation}\label{Clem932}
r_{0}^2\leq4\pi GQ^2l^2/\rm{e}.
\end{equation}
Subjecting the metric (\ref{metricClem93}) and the vector potential
$\bm{A}$ to the transformation at uniform angular velocity
\begin{eqnarray}\label{Clem933}
t&\rightarrow& t-\omega\phi,\,\phi\rightarrow
\phi-\frac{\omega}{\l^2}\,t,\,\alpha=1,\,\beta=-\omega,\,
\gamma=-\frac{\omega}{\l^2},\,\delta=1,
\end{eqnarray}
one arrives at the metric
\begin{eqnarray}\label{metricClem93f}
\bm{g}&&=-(F(r)-\frac{\omega^2}{l^4}r^2)\bm{dt}^2
+2\omega(F(r)-\frac{r^2}{l^2})\bm{dt}\bm{d\phi}
+({r^2-\omega^2F(r)})\bm{d\phi}^2+\frac{\bm{dr}^2}{F(r)},\nonumber\\
F(r)&=&\frac{r^2}{l^2}-4\pi GQ^2\ln{\frac{r^2}{r_0^2}},\nonumber\\
\bm{A}=&Q&\,\sqrt{\pi\,G}\ln{\frac{r^2}{r_{0}^2}}\,\,(\bm{dt}-\omega\bm{d\phi}).
\end{eqnarray}
One could arrive at this result by using the metric components
(\ref{normalmetric}) with transformation coefficients from
(\ref{Clem933}) and setting $C_{1}=2,\,\Delta=1$.

By choosing  the axial symmetry as fundamental, the metric
(\ref{metricClem93f}) can be brought to the form
\begin{eqnarray}\label{metricClem934}
\bm{g}&=&-r^2\,(1-\omega^2/l^2)^2\,\frac{\mathcal{F}(r)}
{\mathcal{H}(r)}\bm{dt}^2+\frac{\bm{dr}^2}{\mathcal{F}(r)^2}
+\mathcal{H}(r)\left(\bm{d\phi}+{\mathcal{W}}(r)\bm{dt}\right)^2,\nonumber\\
\mathcal{F}&=&F=\frac{r^2}{l^2} -4\pi GQ^2\ln{\frac{r^2}{r_0^2}},\,
{\mathcal{W}}= \omega\,\frac{F^2-r^2/l^2}{\mathcal{H}}=-4\pi
GQ^2\frac{\omega}{\mathcal{H}}
\ln{\frac{r^2}{r_0^2}},\nonumber\\
\mathcal{H}&=&r^2-\omega^2\,F=r^2(1-\frac{\omega^2}{l^2})+\omega^2\,4\pi
GQ^2\ln{\frac{r^2}{r_0^2}}.
\end{eqnarray}
The Clement spinning charged BTZ solution is endowed with three
parameters $Q$, $r_{0}$, and $\omega$. It allows for a black hole
interpretation.

Alternatively, introducing the scaling transformation
$r={l}/{\bar{l}}\times\bar{r}$,  the definitions
$\bar{l}^2=l^2-\omega^2,\,|\omega|<l$, and
${\bar{r}_{0}}={\bar{l}}/{l}\times{{r}_{0}}$, the proper Clement
solution, dropping the bar from $r$, is given as
\begin{eqnarray}\label{metricClem93et}
\bm{g}&=& -r^2\,\frac{F(r)}{H(r)}\bm{dt}^2+\frac{\bm{dr}^2}{F(r)}
+H(r)[\bm{d\phi}+W(r)\bm{dt}]^2,\nonumber\\
F(r)&=&\frac{r^2}{l^2}-\frac{{l}^2-\omega^2} {l^2}8\pi
GQ^2\ln{\frac{r}{\bar{r}_{0}}},\, W(r)=-\frac{\omega}{H(r)}8\pi
GQ^2\ln{\frac{r}{\bar{r}_{0}}},\nonumber\\
H(r)&=&r^2+\omega^28\pi GQ^2\ln{\frac{r}{\bar{r}_{0}}},\nonumber\\
\bm{A}&=&2\,Q\,\sqrt{\pi\,G}\ln{\frac{r}{\bar{r}_{0}}}\,\,(\bm{dt}-\omega\bm{d\phi}).\nonumber\\
\end{eqnarray}
The corresponding electromagnetic fields are given by
\begin{eqnarray}\label{Clem936t}
F_{\mu\nu}&=&-\frac{4\,Q}{r}\,\sqrt{\pi\,G}({\delta_{[\mu}}^{t}{\delta_{\nu]}}^{r}
-\omega{\delta_{[\mu}}^{\phi}{\delta_{\nu]}}^{r}),\nonumber\\
{T_{\mu}}^{\nu}&=&-\frac{G\,Q^2}{2\,l^2}\frac{l^2+\omega^2}{r^2}{\delta_{\mu}}^{t}{\delta_{t}}^{\nu}
-\frac{G\,Q^2}{\,l^2}\frac{\omega}{r^2}{\delta_{\mu}}^{t}{\delta_{\phi}}^{\nu}
+\frac{G\,Q^2}{2\,l^2}\frac{l^2+\omega^2}{r^2}{\delta_{\mu}}^{\phi}{\delta_{\phi}}^{\nu}
+\frac{G\,Q^2\omega}{r^2}{\delta_{\mu}}^{\phi}{\delta_{t}}^{\nu}
\nonumber\\
&&
-\frac{G\,Q^2}{2\,l^2}\frac{l^2-\omega^2}{r^2}{\delta_{\mu}}^{r}{\delta_{r}}^{\nu}.
\end{eqnarray}
The length re--scaling was chosen in such a manner that $H\rightarrow
r^2$, at spatial infinity and $
\bm{g}(\ref{metricClem93et})\rightarrow \bm{g}_{\rm{BTZ}}(Q=0).$
According to Clement: one may formally define mass and angular
momentum parameters $M(r_1)$ and $J(r_1)$ by identifying, at a given
scale $r=r_1$, the values of the structural functions with the
corresponding uncharged BTZ values. Nevertheless, the mass and
angular momentum defined in this way occur to be $r_1$--dependent
and diverge logarithmical as $r_1\rightarrow\infty$. This solution
is a black hole if the condition of the form (\ref{Clem932}),
\begin{equation}\label{Clem936}
\bar{r}_{0}^2\leq4\pi GQ^2\bar{l}^2/\rm{e},
\end{equation}
is fulfilled. It possesses two horizons, at which $F(r)$ vanishes,
which are roots of the relation
\begin{equation}\label{Clem937}
{r^2}-{\bar{l}^2}8\pi GQ^2\ln{\frac{r}{\bar{r}_{0}}}=0,
\end{equation}
which are given by the $\rm{LambertW}$ function, see
Eq.~(\ref{Clem931bhroot}). The largest root determines the event
horizon at $r=r_{+}=r_{h}$, while the inner one is a Cauchy horizon
at $r=r_{-}$, with $r_{+}>r_{-}>\bar{r}_{0}$. Since the metric
function $H$ changes sign for a certain value $r=r_{c}<\bar{r}_{0}$,
similarly as the rotating uncharged BTZ solution, thus there are
closed time--like curves in the region inside the radius $r_{c}$. It
is apparent that the metric and the electromagnetic field are
singular at $r=0$.

\subsubsection{Kamata--Koikawa limit}

It should be pointed out that Clement \cite{Clement93} also reported
the so called self--dual solution published later in
\cite{KamataK95}. By accomplishing the limiting transition
$\omega\rightarrow \pm l\Rightarrow \bar{l} \rightarrow 0,$ of the
metric structural functions~(\ref{metricClem93et}), while the other
parameters $Q$ and $\bar{r}_{0}$ remain fixed, one arrives then at
the metric~(\ref{metricClem93et}) with structural functions and
vector field
\begin{eqnarray}\label{ClementKK}
F&=&\frac{r^2}{l^2},\, W=\mp\frac{l}{H}\,8\pi G
Q^2\ln{\frac{r}{\bar{r}_{0}}},\, H=r^2+{l^2}\,8\pi G
Q^2\ln{\frac{r}{\bar{r}_{0}}},\nonumber\\
\bm{A}&=&Q\ln{\frac{r}{\bar{r}_{0}}}\,\,(\bm{dt}\mp
\,l\,\bm{d\phi}),
\end{eqnarray}
Notice that this solution does not possess horizon; in the limiting
transition $\bar {l}\rightarrow 0$ the horizon does not survive
since it disappears below $\bar {l}=(4\pi GQ^2)^{1/2}$, as quoted by
Clement.

The proper KK representation of this one--parameter solution is
achieved by accomplishing the radial transformation and scaling of
parameters
\begin{eqnarray}\label{ClementKK1}
r^2&=&r_{KK}^2-r_{0KK}^2,\, r_{0KK}=(4\pi GQ^2l^2)^{1/2},\,
r_{0}=r_{0KK}/\sqrt{\rm e},
\end{eqnarray}
and the subscripts are self--explanatory. It is worthwhile also to notice
that a derivation and analysis of the KK solution has been
accomplished in~\cite{CataldoS96} too.

\subsubsection{Mart\'\i nez--Teitelboim--Zanelli
solution}{\label{MTZs}}

Mart\'\i nez, Teitelboim and Zanelli, see \cite{MartinezTZ00},
reported a generalization of the BTZ black hole spacetime equipped
with an electric charge $Q$, the mass $M$ and the angular momentum
$J$. The main features of this charged black hole, among others,
following the quoted paper, are: the total $M, J$ and $Q$ which are
boundary terms at infinity, the extreme black hole can be thought of
as a particle moving with the speed of light, and the inner horizon
of the rotating uncharged black hole is unstable under the
perturbation of a small electric charge. According to the quoted
reference, this charged electrically black hole is pathological in
the sense it exists for arbitrary values of the mass and that
there is no upper bound  on the electric charge.

The starting point is the electrostatic metric~(\ref{staticeleCt})
given in terms of the polar coordinate $r$,
\begin{eqnarray}
&&H(r)=C_{1}r+C_{0}\rightarrow{r^2}, C_{1}=2,\,
K_{0}/l^2\rightarrow{-\tilde{M}},b^2\rightarrow{\frac{1}{4}{\tilde{Q}}^2},
\end{eqnarray}
therefore
$F(r)=\frac{r^2}{l^2}-\tilde{M}-\frac{1}{4}{\tilde{Q}}^2\ln{r^2}.$
Using the metric components (\ref{normalmetric}) with
transformation coefficients from the ``rotation boost''
transformation
\begin{eqnarray}
t&\rightarrow&\frac{1}{\sqrt{1-\omega^2/l^2}}\left(t-\omega\phi\right),\,
\phi\rightarrow\frac{1}{\sqrt{1-\omega^2/l^2}}
\left(\phi-\frac{\omega}{l^2}\,t\right),
\end{eqnarray}
one arrives at the metric
\begin{eqnarray}\label{MTZ0}
\bm{g}&=&-\left[\frac{r^2}{\l^2}-\frac{1}{1-\omega^2/l^2}
(\tilde{M}+\frac{{\tilde{Q}}^2}{4}\ln{r^2})\right]\bm{dt}^2
 - 2\frac{\omega}{1-\omega^2/l^2}(\tilde{M}
+\frac{{\tilde{Q}}^2}{4}\ln{r^2})\bm{dt}\bm{d\phi}\nonumber\\
&+&\left[r^2+\frac{\omega^2}{1-\omega^2/l^2}(\tilde{M}+
\frac{{\tilde{Q}}^2}{4}\ln{r^2})\right]\bm{d\phi}^2
+\frac{\bm{dr}^2}{r^2/l^2-\tilde{M}-\frac{1}{4}{\tilde{Q}}^2\ln{r^2}}.
\end{eqnarray}
This metric can be brought to the form
\begin{eqnarray}\label{MTZx}
\bm{g}&=&-r^2\,\frac{{F}(r)}
{{H}(r)}\bm{dt}^2+\frac{\bm{dr}^2}{{F}(r)}
+{H}(r)\left(\bm{d\phi}+{{W}}(r)\bm{dt}\right)^2,\nonumber\\
{F(r)}&=&\frac{r^2}{l^2}-\tilde{M}-\frac{1}{4}{\tilde{Q}}^2\,\ln{r^2},\,
{W(r)}= -\frac{\omega}{1-\omega^2/l^2}\,\frac{\tilde{M}
+\frac{1}{4}{\tilde{Q}}^2\,\ln{r^2}}{{H}(r)},\nonumber\\
{H(r)}&=&r^2+\frac{\omega^2}{1-\omega^2/l^2}(\tilde{M}
+\frac{1}{4}{\tilde{Q}}^2\,\ln{r^2}).
\end{eqnarray}
The electromagnetic field tensor is given by
\begin{eqnarray}\label{MTZx12}
F_{\mu\nu}=\frac{\tilde{Q}}{{r}\sqrt{1-\omega^2/l^2}}
\left({\delta_{[\mu}}^{t}{\delta_{\nu]}}^{r}
+{\omega\,l^2}{\delta_{[\mu}}^{t}{\delta_{\nu]}}^{\phi}\right).
\end{eqnarray}
The angular momentum, charge, and mass can be evaluated via
quasilocal definitions, see Section~\ref{Quasilocal}; the presence
of logarithmic terms in the structural metric functions yields to
divergences at infinity of the energy--momentum quantities. As
pointed out by the authors, the divergence in the mass can be
handled by enclosing the system in a large circle of radius $r_0$ in
which will be bound $M(r_0)$--the energy within $r_0$--and the
electrostatic energy outside $r_0$ given by $-Q^2 \ln{r_0}/2$, thus
the total mass (independent of $r_0$ and finite) is given by
$\tilde{M}=M(r_0)-Q^2 \ln{r_0}/2$.

\subsection{Transformed magnetostatic solution $a\neq0$ solution}

To determine the stationary rotating generalization of the
magnetostatic metric (\ref{eqta7a})
\begin{eqnarray*}\label{staticeleC}
g&=&-\frac{F}{H}dt^2+\frac{1}{F}dr^2+H{d\phi}^2,\nonumber\\
H(r)&=&\frac{4}{
C_{1}^2\,{l}^{2}}\left[K_{0}+h(r)+{a}^{2}{l}^{2}\ln{h(r)}\right],\,
F(r)= {H(r)}\,h(r),\,h(r):=C_{1}\,r+C_{0},
\end{eqnarray*}
one subjects it to $SL(2,R)$--transformations
\begin{eqnarray*}
t&=&
\frac{\alpha}{\sqrt{\Delta}}\tilde{t}+\frac{\beta}{\sqrt{\Delta}}\tilde{\phi},\,
\phi=
\frac{\gamma}{\sqrt{\Delta}}\tilde{t}+\frac{\delta}{\sqrt{\Delta}}\tilde{\phi},
\,\,\Delta=\alpha\delta-\gamma\beta,
\end{eqnarray*}
giving rise to the rotated new metric in the form
\begin{eqnarray*}
g_{\tilde{\mu}\tilde{\nu}}= \left[
\begin {array}{ccc} {-\frac{\alpha^2}{{\Delta}}\frac{F}{H}
+\frac{\gamma^2}{{\Delta}}H}&{0}&
{-\frac{\alpha\beta}{{\Delta}}\frac{F}{H}+\frac{\gamma\delta}{{\Delta}}
H}
\\\noalign{\medskip}{0}&{\frac{1}{F}}&{0}
\\\noalign{\medskip}
{-\frac{\alpha\beta}{{\Delta}}\frac{F}{H}+\frac{\gamma\delta}{{\Delta}}
H}&{0}&{-\frac{\beta^2}{{\Delta}}\frac{F}{H}+\frac{\delta^2}{{\Delta}}H}\end
{array} \right],
\end{eqnarray*}
which is accompanied with the electromagnetic field tensor
\begin{eqnarray}
F^{\tau\sigma}=\left[
\begin{array}{ccc}0&a\frac{\beta}{\sqrt\Delta}&0
\\\noalign{\medskip}-a\frac{\beta}{\sqrt\Delta}&0&a\frac
{\alpha}{\sqrt\Delta}
\\\noalign{\medskip}0&-a\frac{\alpha}{\sqrt\Delta}&0\end{array}\right].
\end{eqnarray}
The corresponding Maxwell energy--momentum tensor becomes
\begin{eqnarray}
{T_{\tilde{\mu}}}^{\tilde{\nu}}
= \left[
\begin {array}{ccc} -{\frac { \left( \alpha\,\delta+\beta\,\gamma
 \right) {a}^{2}}{{8\pi}\Delta}}\frac{H}{F }
 &0&{\frac {\gamma\,
\alpha\,{a}^{2}}{ 4\pi\Delta}}\frac{H }{F}
 \\\noalign{\medskip}0&
\,{\frac {{a}^{2} H}{8\pi\,F }}&0
\\\noalign{\medskip}-{\frac {\delta\,\beta\,{a}^{2}}{
 4\pi\Delta }}\frac{H}{F }&0&{\frac { \left( \alpha\,\delta+\beta\,
\gamma \right) {a}^{2}}{8\pi\, \Delta}\frac{H}{F } } \end {array}
 \right].\nonumber\\
 \end{eqnarray}
Explicitly, the non-zero metric components are
\begin{eqnarray}\label{magmetrici}
g_{\tilde{t}\tilde{t}}&=&-\frac{\alpha^2}{\Delta}\,h(r)
+\frac{\gamma^2}{\Delta}\frac{1}{h(r)\,g_{rr}},\,
g_{\tilde{t}\tilde{\phi}}=-\frac{\alpha\beta}{\Delta}\,h(r)
+\frac{\delta\gamma}{\Delta}\frac{1}{h(r)\,g_{rr}}
,\nonumber\\
g_{\tilde{\phi}\tilde{\phi}}&=&-\frac{\beta^2}{\Delta}\,h(r)
+\frac{\delta^2}{\Delta}\,\frac{1}{h(r)\,g_{rr}},\,
g_{rr}=\frac{1}{4\,h(r)}\frac{C_{1}^2\,{l}^{2}}{\left[K_{0}+h(r)
+{a}^{2}{l}^{2}\ln{h(r)}\right]},\nonumber\\
h(r)&=& C_{1}r+C_{0}.
\end{eqnarray}

\subsubsection{Stationary magneto--electric solution}
In particular, for the $SL(2,R)$--transformation
\begin{eqnarray}
t&=&\tilde{t},\, \phi=-\frac{\omega}{l^2}\tilde{t}+\tilde{\phi},\,
\alpha=1,\,\beta=0,\,\gamma=-\frac{\omega}{l^2},\,\delta=1,
\end{eqnarray}
one obtains a new solution with metric components
\begin{eqnarray}\label{magmetricx}
g_{\tilde{t}\tilde{t}}&=&-h(r)+\frac{\omega^2}{l^4}\frac{1}{h(r)\,g_{rr}}
,\,
g_{\tilde{t}\tilde{\phi}}=-\frac{\omega}{l^2}\,\frac{1}{h(r)\,g_{rr}}
,\nonumber\\
g_{\tilde{\phi}\tilde{\phi}}&=& \frac{1}{h(r)\,g_{rr}} ,\,
g_{rr}=\frac{1}{4\,h(r)}\frac{C_{1}^2\,{l}^{2}}
{\left[K_{0}+h(r)+{a}^{2}{l}^{2}\ln{h(r)}\right]},\nonumber\\
h(r)&=&C_{1}r+C_{0},
\end{eqnarray}
which is  equal to the constant $W=\omega$ stationary magneto-electric
solution (\ref{metricscrho}). Therefore, by means of a $SL(2,R)$--transformation
applied to the magnetostatic cyclic symmetric
$(2+1)$ Einstein--Maxwell solution one can generate a unique
electromagnetic stationary cyclic symmetric solution in the sense of the
structure of the field tensor $F^{\mu\nu}$.

Clement reported a field belonging to this class of solutions, see
~\cite{Clement93}, Eq.~(Cl.23).

\subsubsection{Dias--Lemos
magnetic BTZ--solution counterpart}\label{McscBTZs}

Dias and Lemos \cite{Dias-Lemos-JHEP02} published a rotating
magnetic solution in 2+1 gravity--the magnetic counterpart of the
spinning charged BTZ solution, i.e., a point source generating a
magnetic field. Also, It was established that both the static and
rotating magnetic solutions possess negative mass and that there is
an upper bound for the intensity of the magnetic field
source and for the value of the angular momentum.\\

A simple representation of this solution can be achieved from our
transformed magnetic metric (\ref{magmetrici}) by setting
\begin{eqnarray*}
t&=&\tilde{t},\,
\phi=-\frac{\omega}{l^2}\tilde{t}+\tilde{\phi},\,\alpha=1,\,\beta=-\omega,\,\gamma=
-\frac{\omega}{l^2},\,\delta=1,\,\,\Delta=1,
\nonumber\\
C_{1}&=&2,\,h(r)=C_{1}r+C_{0}\rightarrow{r^2}
\end{eqnarray*}
obtaining
\begin{eqnarray}\label{diaslemos02}
\bm{g}&=&-(r^2-\frac{\omega^2}{l^4}F)\bm{dt}^2+({F-\omega^2\,r^2})\bm{d\phi}^2
-2\omega(\frac{F}{l^2}-r^2)\bm{dt}\bm{d\phi}
+\frac{\bm{dr}^2}{F(r)},\nonumber\\
F&=&\frac{K_{0}}{l^2}+\frac{r^2}{l^2}+a^2\ln{r^2}.
\end{eqnarray}
The proper Dias--Lemos representation uses a more involved
definitions of the transformed $r$--coordinate and parameterizations
of the $SL(2,R)$--transformations:
\begin{eqnarray*}
h(r)&=&C_{1}r+C_{0}\rightarrow(\rho^2+r_{+}^2-ml^2)/l^2,
\,C_{1}=2/l^2,\,\chi^2=a^2\,l^4,
\end{eqnarray*}
\begin{eqnarray*}
t&=&\sqrt{1+\omega^2}\tilde{t}-l\,\omega\tilde{\phi},\nonumber\\
\phi&=&-\frac{\omega}{l}\tilde{t}+\sqrt{1+\omega^2}\,\tilde{\phi},\nonumber\\
\end{eqnarray*}
dropping tildes, one has
\begin{eqnarray}\label{metricDL}
&&\bm{g}=-\left[h -\frac{\omega^2}{l^2}\left(m\,l^2
+\chi^2\ln|h|\right)\right]\bm{dt^}2
-2\frac{\omega}{l}\sqrt{\omega^2+1}\left[m\,l^2
+\chi^2\ln|h|\right]\bm{dt}\,{\bm {d\phi}}\nonumber\\
&&+[h\,l^2+(\omega^2+1)\left(\,m\,l^2 +\chi^2\ln|h|\right)]{\bm
{d\phi}}^2 +\frac{l^2\,\rho^2\,\bm{d\rho}^2}{(\rho^2+r_{+}^2-ml^2)
[\rho^2+r_{+}^2+\chi^2\ln|h|]},\nonumber\\
&&h=(\rho^2+r_{+}^2-ml^2)/l^2,
\end{eqnarray}
with vector potential
\begin{eqnarray}\label{fieldDL}
\bm{A}&=&\frac{1}{2}\chi\ln|h(\rho)|[-\frac{\omega}{l}\bm
{dt}+\sqrt{1+\omega^2}\,\bm {d\Phi}].\nonumber\\
\end{eqnarray}
Notice that in the above--mentioned representation, the equation
\begin{eqnarray}\label{horDL}
r_{+}^2+\chi^2\ln|(\rho^2+r_{+}^2)/l^2-m|=0,
\end{eqnarray}
used in the Eq.~(3.2) of \cite{Dias-Lemos-JHEP02} was not used here. According to
these authors, this rotating magnetic spacetime is null and time--like
geodesically complete and as such horizonless.

\subsection{Transformed hybrid static $c\neq0$ solution}

To determine the stationary rotating generalization of the hybrid
static solution (\ref{hybridstatic})
\begin{eqnarray*}\label{hybridstaticx}
\bm{g}&=&-\frac{F}{H}{\bm{dt}}^2+
\frac{1}{F}{\bm{d{r}}}^2+H\,{\bm{d\phi}}^2,\nonumber\\
F&=&\frac{4}{l^2}(r-r_{1})(r-r_{2}),\, H=\frac{2}{l}\,K_{0}^2\,
(r-r_{1})^{(1\pm\sqrt{\alpha_{0}})/2}(r-r_{2})^{(1\mp\sqrt{\alpha_{0}})/2},\nonumber\\
\bm{A}&=& \frac{c}{2}(t\bm{d\phi}-\phi \bm{dt}),
\end{eqnarray*}
one subjects it to the $SL(2,R)$--transformations
\begin{eqnarray*}
t&=& \frac{\alpha}{\sqrt{\Delta}}\tilde{t}
+\frac{\beta}{\sqrt{\Delta}}\tilde{\phi}, \phi=
\frac{\gamma}{\sqrt{\Delta}}\tilde{t}
+\frac{\delta}{\sqrt{\Delta}}\tilde{\phi},
\,\,\Delta=\alpha\delta-\gamma\beta,
\end{eqnarray*}
arriving at the stationary rotating solution, omitting tildes, given
by
\begin{eqnarray}\label{hybridstaticxs}
\bm{g}&=&-({\frac{\alpha^2}{{\Delta}}\frac{F}{H}
-\frac{\gamma^2}{{\Delta}}H}){\bm{dt}}^2
+2({-\frac{\alpha\beta}{{\Delta}}\frac{F}{H}
+\frac{\gamma\delta}{{\Delta}} H}){\bm{dt}}{\bm{d\phi}}
+({-\frac{\beta^2}{{\Delta}}\frac{F}{H}
+\frac{\delta^2}{{\Delta}}H}){\bm{d\phi}}^2
+\frac{1}{F}{\bm{d{r}}}^2,\nonumber\\
F&=&\frac{4}{l^2}(r-r_{1})(r-r_{2}), \, H=\frac{2}{l}\,K_{0}^2\,
(r-r_{1})^{(1\pm\sqrt{\alpha_{0}})/2}(r-r_{2})^{(1\mp\sqrt{\alpha_{0}})/2},\nonumber\\
\bm{A}&=&\frac{c}{2}(t\bm{d\phi}-\phi \bm{dt}).
\end{eqnarray}
In the seed hybrid static metric has been used $\alpha_{0}$ instead
of the original $\alpha$ to avoid confusion with the parameters
appearing in the $SL(2,R)$--transformations.

Therefore the transformed electromagnetic tensors stay unchanged
\begin{eqnarray*}
&&F_{\mu\nu}={2c} {\delta_{[\mu}}^{t}{\delta_{\nu]}}^{\phi},
\nonumber\\ &&{T_{\mu}}^{\nu} =\frac{1}{8\pi}\frac{c^2}{F}
(-{\delta_{\mu}}^{t}{\delta_{t}}^{\nu}
+{\delta_{\mu}}^{r}{\delta_{r}}^{\nu}
-{\delta_{\mu}}^{\phi}{\delta_{\phi}}^{\nu}).
\end{eqnarray*}
Hence, by means of a $SL(2,R)$--transformation applied to the hybrid
static cyclic symmetric $2+1$ Einstein--Maxwell solution, one
generates a family of hybrid, stationary cyclic symmetric solutions
with structurally unique field tensors $F_{\mu\nu}$ and
${T_{\mu}}^{\nu}$.

\subsubsection{Cataldo azimuthal
rotating solution}{\label{CAThybridSt}} Applying the above--mentioned $SL(2,R)$
transformation to the Cataldo azimuthal electrostatic
solution~(\ref{metricsc3}) one arrives at a new stationary solution
having the static BTZ solution as a limit, namely
\begin{eqnarray}\label{hybridstaticxd}
\bm{g}&&=-(\frac{\alpha^2}{\Delta}\,H_{(-)}
-\frac{\gamma^2}{\Delta}\,H_{(+)}){\bm{dt}}^2+2(-\frac{\alpha\beta}{{\Delta}}H_{(-)}
\nonumber\\&&+\frac{\gamma\delta}{{\Delta}}\,H_{(+)}){\bm{dt}}{\bm{d\phi}}
+({-\frac{\beta^2}{{\Delta}}H_{(-)}
+\frac{\delta^2}{\Delta}H_{(+)}}){\bm{d\phi}}^2 +
\left(\frac{\rho^2}{l^2}-M\right)^{-1}{\bm{d{r}}}^2,\nonumber\\
&&H_{(+)}:= \rho^{1+\sqrt{\alpha_{0}}}
\left(\frac{\rho^2}{l^2}-M\right)^{(1-\sqrt{\alpha_{0}})/2},\,
H_{(-)}:= \rho^{1-\sqrt{\alpha_{0}}}
\left(\frac{\rho^2}{l^2}-M\right)^{(1+\sqrt{\alpha_{0}})/2}.\nonumber\\
\end{eqnarray}
The electromagnetic tensors are given by
\begin{eqnarray}
F_{\mu\nu}&&=\frac{M\sqrt{1-\alpha_{0}}}{2}
\,\,{\delta_{[\mu}}^{t}{\delta_{\nu]}}^{\phi},\nonumber\\{T_{\mu}}^{\nu}
&&=\frac{M^2}{32\pi\rho^2}\frac{(1-\alpha_{0})}{\rho^2/l^2-M}
(-{\delta_{\mu}}^{t}{\delta_{t}}^{\nu}
+{\delta_{\mu}}^{\rho}{\delta_{\rho}}^{\nu}
-{\delta_{\mu}}^{\phi}{\delta_{\phi}}^{\nu}).
\end{eqnarray}
In particular, one could choose the rotation boost transformation
\begin{eqnarray*}
&&
t\rightarrow\frac{1}{\sqrt{1-\omega^2/l^2}}\left(t-\omega\phi\right),\,
\, \phi\rightarrow\frac{1}{\sqrt{1-\omega^2/l^2}}
\left(\phi-\frac{\omega}{l^2}\,t\right),
\end{eqnarray*}
where the
parameter $\omega$ can be related to the angular momentum constant.

\section{Concluding remarks}\label{Remarks}
In the framework of the (2+1)--dimensional Einstein--Maxwell theory
with cosmological constant different families of exact solutions for
cyclic symmetric stationary (static) metrics have been derived. For
the static classes and hybrid, static and stationary as well,
families their uniqueness is proven by the used integration
procedure. The completeness and relationship of all uniform
electromagnetic $F_{\mu\mu;\alpha}=0$, and constant invariant
$F_{\mu\nu}F^{\mu\nu}=2 \gamma$ solutions is achieved. The
uniqueness of the stationary families of solutions has been
partially established;  various specific branches of solutions in
the general case are determined via a straightforward integration.
In this systematic approach all known electromagnetic stationary
cyclic symmetric solutions are properly identified. It seems to be a
rule that electrically charged solutions allow for a black
interpretation while for the magnetic classes such black hole
feature seems to be absent; a research on this respect is undertaken
and will be published elsewhere.

\section*{Acknowledgments}\label{Acknowledge}
This work has been partially supported by UC MEXUS--CONACyT for
Mexican and UC Faculty Fellowships 09012007, and grants CONACyT
57195, and 82443. The author thanks S. Carlip and acknowledges the
hospitality of the Physics Department, University of California,
Davis, at which most of the related work has been accomplished.

\end{document}